\documentclass[fleqn,usenatbib]{mnras}

\usepackage[T1]{fontenc}
\usepackage{ae,aecompl}

\usepackage{newtxtext,newtxmath}

\usepackage{graphicx}
\usepackage{amsmath}
\usepackage{amssymb}
\usepackage{aas_macros}
\usepackage{color}
\usepackage{hyperref}

\title[The structure of PSBs at $0.5 < z < 2$]{The structure of post-starburst galaxies at $0.5 < z < 2$:
evidence for two distinct quenching routes at different epochs}

\author[D.~T.~Maltby et al.]
{David~T.~Maltby,$^{1}$\thanks{E-mail: david.maltby@nottingham.ac.uk}
 Omar~Almaini,$^{1}$ Vivienne~Wild,$^{2}$ Nina~A.~Hatch,$^{1}$
 \newauthor William~G.~Hartley,$^{3}$ Chris~Simpson,$^{4}$ Kate~Rowlands$^{5}$ and Miguel~Socolovsky$^{1}$\\
$^{1}$School of Physics and Astronomy, University of Nottingham, University Park, Nottingham NG7 2RD, UK\\
$^{2}$School of Physics and Astronomy, University of St Andrews, North Haugh, St Andrews KY16 9SS, UK\\
$^{3}$Department of Physics and Astronomy, University College London, 3rd Floor, 132 Hampstead Road, London NW1 2PS, UK\\
$^{4}$Gemini Observatory, Northern Operations Center, 670 N.~A`ohoku Place, Hilo, HI 96720, USA\\
$^{5}$Department of Physics and Astronomy, Johns Hopkins University, Bloomberg Center, 3400 N.\ Charles St., Baltimore, MD 21218, USA
\vspace{-0.4cm}}

\begin{document}

\date{Accepted 2018 June 28. Received 2018 June 07; in original form 2017 December 07}

\pagerange{\pageref{firstpage}--\pageref{lastpage}} \pubyear{0000}

\maketitle

\label{firstpage}

%%%%%%%%%%%%%%%%%%%%%%%%%%%%%%%%%%%%%%%%%%%%%%%%%%%%%%%%%%%%%%%%%%%%%%%%%%%%%%%%%%%%%%%%%%%%%%%%%%%%%%%%%%%%%

\begin{abstract}
We present an analysis of the structure of post-starburst (PSB) galaxies in the redshift range $0.5 < z < 2$,
using a photometrically-selected sample identified in the Ultra Deep Survey (UDS) field.  We examine the
structure of $\sim80$ of these transient galaxies using radial light $\mu(r)$ profiles obtained from CANDELS
{\em Hubble Space Telescope} near-infrared/optical imaging, and compare to a large sample of $\sim2000$
passive and star-forming galaxies.  For each population, we determine their typical structural properties
(effective radius~$r_{\rm e}$, S{\'e}rsic index~$n$) and find significant differences in PSB structure at
different epochs.  At high redshift ($z > 1$), PSBs are typically massive ($M_* > 10^{10}\rm\,M_{\odot}$),
very compact and exhibit high S{\'e}rsic indices, with structures that differ significantly from their
star-forming progenitors but are similar to massive passive galaxies.  In~contrast, at lower redshift
($0.5 < z < 1$), PSBs are generally of low mass ($M_* < 10^{10}\rm\,M_{\odot}$) and exhibit compact but less
concentrated profiles (i.e.~lower S{\'e}rsic indices), with structures similar to low-mass passive discs.
Furthermore, for both epochs we find remarkably consistent PSB structure across the optical/near-infrared
wavebands (which largely trace different stellar populations), suggesting that any preceding starburst and/or
quenching in PSBs was not strongly centralized.  Taken together, these results imply that PSBs at $z > 1$
have been recently quenched during a major disruptive event (e.g.\ merger or protogalactic collapse) which
formed a compact remnant, while at $z < 1$ an alternative less disruptive process is primarily responsible.
Our results suggest that high-$z$ PSBs are an intrinsically different population to those at lower redshifts,
and indicate different quenching routes are active at different epochs.
\end{abstract}

\begin{keywords}
galaxies: evolution --- galaxies: fundamental parameters --- galaxies: high-redshift --- galaxies: structure
\vspace{-0.8cm}
\end{keywords}

%%%%%%%%%%%%%%%%%%%%%%%%%%%%%%%%%%%%%%%%%%%%%%%%%%%%%%%%%%%%%%%%%%%%%%%%%%%%%%%%%%%%%%%%%%%%%%%%%%%%%%%%%%%%%
\section[]{Introduction}

\label{Introduction}

%% Motivation %%-------------------------------------------------------------------------------------------%%
In the local Universe, a strong bimodality is observed in several galaxy properties, e.g.\ optical colour,
morphology and spectral-type \citep[e.g.][]{Strateva_etal:2001}.  In general, massive galaxies tend to be
red, passive and of early-type morphology (i.e.\ elliptical, S0), while lower mass galaxies tend to be blue,
star-forming and of late-type morphology (i.e.\ spiral).  These two populations are now commonly called the
{\em red-sequence} and the {\em blue cloud}, respectively.  In recent years, large-scale photometric surveys
have enabled the evolution of this bimodality and the formation/build-up of the red-sequence to be traced out
to $z > 2$ \citep[e.g.][]{Bell_etal:2004, Cirasuolo_etal:2007, Faber_etal:2007, Brammer_etal:2011,
Muzzin_etal:2013}.  However, despite significant progress, we still do not fully understand how star
formation is quenched at high redshift, as required to transfer galaxies from the blue cloud on to the
red-sequence.

%% Quenching mechanisms %%---------------------------------------------------------------------------------%%
Although the principal drivers for quenching star formation in galaxies remain uncertain, various physical
mechanisms have been proposed.  For example, the stripping of the interstellar medium \citep[e.g.][]
{Gunn&Gott:1972}, gas-removal by AGN or starburst-driven superwinds \citep[e.g.][]{Silk&Rees:1998,
Hopkins_etal:2005,Diamond-Stanic_etal:2012}, or an exhaustion of the gas supply via strangulation
\cite[e.g.][]{Larson_etal:1980}.  Other possible processes include morphological quenching \citep[e.g.][]
{Martig_etal:2009}, and the shock heating of infalling cold gas by the hot halo \citep[e.g.][]
{Dekel&Birnboim:2006}.  Furthermore, in addition to these `initial' quenching processes, radio-mode AGN
feedback may also be required to prevent further gas accretion and keep star formation suppressed on longer
timescales \citep{Best_etal:2005, Best_etal:2006}.

%% Structural transformations %%---------------------------------------------------------------------------%%
For massive galaxies, the quenching of star formation is also accompanied by a significant evolution in their
structural properties.  Massive galaxies at high redshift ($z\sim2$) are typically disc-dominated, while in
the local Universe they are generally spheroidal \citep[e.g.][]{vanderWel_etal:2011, Buitrago_etal:2013}.
This structural transition appears to occur at $z > 1$ for most galaxies with $M_* > 10^{10.5}\rm\,M_{\odot}$
\citep{Mortlock_etal:2013}.  However, it is currently unclear whether this transition occurs during the same
event that quenched the star formation.  In addition, massive passive galaxies in the early Universe also
appear to be significantly more compact than their local counterparts \citep[e.g.][]{Trujillo_etal:2006}.
This implies a dramatic size growth via e.g.\ minor mergers \citep{Naab_etal:2009}, although other scenarios
are possible \citep[e.g.\ progenitor bias;][]{Carollo_etal:2013}.  Possible mechanisms for the formation of
these compact high-$z$ galaxies (i.e.\ {\em red nuggets}) include: i)~central starbursts triggered by either
a gas-rich merger \citep{Hopkins_etal:2009, Wellons_etal:2015} or dissipative `protogalactic collapse'
\citep{Dekel_etal:2009, Zolotov_etal:2015}, which is followed by a rapid quenching through e.g.\ AGN or
starburst-driven superwinds \citep[e.g.][]{Hopkins_etal:2005}; and ii)~a formation at very early times when
the Universe itself was much denser \citep{Wellons_etal:2015}.

%% Post-starburst galaxies %%------------------------------------------------------------------------------%%
To identify the processes driving quenching and structural evolution at high redshift, it is useful to
consider galaxies that have been recently quenched (i.e.\ caught in transition).  The rare class of
post-starburst (PSB) galaxies is one such example, as they represent systems in which a major burst of star
formation was rapidly quenched within the last few hundred Myr.  Spectroscopically, these galaxies are
identified from the characteristic strong Balmer absorption lines related to an enhanced A-star population,
combined with a general lack of strong emission lines \citep{Dressler&Gunn:1983, Wild_etal:2009}.  However,
due to their intrinsic short-lived nature, until recently only a handful of these rare galaxies had been
spectroscopically identified at $z > 1$ \citep[e.g.][]{Vergani_etal:2010, Bezanson_etal:2013,
vandeSande_etal:2013, Newman_etal:2015, Belli_etal:2015, Williams_etal:2017}.

%% Supercolour technique %%--------------------------------------------------------------------------------%%
To identify PSBs at high redshift in greater numbers, two photometric methods have recently been developed.
\cite{Whitaker_etal:2012a} used medium-band near-infrared photometry to identify `young red-sequence'
galaxies from rest-frame {\em UVJ} colour--colour diagrams.  Alternatively, \cite{Wild_etal:2014} established
a classification scheme based on a Principal Component Analysis (PCA) of broad-band galaxy SEDs.
\cite{Wild_etal:2014} apply their technique to the multiwavelength photometry of the Ultra Deep Survey (UDS;
Almaini et al., in preparation), and find that just three shape parameters (`supercolours') provide a compact
representation of a wide variety of SED shapes.  This enables the separation of a tight red-sequence from
star-forming galaxies, and also the identification of several unusual populations, e.g.\ PSBs, which are
identified as galaxies that have formed a significant fraction of their mass in a recently-quenched
starburst.  This PCA technique has now led to the identification of $>900$ PSBs in the UDS field at
$0.5 < z < 2$ \citep[see][]{Wild_etal:2016}.  The validity of this method has also been confirmed using deep
optical spectroscopy \citep{Maltby_etal:2016}.  Of the photometrically-selected PSBs targeted for
spectroscopic follow-up, $\sim80$ per cent show the expected strong Balmer absorption, (i.e.~$\rm H\,\delta$
equivalent width $W_{\rm{H\,\delta}} > 5$\,\AA, a general PSB diagnostic; see e.g.~\citealt{Goto:2007}).
Furthermore, the confirmation rate remains high ($\sim60$ per cent), even when stricter criteria are used to
exclude cases with significant [O\,{\sc ii}] emission.  This is a more robust classification that ensures
fewer star-forming contaminants, but excludes PSBs hosting AGN or with low levels of residual star formation.

%% Morphology of PSB galaxies %%---------------------------------------------------------------------------%%
For PSB galaxies, structural analyses can provide useful constraints on their evolutionary history and the
likely mechanisms responsible for quenching their star formation.  However, until recently, these analyses
have largely been restricted to the Hubble-type morphologies of spectroscopic PSBs at $z < 1$ \citep[e.g.][]
{Dressler_etal:1999, Caldwell_etal:1999, Tran_etal:2003, Poggianti_etal:2009, Vergani_etal:2010}.  In
general, these studies find that, although PSBs are a morphologically heterogeneous population, they
typically exhibit disc-like morphologies (e.g.\ S0/Sa).  Fortunately, the recent development of photometric
selection techniques has allowed the structure of these galaxies to be explored at $z > 1$, for the first
time.  For example, \cite{Almaini_etal:2017} examine the structure of massive ($M_* > 10^{10}\rm\,M_{\odot}$)
PSBs in the UDS at $z > 1$.  They find that, in contrast to observations at lower redshift, these PSBs are
spheroidally-dominated and exceptionally compact, with sizes typically smaller than older passive galaxies.
They conclude that for massive PSBs at this epoch: i) morphological transformation has already taken place,
occurring either before (or during) the quenching event; and ii) their results are consistent with the PSB
phase being triggered by a gas-rich dissipative collapse, which quenched star formation and formed a compact
remnant.  Similar results have also been reported by \cite{Whitaker_etal:2012a} and \cite{Yano_etal:2016},
with young passive galaxies at $z > 1$ being more compact than their older counterparts.  However, at
intermediate redshifts ($z\sim1$) there are currently conflicting results on the relationship between stellar
age and the compactness of passive/recently-quenched galaxies \citep[see  e.g.][]{Keating_etal:2015,
Williams_etal:2017}, and further study is required at this epoch.

%% Aims %%-------------------------------------------------------------------------------------------------%%
In this paper, we build on previous results by using the PSB sample of \cite{Wild_etal:2016} to explore the
structure of these galaxies, self-consistently, across a wide redshift range \mbox{($0.5 < z < 2$)}.  For
this we mainly use average (i.e.\ stacked) one-dimensional radial light $\mu(r)$ profiles obtained from the
{\em Hubble Space Telescope} ({\em HST}) optical/near-infrared imaging available from the CANDELS survey
\citep{Grogin_etal:2011, Koekemoer_etal:2011}.  This work directly complements the study by
\cite{Almaini_etal:2017}, which uses the same parent PSB sample.  We build on their recent results by
extending the PSB structural analyses to: i)~lower redshifts $z < 1$; ii)~include {\em HST} optical imaging
to probe younger stellar populations; and iii)~consider galaxies with more complex structures (i.e.~multiple
components).  Taken together, these structural analyses will aid in our understanding of the triggering
mechanisms for the PSB phase, and of the mechanisms driving both the quenching and structural transformation
of galaxies in the distant Universe.

%% Overview %%---------------------------------------------------------------------------------------------%%
The structure of this paper is as follows.  In Section~\ref{Data and sample selection}, we give a brief
description of the data relevant to this work, and outline the PCA method used for identifying PSBs at high
redshift.  In Section~\ref{Radial light profiles}, we describe the isophotal fitting technique used to obtain
the one-dimensional radial light $\mu(r)$ profiles for our galaxies from the CANDELS optical/near-infrared
imaging.  Through \mbox{Sections~\ref{Profile fitting}--\ref{Optical imaging}}, we present various structural
analyses using our $\mu(r)$ profiles for passive, star-forming and PSB galaxies at two different epochs
($0.5 < z < 1$ and $1 < z < 2$).  We include a discussion of our results in Section~\ref{Discussion}, and
draw our conclusions in Section~\ref{Conclusions}.  Throughout this paper, we use AB magnitudes and adopt a
cosmology of $H_{0} = 70\rm\,km\,s^{-1}\,Mpc^{-1}$, $\Omega_{\Lambda} = 0.7$ and $\Omega_{\rm{m}} = 0.3$.

%%%%%%%%%%%%%%%%%%%%%%%%%%%%%%%%%%%%%%%%%%%%%%%%%%%%%%%%%%%%%%%%%%%%%%%%%%%%%%%%%%%%%%%%%%%%%%%%%%%%%%%%%%%%%
%%---------------------------------------------------------------------------------------------------------%%
%%%%%%%%%%%%%%%%%%%%%%%%%%%%%%%%%%%%%%%%%%%%%%%%%%%%%%%%%%%%%%%%%%%%%%%%%%%%%%%%%%%%%%%%%%%%%%%%%%%%%%%%%%%%%
\section[]{Data and sample selection}

\label{Data and sample selection}

%%%%%%%%%%%%%%%%%%%%%%%%%%%%%%%%%%%%%%%%%%%%%%%%%%%%%%%%%%%%%%%%%%%%%%%%%%%%%%%%%%%%%%%%%%%%%%%%%%%%%%%%%%%%%
\subsection[]{The Ultra Deep Survey (UDS)}

\label{The UDS}

%% Overview %%---------------------------------------------------------------------------------------------%%
This study is based on galaxy populations identified using the multiwavelength photometric data of the Ultra
Deep Survey (UDS; Almaini et al., in preparation)\footnote{http://www.nottingham.ac.uk/astronomy/UDS/}.  This
survey is the deepest component of the UKIRT (United Kingdom Infra-Red Telescope) Infrared Deep Sky Survey
\citep[UKIDSS;][]{Lawrence_etal:2007} and comprises extremely deep {\em JHK} photometry covering an area of
$0.77\rm\,deg^{2}$.  In this work, we use the eighth UDS data release (DR8) where the limiting depths are
$J = 24.9$, $H = 24.2$ and  $K = 24.6$ (AB; $5\sigma$ in $2$-arcsec apertures).  The final UDS data release
(DR11; June 2016), which achieved depths of $J = 25.4$, $H = 24.8$ and $K = 25.3$, will be used to
extend our PSB studies in future work.

%% Multiwavelenth data %%----------------------------------------------------------------------------------%%
The UDS is complemented by extensive multiwavelength observations.  These include, e.g.\ deep optical
{\em BVRi$'$z$'$} photometry from the Subaru--{\em XMM-Newton} Deep Survey \citep[SXDS;][]
{Furusawa_etal:2008} and mid-infrared observations ($3.6$ and $4.5\rm\,\mu{m}$) from the {\em Spitzer} UDS
Legacy Program (SpUDS; PI: Dunlop).  Deep optical $u'$-band photometry is also available from Megacam on the
Canada-France-Hawaii Telescope (CFHT).  The extent of the UDS field with full multiwavelength coverage
(optical--mid-infrared) is $\sim0.62\rm\,deg^{2}$.  For a complete description of these data, including a
description of the catalogue construction, see \cite{Hartley_etal:2013} and \cite{Simpson_etal:2012}.

%% Photometrc redshifts and stellar masses %%--------------------------------------------------------------%%
In this work, we use the photometric redshifts and stellar masses described in \cite{Simpson_etal:2013}.
These photometric redshifts ($z_{\rm\,phot}$) were determined by fitting the $11$-band UDS photometry
({\em u$'$BVRi$'$z$'$JHK}, $3.6\,\rm\mu{m}$ and $4.6\,\rm\mu{m}$) using a grid of galaxy templates built from
\cite{Bruzual&Charlot:2003} stellar population models.  The templates used had ages spaced logarithmically
between $30\rm\,Myr$ and $10\rm\,Gyr$, and included additional younger templates with dust-reddened SEDs.
The quality of these $z_{\rm\,phot}$ measurements was confirmed by comparison to over $3000$ secure
spectroscopic redshifts $z_{\rm\,spec}$, with a normalized median absolute deviation
$\sigma_{\rm\,NMAD} = 0.027$.  Stellar masses were also determined by fitting the $11$-band UDS photometry.
This fitting used a large grid of synthetic SEDs from the stellar population models of
\cite{Bruzual&Charlot:2003} and assumed a \cite{Chabrier:2003} initial mass function (IMF).  Random errors in
these stellar masses are typically $\pm0.1\rm\,dex$ \citep[see][for further details]{Simpson_etal:2013}.

%%%%%%%%%%%%%%%%%%%%%%%%%%%%%%%%%%%%%%%%%%%%%%%%%%%%%%%%%%%%%%%%%%%%%%%%%%%%%%%%%%%%%%%%%%%%%%%%%%%%%%%%%%%%%
\subsection[]{CANDELS--UDS}

\label{CANDELS-UDS}

%% Overview %%---------------------------------------------------------------------------------------------%%
For our morphological analyses, we use the deep {\em{HST}} near-infrared/optical imaging from the CANDELS
survey \citep{Grogin_etal:2011, Koekemoer_etal:2011}.  This $902$-orbit survey comprises Wide Field Camera~3
(WFC3) and parallel Advanced Camera for Surveys (ACS) imaging covering a total area of
$\sim800\rm\,arcmin^{2}$ spread across five survey fields.  One of these fields was selected to target a
sub-region of the UDS (CANDELS--UDS) and covers an area of $\sim210\rm\,arcmin^{2}$ ($\sim7$ per cent of the
UDS field).  In this study, we focus mainly on the WFC3 near-infrared imaging ($J_{\rm\,F125W}$,
$H_{\rm\,F160W}$) but also extend our analysis using the optical imaging from the ACS ($V_{\rm\,F606W}$,
$I_{\rm\,F814W}$).

%%%%%%%%%%%%%%%%%%%%%%%%%%%%%%%%%%%%%%%%%%%%%%%%%%%%%%%%%%%%%%%%%%%%%%%%%%%%%%%%%%%%%%%%%%%%%%%%%%%%%%%%%%%%%
\subsection[]{Sample selection}

\label{Sample selection}

%% Overview %%---------------------------------------------------------------------------------------------%%
In this work, we use the large sample of UDS galaxies ($z > 0.5$) recently classified by
\cite{Wild_etal:2016}.  These galaxies were classified using a photometric technique, developed by
\cite{Wild_etal:2014}, which is based on a Principal Component Analysis (PCA) of galaxy SEDs.  We provide a
brief overview of this method below.

%% Supercolour technique %%--------------------------------------------------------------------------------%%
The aim of the PCA technique is to describe a large variety of SED shapes through the linear combination of
only a small set of principal components (i.e.\ shape parameters).   In \cite{Wild_etal:2014,Wild_etal:2016},
these components were derived from a large library of `stochastic burst' model SEDs generated from
\cite{Bruzual&Charlot:2003} stellar population synthesis models with stochastic star-formation histories.
The result is a mean SED ($m_{\lambda}$) and a series of $p$ eigenspectra $e_{i\lambda}$ (i.e.\ principal
components) from which any normalised SED ($f_{\lambda}/{n}$) can be approximately reconstructed:
\begin{equation}
\frac{f_{\lambda}}{n} = m_{\lambda} + \sum_{i=1}^{p} a_i e_{i\lambda}.
\end{equation}

The amplitudes of each component ($a_{i}$) indicate its contribution to the overall shape of the galaxy SED,
and are referred to as `{\em supercolours}' (SCs).  These SCs can be used to uniquely and succinctly define
the shape of an SED, while retaining all the key information available from multiwavelength photometry.  In
fact, only the first three SCs are required to account for $>99.9$ per cent of the variance in the models of
\cite{Wild_etal:2014}.  Consequently, these three SCs alone can be used to provide a compact representation
of a wide variety of SED shapes.

%% SC correlations %%--------------------------------------------------------------------------------------%% 
Various correlations exist between these SCs and the properties of the model SEDs, e.g.\ mean stellar age,
dust content, metallicity and the fraction of mass formed in bursts in the last~Gyr.  These correlations
enable the separation of a tight red-sequence from star-forming galaxies, as well as the identification of
several unusual populations, e.g.\ i)~very dusty star-forming galaxies; ii)~metal poor quiescent dwarf
galaxies, and iii)~PSBs, which are selected as recently and rapidly quenched galaxies that have formed $>10$
per cent of their mass within the last Gyr (see \citealt{Maltby_etal:2016}, for spectroscopic verification).
We also separate the star-forming population into three sub-classes broadly reflecting an increase in
luminosity-weighted mean stellar age, or decrease in specific star-formation rate (SF1 $\rightarrow$ SF2
$\rightarrow$ SF3).  Although, note that these SF classes will suffer from the usual degeneracies between age
and moderate amounts of dust and metallicity.  With respect to the PSB selection, one important caveat is
that not all those identified will necessarily have undergone the implied short-lived `burst' of star
formation prior to quenching, and that some may have experienced a more extended ($\leq3\rm\,Gyr$) period of
star formation that was rapidly quenched \citep[see][for further details]{Wild_etal:2016}.  Nonetheless, this
population as a whole does represent transient galaxies that have been recently and rapidly quenched (which
is what we are primarily interested in), and we simply retain the PSB nomenclature here for consistency with
previous works in the series \citep{Wild_etal:2014, Wild_etal:2016, Maltby_etal:2016, Almaini_etal:2017,
Socolovsky_etal:2018}.

%% Parent sample %%----------------------------------------------------------------------------------------%%
To classify real galaxies, SCs are calculated by projecting their SEDs onto the PCA eigenspectra.  These SCs
are then compared with those of the model SEDs in order to determine the galaxy's most probable nature
(e.g.~red-sequence, star-forming, PSB).  The benefit of this approach is that the SCs of real galaxies are
independent of model-fitting and free to have values that differ substantially from those of the input model
library.  In the UDS field, this PCA analysis utilises $8$ UDS filters ({\em VRi$'$z$'$JHK}, $3.6\rm\,\mu$m;
a filter set which optimises the principal components for PSB identification), and is performed on all
galaxies with $K_{\rm AB} < 24$ and $0.5 < z < 2.0$ ($48\,713$ galaxies; \citealt{Wild_etal:2016}).  This
resulted in a large parent sample of $4249$ red-sequence (or `passive') galaxies, $39\,970$ star-forming
galaxies and $921$ PSBs.

%% CANDELS sample %%---------------------------------------------------------------------------------------%%
In this work, we use CANDELS {\em HST} imaging for our morphological analyses.  Approximately $10$ per cent
of our parent galaxy sample lies within the CANDELS--UDS field and has available {\em HST} imaging.  This
provides a final sample of $429$ passive galaxies, $3579$ star-forming galaxies ($2278$~SF1, $761$ SF2, $540$
SF3) and $98$~PSB galaxies.  We use this sample of CANDELS galaxies throughout this study.

%%%%%%%%%%%%%%%%%%%%%%%%%%%%%%%%%%%%%%%%%%%%%%%%%%%%%%%%%%%%%%%%%%%%%%%%%%%%%%%%%%%%%%%%%%%%%%%%%%%%%%%%%%%%%
\subsection{Stellar mass distributions}

\label{Mass distributions}

%% Mass-vs-redshift %%-------------------------------------------------------------------------------------%%
\begin{figure}
\includegraphics[width=0.49\textwidth]{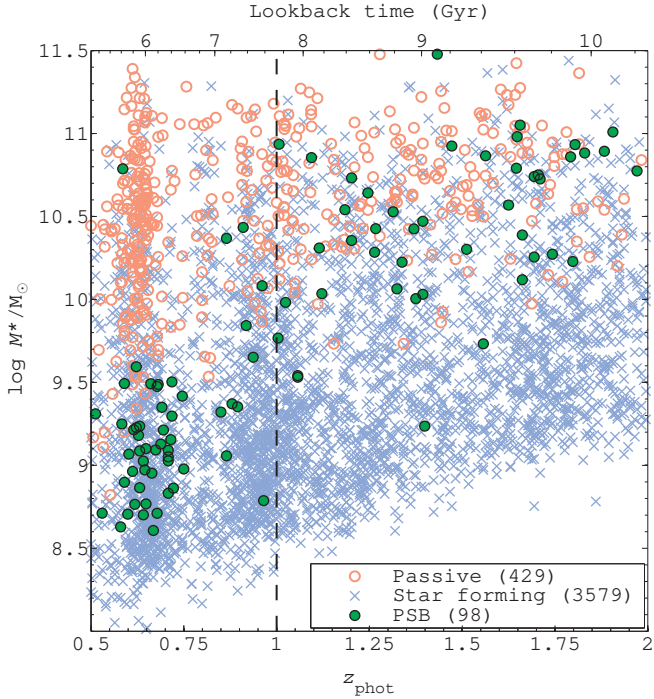}
\centering
\vspace{-0.4cm}
\caption{\label{mass-vs-z} The stellar mass $M_*$ distribution as a function of photometric redshift for
passive (red circles), star-forming (blue crosses) and PSB galaxies (green points), within the CANDELS--UDS
field.  Respective sample sizes are shown in the legend.  For PSB galaxies, there is significant evolution in
the stellar mass distribution over $0.5 < z < 2$ (see \protect{\citealt{Wild_etal:2016}}, for further
details).  In this paper, we take this evolution into account by considering the structural properties of
these galaxies at two different epochs: $0.5 < z < 1$ and $1 < z < 2$ (separated by the black dashed line).}
\end{figure}

%% Mass distributions %%-----------------------------------------------------------------------------------%%
In Fig.~\ref{mass-vs-z}, we present the stellar mass $M_*$ distribution as a function of redshift for
passive, star-forming and PSB galaxies, within the CANDELS--UDS field.  This clearly indicates that for PSBs,
there is a strong evolution in the $M_*$ distribution across \mbox{$0.5 < z < 2$}.  PSBs at $z < 1$ are
generally of low stellar mass \mbox{($M_* < 10^{10}\rm\,M_{\odot}$)}, while at higher redshift ($z > 1$) they
are typically of high stellar mass \mbox{($M_* > 10^{10}\rm\,M_{\odot}$)}.  \cite{Wild_etal:2016} recently
reported this evolution in the PSB mass function using the SC-classified galaxies from the entire UDS field
(i.e.~our parent galaxy sample; see Section~\ref{Sample selection}).  Their results indicate that the
comoving space density of massive PSBs ($M_* > 10^{10}\rm\,M_{\odot}$) is $\sim10\times$ higher at $z\sim2$
than at \mbox{$z\sim0.5$} \citep[see][for a similar result]{Whitaker_etal:2012a}.  Furthermore, at \mbox{$z > 1$}
the clear turnover in the PSB mass function towards low $M_*$ (see~\citealt{Wild_etal:2016}, figure 4),
suggests that the absence of low-mass PSBs at $z > 1$ is likely to be genuine, and not just an effect of
mass-incompleteness.  However, this issue will be explored in more depth in a future study, using the deeper
UDS DR11 data (Wilkinson et al., in preparation).  Taken together, these results suggest that PSBs at $z > 1$
are likely to be a different population to those observed at lower redshifts, potentially with different
evolutionary histories, and with the PSB phase being triggered by different mechanisms.  Consequently, in
this study, we account for this evolution in the $M_*$ distribution by examining PSB structure in two
separate epochs: intermediate-$z$ ($0.5 < z < 1$) and high-$z$ ($1 < z < 2$).

%% Comparison to UDS field %%------------------------------------------------------------------------------%%
For both epochs, we also assess whether our sample of CANDELS galaxies is representative of those from the
wider UDS field (see Section~\ref{Sample selection}).  For each galaxy population, we compare the $M_*$ and
redshift distributions between these two fields, and find no significant differences in most cases (based on
Kolmogorov-Smirnov tests; $p > 0.01$).  The only exceptions are for the redshift distributions of passive and
PSB galaxies at $0.5 < z < 1$, where we find an excess in the CANDELS samples at $z\sim0.65$.  This is due to
a known supercluster in this field \citep[see e.g.][]{vanBreukelen_etal:2006, Galametz_etal:2018}, and
therefore environmental effects may be particularly relevant for our galaxy samples at this epoch.  We return
to this point in Section~\ref{Discussion: intermediate redshift}.

%% Completeness %%-----------------------------------------------------------------------------------------%%
In this work, our galaxy samples are derived from the SC analysis of \cite{Wild_etal:2016}, which was
performed on UDS galaxies with $K < 24$ (see Section~\ref{Sample selection}).  Using this $K$ limit, the
equivalent mass-completeness limits were determined as a function of redshift using the method of
\cite{Pozzetti_etal:2010}.  For galaxies in the CANDELS--UDS field, we find that completeness is
$\gtrsim95$~per~cent for the following mass ranges: $M_* > 10^9\rm\,M_\odot$ ($0.5 < z < 1$) and
$M_* > 10^{10}\rm\,M_\odot$ ($1 < z < 2$).  These mass limits are used throughout this work.  Note, that
since mass--completeness varies smoothly across redshift, these limits are conservative and strictly only
applicable at the upper-$z$ limit of each epoch.  This is particularly important when considering the
high-$z$ epoch, where a lower mass limit of $M_* > 10^{9.5}\rm\,M_\odot$ would actually yield equivalent
completeness over $1 < z < 1.5$.  The sizes of the final mass-limited galaxy samples used throughout this
work, are presented in Table~\ref{galaxy samples}.

%% Mass-limited galaxy samples %%--------------------------------------------------------------------------%%
\begin{table}
\centering
\begin{minipage}{70mm}
\centering
\caption{\label{galaxy samples} The mass-limited galaxy samples used throughout this work, including the
sub-samples for the star-forming population (SF1, SF2, SF3).}
\begin{tabular}{lcccccc}
\hline
{Sample}	&{$N$(intermediate-$z$)}	&{}	&{$N$(high-$z$)}		\\
{}		&{$0.5 < z < 1.0$}		&{}	&{$1.0 < z < 2.0$}		\\
{}		&{$M_* > 10^9\rm\,M_\odot$}	&{}	&{$M_* > 10^{10}\rm\,M_\odot$}	\\
\hline
Passive		&{$256$}			&{}	&{$165$}			\\[1ex]
Star forming	&{$883$}			&{}	&{$536$}			\\
\ \ SF1		&{$404$}			&{}	&{$54$}				\\
\ \ SF2		&{$265$}			&{}	&{$192$}			\\
\ \ SF3		&{$214$}			&{}	&{$290$}			\\[1ex]
PSB		&{$36$}				&{}	&{$39$}				\\
\hline
\end{tabular}
\end{minipage}
\end{table}

%%%%%%%%%%%%%%%%%%%%%%%%%%%%%%%%%%%%%%%%%%%%%%%%%%%%%%%%%%%%%%%%%%%%%%%%%%%%%%%%%%%%%%%%%%%%%%%%%%%%%%%%%%%%%
%-----------------------------------------------------------------------------------------------------------%
%%%%%%%%%%%%%%%%%%%%%%%%%%%%%%%%%%%%%%%%%%%%%%%%%%%%%%%%%%%%%%%%%%%%%%%%%%%%%%%%%%%%%%%%%%%%%%%%%%%%%%%%%%%%%
\section{Radial light profiles}

\label{Radial light profiles}

%% Overview %%---------------------------------------------------------------------------------------------%%
In this section, we describe the measurement of galaxy radial light $\mu(r)$ profiles from the CANDELS
{\em HST} imaging and the production of stacked $\widetilde\mu(r)$ profiles.  The relevant profile fitting
and structural analyses are presented in Sections~\ref{Profile fitting}--\ref{Optical imaging}.

% Justification of the method %%---------------------------------------------------------------------------%%
In this work, we stack one-dimensional radial $\mu(r)$ profiles, which enables us to maximise
signal-to-noise, particularly for the outer galactic regions.  This is desirable for the reliable
multi-component decomposition of our faint galaxies, and necessary for the identification of faint components
(e.g.~outer discs) that may not be detected in individual profiles.  This one-dimensional approach has the
advantage of providing a simple visualisation of the true galactic structure (i.e.\ non-parametrised) for
comparison to fitted $\mu(r)$ profiles.  This can be useful for determining whether an extra component
(e.g.\ outer disc) is really present.  We are aware that the inherent loss of azimuthal information could
introduce some uncertainty to the fitted structural parameters, and an alternative approach would be to use
two-dimensional analyses (i.e.\ stack galaxy images).  However, the differences in results between a one- and
two-dimensional analysis are minimal.  For face-on galaxies the two methods will yield the same results, and
for inclined galaxies, single S{\'e}rsic and disc parameters (and most bulge parameters) will be consistent
within fitting errors in most cases, with a characteristic scatter at the $10$--$20$ per cent level
\citep[see e.g.][]{deJong:1996, MacArthur_etal:2003, McDonald_etal:2011}.  In~this work, we choose to adopt
the one-dimensional approach, but we also independently confirm that these two methods would lead to
consistent results for our samples (see Section~\ref{Further robustness tests}).

%%%%%%%%%%%%%%%%%%%%%%%%%%%%%%%%%%%%%%%%%%%%%%%%%%%%%%%%%%%%%%%%%%%%%%%%%%%%%%%%%%%%%%%%%%%%%%%%%%%%%%%%%%%%%
\subsection{Isophotal fitting}

\label{Isophotal fitting}

%% Introduction %%-----------------------------------------------------------------------------------------%%
For each galaxy, we use the {\sc iraf} task ellipse\footnote{{\sc stsdas} package -- version 3.14} in order
to obtain their azimuthally-averaged radial light $\mu(r)$ profiles from the CANDELS ACS/WFC3 imaging
\citep[see][]{Jedrzejewski:1987}.  This isophotal fitting was performed independently in each of the four
CANDELS wavebands ($V_{\rm\,F606W}$, $I_{\rm\,F814W}$, $J_{\rm\,F125W}$ and $H_{\rm\,F160W}$).

\begin{figure*}
\hspace{0.1cm}
\includegraphics[width=1.02\textwidth]{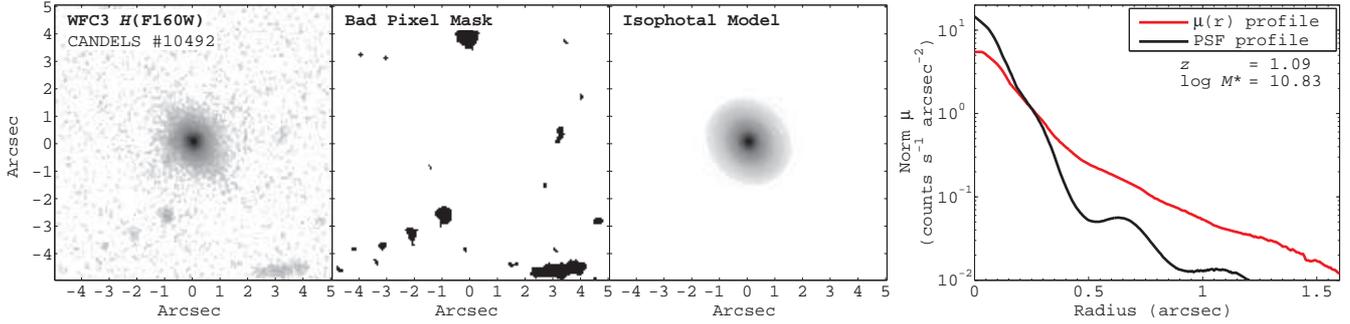}
\vspace{-0.4cm}
\caption{\label{Fit example} An example isophotal fit for a typical post-starburst galaxy.  Left-hand panel:
{\em HST}/WFC3 $H_{\rm\,F160W}$ image.  Centre-left panel: the bad pixel mask used for the isophotal fit.
Centre-right panel: the isophotal model generated by ellipse.  Right-hand panel: the resultant
azimuthally-averaged radial light $\mu(r)$ profile (red line), with the {\em HST}/WFC3 $H_{\rm\,F160W}$ PSF
profile (black line) also shown for reference.  Both profiles have been normalised using the flux contained
within a $2.828$ arcsec aperture.}
\end{figure*}

%% Masking %%----------------------------------------------------------------------------------------------%%
In our isophotal fitting, bad pixel masks were used to remove all potential sources of contamination, e.g.\
background/companion galaxies and foreground stars (anything not associated with the subject galaxy itself).
These masks were created from {\sc SExtractor} segmentation maps, and a separate mask was generated for each
CANDELS waveband.  A validation of these masks was also performed by visual inspection to ensure that all
potential sources of contamination were adequately masked; see Fig.~\ref{Fit example} for a typical example.

%% Ellipse fitting %%--------------------------------------------------------------------------------------%%
To determine the radial $\mu(r)$ profiles, free-parameter isophotal fits were performed for each galaxy
(fixed centre, free ellipticity~$e$ and position angle~PA).  These fits tend to follow significant
morphological features (e.g.~bulges, bars, spiral arms) and are consequently suitable for tracing a galaxy's
principal structural component (i.e.~bulge and/or disc).  In these isophotal fits, we use linear radial
sampling ($\Delta{r} = 0.1\rm\,pixel$, where $r$ is the radius along the semi-major axis) and a fixed
isophotal centre determined for each galaxy by \mbox{{\sc SExtractor}}.  In isophotal fitting, it is often
advisable to begin the fitting procedure from a good initial estimate for an inner isophote.  To provide this
initial isophote, we use the shape parameters ($e$, PA) obtained for our galaxies from {\sc SExtractor}.  In
our isophotal fits, four iterations of a $3\sigma$ rejection are also applied to deviant points along each
isophote to remove the influence of non-axisymmetric features on the resultant $\mu(r)$ profile
(i.e.~star-forming regions and supernovae).  A typical example of an isophotal fit for one of the PSB
galaxies is presented in Fig.~\ref{Fit example}.

%% Example isophotal fit %%--------------------------------------------------------------------------------%%

%%%%%%%%%%%%%%%%%%%%%%%%%%%%%%%%%%%%%%%%%%%%%%%%%%%%%%%%%%%%%%%%%%%%%%%%%%%%%%%%%%%%%%%%%%%%%%%%%%%%%%%%%%%%%
\subsection{Sky subtraction}

\label{sky subtraction}

%% Sky subtraction %%--------------------------------------------------------------------------------------%%
With isophotal analyses, it is very important to perform a careful sky subtraction to remove the effect of
the sky/background on the resultant $\mu(r)$ profile.  The slight under-/over-subtraction of the sky can
easily lead to an incorrect profile shape, particularly in the outer regions of the profile (see
\citealt{Maltby_etal:2012a, Maltby_etal:2012b, Maltby_etal:2015}, for some recent studies).  The publicly
available CANDELS {\em HST} WFC3/ACS imaging have already undergone a careful sky subtraction
\citep{Koekemoer_etal:2011} and residual sky in these images is not expected to be significant.  Nonetheless,
we measure the residual sky level in the WFC3/ACS images in order to assess the potential influence on the
shape of our $\mu(r)$ profiles.

%% Residual sky background %%------------------------------------------------------------------------------%%
For each galaxy in our sample, we obtain an estimate of the local sky background ($n_{\rm sky}$) by using
pixels obtained from the four corners of the galaxy WFC3/ACS image (i.e.\ postage stamp).  The sizes of these
postage stamps are variable and designed to optimally contain the subject galaxy.  The size is based on a
multiple of the \cite{Kron:1980} radius ($\sim4\times$), and therefore from theoretical light profiles these
postage stamps should contain $>98$ per cent of the subject galaxy's light \citep[see e.g.][]
{Bertin&Arnouts:1996}.  As a consequence, in sampling the corners of these postage stamps we have a
reasonable expectation of probing the actual sky background.  The corner image pixels were selected using
quarter-circle wedges of side equal to 10 per~cent of the smallest image dimension (corresponding to a region
$> 3.6$ Kron radii from the galaxy's centre).  We then apply our bad pixel masks to ensure only `dark' pixels
are used and obtain the median pixel value $\widetilde{n_{\rm e}}$ (or `sky level') in each wedge.  The mean
of these sky levels from the four corners of the galaxy postage stamp is then used as a local estimate of the
residual sky background ($n_{\rm sky}$).  For each galaxy postage stamp, the corner-to-corner rms in their
four $\widetilde{n_{\rm e}}$ measurements is also determined and taken as an estimate for the $1\sigma$ error
in the local sky background $\sigma_{\rm sky}$ (i.e.\ the local error in the sky subtraction).

For the near-infrared imaging ($J_{\rm\,F125W}$, $H_{\rm\,F160W}$), the residual sky level was determined to
be well below tolerance levels, with the average sky level at least two orders of magnitude below a typical
$\mu(r)$ profile at the limiting galactocentric radius used in this study ($r_{\rm lim} = 1.6$ arcsec).  This
limiting radius is defined as the threshold at which a typical $\mu(r)$ profile enters the region dominated
by uncertainly in the sky background (i.e.\ the flux limit corresponding to the average sky subtraction error
$\widetilde\sigma_{\rm sky}$).  However, for the optical imaging ($V_{\rm\,F606W}$, $I_{\rm\,F814W}$), the
residual sky level is much higher than in the near-infrared, and potentially a significant component of the
$\mu(r)$ profile at $r_{\rm lim}$ (possibly accounting for up to $\sim20$ per cent of the flux).  To address
this issue, we correct all our optical/near-infrared $\mu(r)$ profiles by subtracting the corresponding local
residual sky background $n_{\rm sky}$ on a galaxy-galaxy basis.

%%%%%%%%%%%%%%%%%%%%%%%%%%%%%%%%%%%%%%%%%%%%%%%%%%%%%%%%%%%%%%%%%%%%%%%%%%%%%%%%%%%%%%%%%%%%%%%%%%%%%%%%%%%%%
\subsection{PSF determination}

\label{PSF determination}

%% Motivation %%-------------------------------------------------------------------------------------------%%
Point spread functions (PSFs) for the CANDELS {\em HST} imaging are well determined and have FWHM varying
between $0.08$--$0.18$ arcsec \citep{Koekemoer_etal:2011}.  However, at the redshifts studied in this work
($z > 0.5$), the half-light radii of galaxies are typically $< 1$ arcsec \citep[see e.g.][]
{Almaini_etal:2017}.  Consequently, the {\em HST} PSF can be a considerable factor in the $\mu(r)$ profiles
of our galaxies (see Fig.~\ref{Fit example}, for an example).  The determination of an accurate PSF and its
influence on our $\mu(r)$ profiles is therefore critical to the measurement of reliable structural properties
($r_{\rm e}$, $n$) in this work.

\begin{figure}
\includegraphics[width=0.45\textwidth]{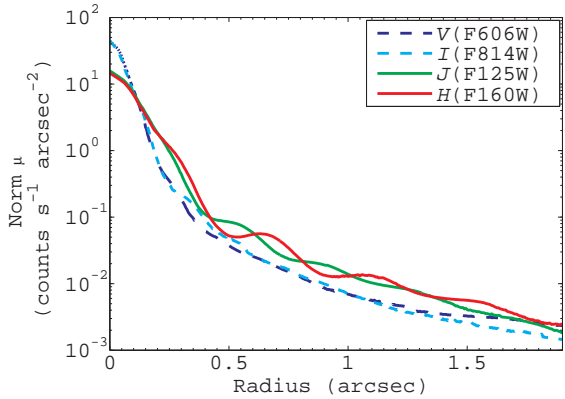}
\centering
\vspace{0.0cm}
\caption{\label{PSFs} The CANDELS {\em HST} PSFs.  Radial light $\mu(r)$ profiles for the CANDELS PSFs,
showing the differences in structure between the four wavebands.  These PSFs have been empirically determined
from isolated stars in the CANDELS {\em HST} images.  All profiles have been normalised using the flux
contained within a $2.828$ arcsec aperture.}
\end{figure}

%% PSF construction %%-------------------------------------------------------------------------------------%%
To construct our PSFs, we use isolated stars identified in the UDS field \citep[see][]{Lani_etal:2013,
Almaini_etal:2017} that reside within the CANDELS--UDS region ($\sim150$ stars).  For each CANDELS waveband,
we create postage stamps for these stars (stamp size: $4\times4\rm\,arcsec^2$), which are then normalised
in total flux (aperture diameter $= 2.828$ arcsec) and combined in a median stack.  The resultant PSF images
reveal significant structural features that could easily affect our galaxy $\mu(r)$ profiles
(e.g.\ diffraction spikes and Airy rings).  To illustrate this, we use isophotal fitting to generate
radial $\mu(r)$ profiles from our PSF images (see Fig.~\ref{PSFs}).  These profiles also show that the PSF
structure changes considerably between the different CANDELS wavebands.  As expected, the WFC3 PSFs
($J_{\rm\,F125W}$, $H_{\rm\,F160W}$) are broader than the ACS PSFs ($V_{\rm\,F606W}$, $I_{\rm\,F814W}$), but
they also exhibit more prominent Airy rings that manifest as significant bumps in their radial $\mu(r)$
profiles.  In this work, we use the PSF images determined here to account for the nature of the CANDELS PSFs
in our structural analyses (see Section~\ref{Profile fitting}).

%% CANDELS PSFs %%-----------------------------------------------------------------------------------------%%

%%%%%%%%%%%%%%%%%%%%%%%%%%%%%%%%%%%%%%%%%%%%%%%%%%%%%%%%%%%%%%%%%%%%%%%%%%%%%%%%%%%%%%%%%%%%%%%%%%%%%%%%%%%%%
\subsection{Stacked light profiles $\widetilde{\mu}(r)$}

\label{Stacked light profiles}

%% Average H-band profiles %%------------------------------------------------------------------------------%%
\begin{figure*}
\includegraphics[width=0.99\textwidth]{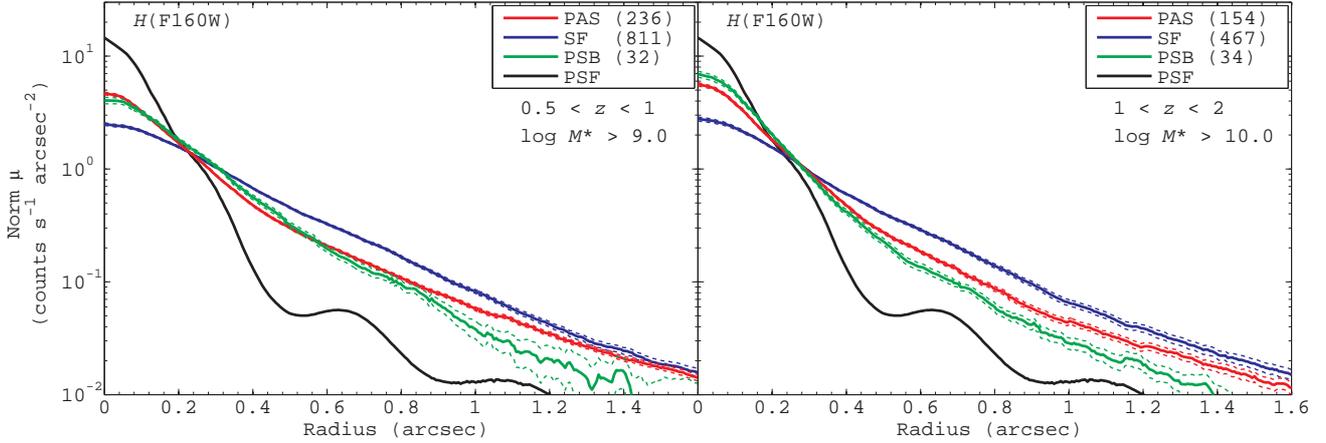}
\centering
\caption{\label{Median profiles} Median-stacked $H_{\rm\,F160W}$ light profiles $\widetilde\mu(r)$ for
different galaxy populations and at different epochs.  Left-hand panel: $\widetilde\mu(r)$ profiles for
passive (PAS; red line), star-forming (SF; blue line) and PSB galaxies (green line) at $0.5 < z < 1$.
Right-hand panel: analogous $\widetilde\mu(r)$ profiles at $1 < z < 2$.  The errors in the $\widetilde\mu(r)$
profiles (dashed lines) are $1\sigma$ confidence limits, which are determined from the mean of the standard
errors from $100$ simulated stacks generated via a bootstrap method.  The respective sample sizes used to
generate the $\widetilde\mu(r)$ profiles are shown in the legend.  The {\em HST} $H_{\rm\,F160W}$ PSF profile
(black line) is also shown for reference. Virtually identical $\widetilde\mu(r)$ profiles are also obtained
from the $J_{\rm\,F125W}$ imaging.
\vspace{0.1cm}}
\end{figure*}

%% Overview %%---------------------------------------------------------------------------------------------%%
To assess the general structure of PSB galaxies, we combine their individual $\mu(r)$ profiles in median
stacks.  This is performed separately in each CANDELS waveband and for our samples at the two different
epochs, $0.5 < z < 1$ and $1 < z < 2$.  Analogous median stacks are also generated for the passive and
star-forming galaxies.  We give a brief description of the stacking procedure below.

%% Profile stacking %%-------------------------------------------------------------------------------------%%
To generate our median stacked profiles $\widetilde\mu(r)$, we take the median flux of the respective sample
of individual $\mu(r)$ profiles as a function of radius.  The individual $\mu(r)$ profiles were normalised in
flux prior to stacking (using the flux within a $2.828$ arcsec aperture).  During the stacking process, we
also perform one iteration of a $3\sigma$ clip to individual $\mu(r)$ profiles that deviate from the median
flux within the limiting galactocentric radius ($r_{\rm lim} = 1.6$ arcsec; see
Section~\ref{sky subtraction}).  Clipping is not performed beyond $r_{\rm lim}$ due to the increased level of
uncertainty in the individual $\mu(r)$ profiles as they approach the limit of the background noise.  This
clipping improves the $1\sigma$ error boundaries on our $\widetilde\mu(r)$ profiles, but has no significant
effect on their overall shape (i.e.\ structural parameters).  Note that we do not normalise for apparent
size (i.e.\ angular extent) in our stacking analysis, since this would also destandardize the significant
effect of the PSF (see Section~\ref{PSF determination}).  However, from an assessment of simulated
$\widetilde\mu(r)$ stacks, we find this normalisation to be unnecessary.  We also note that for the two
epochs studied, the change in angular scale ($\rm kpc\,arcsec^{-1}$) with redshift has no significant
influence on the shape of the resultant $\widetilde\mu(r)$ profiles (see
Section~\ref{Profile fitting: stacked light profiles}).

%% Comparisions %%-----------------------------------------------------------------------------------------%%
For each galaxy population, the $\widetilde\mu(r)$ profiles from the CANDELS $H_{\rm\,F160W}$ imaging are
presented in Fig.~\ref{Median profiles}.  Random errors in these $\widetilde\mu(r)$ profiles ($1\sigma$) are
the error in the median flux as a function of radius, which is determined from the mean of the standard
errors from $100$ simulated stacks generated via a bootstrap method.  Virtually identical $\widetilde\mu(r)$
profiles were also obtained from the $J_{\rm\,F125W}$ imaging.  For the optical imaging ($V_{\rm\,F606W}$,
$I_{\rm\,F814W}$), the $\widetilde\mu(r)$ profiles are presented in Section~\ref{Optical imaging}.

An inspection of our $H_{\rm\,F160W}$ $\widetilde\mu(r)$ profiles reveals some significant differences in
structure between the different populations (see Fig.~\ref{Median profiles}).  For both epochs, the passive
and PSB populations have stellar distributions that appear more compact and centrally concentrated than the
star-forming population (see \citealt{Williams_etal:2010} and \citealt{vanderWel_etal:2014}, for similar
results).  The PSB population also appears to have stellar distributions that are marginally more compact
than the passive population, particularly at high redshift ($1 < z < 2$).  In the following sections, we
perform profile fitting on these profiles to analyse their structure in more detail.

%%%%%%%%%%%%%%%%%%%%%%%%%%%%%%%%%%%%%%%%%%%%%%%%%%%%%%%%%%%%%%%%%%%%%%%%%%%%%%%%%%%%%%%%%%%%%%%%%%%%%%%%%%%%%
\section{Profile fitting}

\label{Profile fitting}

%% Overview %%---------------------------------------------------------------------------------------------%%
To determine structural properties from the average $\widetilde\mu(r)$ profiles (e.g.\ effective radius
$r_{\rm e}$; S{\'e}rsic index $n$), we perform one-dimensional profile fitting via the comparison of these
measured profiles to a large library of $\sim22\,000$ model galaxy profiles.

%% The model library %%------------------------------------------------------------------------------------%%
To build the model library, we begin by generating mock galaxy images using two-dimensional S{\'e}rsic models
that cover a wide range of profile shapes ($0.7 < n < 8$; \mbox{$0.01 < r_{\rm e} < 1.5\rm\,arcsec$}).  Each
image was then convolved with the relevant {\em HST} PSF (see Section~\ref{PSF determination}) and normalised
in total flux.  Free-parameter isophotal fits (fixed centre, free ellipticity $e$ and position angle PA) were
then performed to generate the azimuthally-averaged radial light profiles which comprise the model library
$\{\mu_{\rm mock}(r)\}$.  These isophotal fits are analogous to those described for our measured light
$\mu(r)$ profiles in Section~\ref{Isophotal fitting}.  A separate model library is generated for each CANDELS
waveband ($V_{\rm\,F606W}$, $I_{\rm\,F814W}$, $J_{\rm\,F125W}$, $H_{\rm\,F160W}$), in order to take account
of the significant differences observed in the structure of their PSFs (see Section~\ref{PSF determination};
Fig.~\ref{PSFs}).

%% Profile fitting %%--------------------------------------------------------------------------------------%%
For profile fitting (single S{\'e}rsic), a measured light profile is compared to every profile in the
relevant model library and the best-fit is obtained by $\chi^2$ minimisation.   The measured/model profiles
are resampled ($0.1\times\rm\,sample\,rate$) to ensure the data points used in the $\chi^2$ minimisation are
radially independent.  The full library for the relevant waveband is also used in each fit, to ensure a
global minimum solution is obtained.  During this process the normalisation of the model profiles are allowed
to vary.  This is necessary to ensure the best-fit is only defined by the `shape' of the profile, and not by
slight differences in the normalisation between the measured and model profiles.  For fitting purposes, we
only use data from $r < r_{\rm lim}$ ($r_{\rm lim} = 1.6$~arcsec; see Section~\ref{sky subtraction}) to
ensure the fit is driven by the main structural components (bulge/disc) and not affected by any uncertainty
in the sky subtraction.

In the following Sections (\ref{Profile fitting: stacked light profiles}--\ref{Two-component fits}), we focus
in detail on the profile fitting results for the near-infrared $\mu(r)$ profiles (WFC3 -- $J_{\rm\,F125W}$,
$H_{\rm\,F160W}$).  The profile fitting results for the optical $\mu(r)$ profiles (ACS -- $V_{\rm\,F606W}$,
$I_{\rm\,F814W}$) will be discussed in Section~\ref{Optical imaging}.  In~this work, we are mainly interested
in the general structure of our galaxies stellar distributions.  For our galaxies ($z > 0.5$), the
near-infrared directly probes the old stellar component \mbox{($\lambda_{\rm rest} > 4000$\rm\,\AA)}, which
comprises the bulk of the stellar mass.  Therefore, the near-infrared $\mu(r)$ profiles are the principal
focus of this study.

%%%%%%%%%%%%%%%%%%%%%%%%%%%%%%%%%%%%%%%%%%%%%%%%%%%%%%%%%%%%%%%%%%%%%%%%%%%%%%%%%%%%%%%%%%%%%%%%%%%%%%%%%%%%%
\subsection{Stacked light profiles $\widetilde\mu(r)$}

\label{Profile fitting: stacked light profiles}

\begin{figure*}
\includegraphics[width=0.76\textwidth]{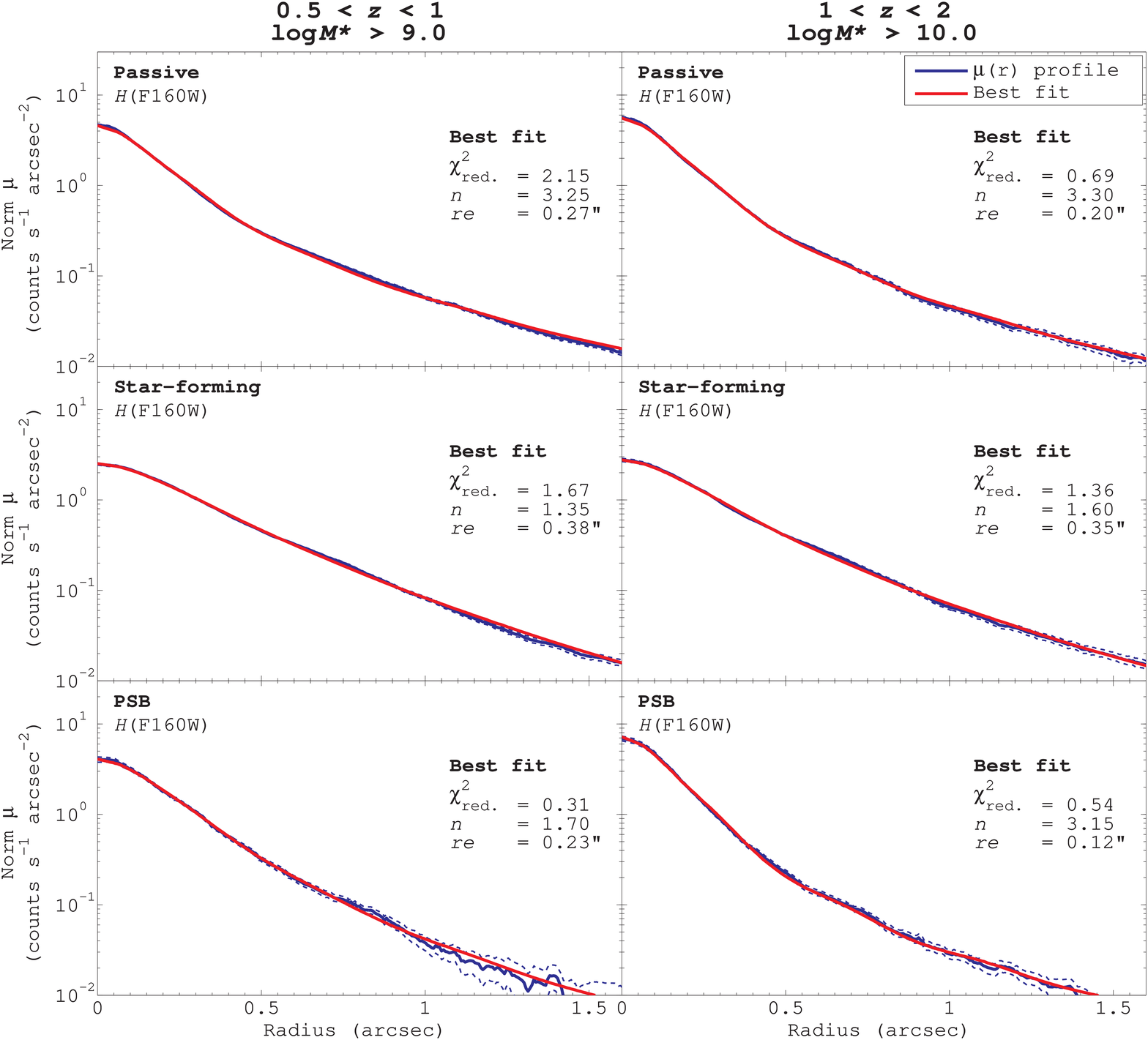}
\centering
\caption{\label{Sersic profiles} Single S{\'e}rsic fits to our median-stacked $H_{\rm\,F160W}$ light profiles
$\widetilde\mu(r)$ from two separate epochs: $0.5 < z < 1$ (left-hand column) and \mbox{$1 < z < 2$}
(right-hand column).  In all cases (passive, star-forming, PSB), our $\widetilde\mu(r)$ profiles are well
described by a single S{\'e}rsic profile.  Virtually identical fits are also obtained from our
$J_{\rm\,F125W}$ $\widetilde\mu(r)$ profiles (see Table~\ref{Sersic results}).}
\vspace{-0.35cm}
\end{figure*}

%% Profile fitting %%--------------------------------------------------------------------------------------%%
For each galaxy population, we perform single S{\'e}rsic fits on the median-stacked light profiles
$\widetilde\mu(r)$ and obtain their typical structural properties (i.e.\ $r_{\rm e}$, $n$).  The resultant
fits for the $H_{\rm\,F160W}$ $\widetilde\mu(r)$ profiles are shown in Fig.~\ref{Sersic profiles}.  In all
cases, the $\widetilde\mu(r)$ profiles are well described by a single S{\'e}rsic profile, with the best fit
having a reduced chi-squared $\chi_{\rm red}^2\sim1$.  Furthermore, the $\chi^2$ distribution for fits across
the full $r_{\rm e}$--$n$ parameter space shows that these best-fits are stable, well defined and that no
degeneracies are present.  Very similar results are also obtained for the $J_{\rm\,F125W}$ $\widetilde\mu(r)$
profiles.  The resultant structural parameters ($r_{\rm e}$, $n$) for each galaxy population in both
$J_{\rm\,F125W}$ and $H_{\rm\,F160W}$, are shown in Table~\ref{Sersic results}.  The uncertainty in these
structural parameters ($1\sigma$) is estimated, independently for each galaxy population, using the variance
between analogous fits performed on $100$ simulated $\widetilde\mu(r)$ profile stacks generated via a
bootstrap analysis.  For estimates of the effect of the PSF and sky subtraction error
$\widetilde\sigma_{\rm sky}$ on these measurements, see Section~\ref{Further robustness tests}.

%% Best-fit profiles %%------------------------------------------------------------------------------------%%

%% Best-fit structural parameters %%-----------------------------------------------------------------------%%
\begin{table*}
\centering
\begin{minipage}{175mm}
\centering
\caption{\label{Sersic results} Near-infrared single S{\'e}rsic fits: the structural properties for our
median-stacked $J_{\rm\,F125W}$ and $H_{\rm\,F160W}$ light profiles $\widetilde\mu(r)$.  Structural
properties \mbox{($r_{\rm e}$, $n$)} are shown for different galaxy populations (passive, star-forming, PSB)
at two different epochs, $0.5 < z < 1$ and $1 < z < 2$.  Errors in the structural parameters ($1\sigma$)
are determined from the variance between fits performed on $100$ simulated $\widetilde\mu(r)$ profile stacks
generated via a bootstrap method.}
\begin{tabular}{lccccccccccc}
\hline
Galaxy		&\multicolumn{5}{c}{$0.5 < z < 1$}							&{}	&\multicolumn{5}{c}{$1 < z < 2$}							\\
Population	&\multicolumn{2}{c}{$J_{\rm\,F125W}$}	&{}	&\multicolumn{2}{c}{$H_{\rm\,F160W}$}	&{}	&\multicolumn{2}{c}{$J_{\rm\,F125W}$}	&{}	&\multicolumn{2}{c}{$H_{\rm\,F160W}$}	\\
{}		&{$n$}		&{$r_{\rm e}$}		&{}	&{$n$}		&{$r_{\rm e}$}		&{}	&{$n$}		&{$r_{\rm e}$}		&{}	&{$n$}		&{$r_{\rm e}$}		\\
{}		&{}		&{(arcsec)}		&{}	&{}		&{(arcsec)}		&{}	&{}		&{(arcsec)}		&{}	&{}		&{(arcsec)}		\\[0.5ex]
\hline																									\\[-1.5ex]
Passive		&{$3.25\pm0.15$}&{$0.28\pm0.01$}	&{}	&{$3.25\pm0.15$}&{$0.27\pm0.01$}	&{}	&{$3.20\pm0.18$}&{$0.21\pm0.01$}	&{}	&{$3.30\pm0.19$}&{$0.20\pm0.01$}	\\[1ex]	
Star-forming	&{$1.30\pm0.05$}&{$0.39\pm0.01$}	&{}	&{$1.35\pm0.04$}&{$0.38\pm0.01$}	&{}	&{$1.65\pm0.10$}&{$0.37\pm0.01$}	&{}	&{$1.60\pm0.06$}&{$0.35\pm0.01$}	\\[1ex]
\ \ SF1		&{$1.10\pm0.04$}&{$0.40\pm0.01$}	&{}	&{$1.15\pm0.05$}&{$0.38\pm0.01$}	&{}	&{$1.20\pm0.16$}&{$0.52\pm0.06$}	&{}	&{$1.20\pm0.14$}&{$0.47\pm0.06$}	\\
\ \ SF2		&{$1.35\pm0.08$}&{$0.39\pm0.02$}	&{}	&{$1.35\pm0.07$}&{$0.38\pm0.02$}	&{}	&{$1.45\pm0.12$}&{$0.43\pm0.02$}	&{}	&{$1.40\pm0.10$}&{$0.41\pm0.02$}	\\
\ \ SF3		&{$1.70\pm0.12$}&{$0.38\pm0.02$}	&{}	&{$1.70\pm0.11$}&{$0.36\pm0.02$}	&{}	&{$1.95\pm0.13$}&{$0.32\pm0.01$}	&{}	&{$1.85\pm0.13$}&{$0.30\pm0.01$}	\\[1ex]
Post-starburst	&{$1.70\pm0.20$}&{$0.22\pm0.01$}	&{}	&{$1.70\pm0.24$}&{$0.23\pm0.01$}	&{}	&{$3.65\pm0.32$}&{$0.13\pm0.01$}	&{}	&{$3.15\pm0.34$}&{$0.12\pm0.01$}	\\
\hline
\end{tabular}
\end{minipage}
\end{table*}

\begin{figure*}
\includegraphics[width=1\textwidth]{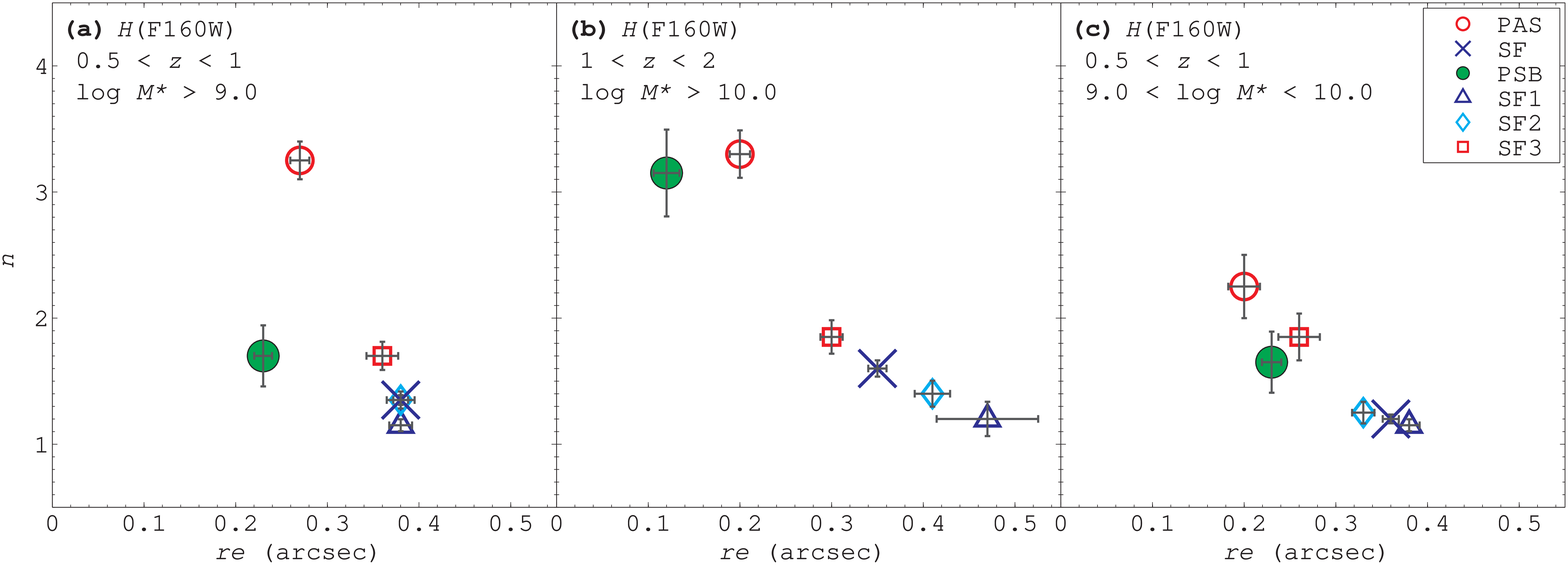}
\vspace{-0.25cm}
\centering
\caption{\label{Sersic fits} Structural properties (effective radius $r_{\rm e}$, S{\'e}rsic index $n$) from
the single S{\'e}rsic fits to our median-stacked $H_{\rm\,F160W}$ light profiles $\widetilde\mu(r)$ -- see
Table~\ref{Sersic results}.  Left-hand panel (a): the typical structural properties for passive (PAS),
star-forming (SF) and PSB galaxies at $0.5 < z < 1$.  The structural properties for the star-forming
sub-populations (SF1 $\rightarrow$ SF3; based on decreasing sSFR), are also shown.  Centre panel (b):
analogous results for $1 < z < 2$.  These results reveal significant differences in the structure of PSBs at
different epochs.  At $z > 1$, PSBs are extremely compact and of high $n$, with structures similar to the
passive population but more compact.  However, at $z < 1$, PSBs are relatively compact but exhibit much lower
$n$ than the PSBs at $z > 1$.  Right-hand panel (c): the results for low-$z$ ($0.5 < z < 1$) when the galaxy
samples have been limited to a mass range comparable to the PSB population (i.e.\ low-mass;
$10^9 < M_* < 10^{10}\rm\,M_{\odot}$).  This shows that PSBs at this epoch have structures that are very
similar to both the low-mass passive and SF3 populations.  Errors in the structural parameters ($1\sigma$)
are determined from the variance between fits performed on $100$ simulated $\widetilde\mu(r)$ profile stacks
generated via a bootstrap method.  For all populations studied, the $H_{\rm\,F160W}$ structural parameters
are entirely consistent with those obtained from the $J_{\rm\,F125W}$ $\widetilde\mu(r)$ profiles (see
Table~\ref{Sersic results}).}
\end{figure*}

%% Fit results %%------------------------------------------------------------------------------------------%%
Fig.~\ref{Sersic fits}\,(a/b) shows a comparison of the $H_{\rm\,F160W}$ $\widetilde\mu(r)$ structural
properties ($r_{\rm e}$,~$n$) for each galaxy population and for the two epochs ($0.5 < z < 1$ and
$1 < z < 2$).  A similar comparison of the $J_{\rm\,F125W}$ $\widetilde\mu(r)$ profiles yields entirely
consistent results, with respect to the errors (see Table~\ref{Sersic results}).  For both epochs, the
general structure of the passive and star-forming populations are as expected.  Passive galaxies are compact
[$0.2 < r_{\rm e} < 0.3$ arcsec ($1.7$--$2\rm\,kpc$)] and have high S{\'e}rsic indices ($n\sim3.3$),
indicating their spheroidal nature.  In contrast, star-forming galaxies have significantly more extended
stellar distributions [$r_{\rm e} > 0.3$ arcsec ($>2.7\rm\,kpc$)] with the lower S{\'e}rsic indices
($1 < n < 2$) typical of their disc-dominated structures.  For the star-forming sub-populations (SF1, SF2,
SF3) at both epochs, we observe a slight increase in S{\'e}rsic index from SF1~$\rightarrow$~SF3 (i.e.\ with
increasing mean stellar age), while at $z > 1$ we also observe a significant decrease in size
[$r_{\rm e}\sim0.5\rightarrow0.3$~arcsec ($4.2\rightarrow2.6\rm\,kpc$)].  This trend suggests an increase in
the dominance of the bulge component towards older star-forming galaxies.  Interestingly, for PSB galaxies we
observe significant differences in their structure at different epochs.  At $z > 1$, the PSBs are extremely
compact [$r_{\rm e}\sim0.13$ arcsec ($\sim1.1\rm\,kpc$)] and of high S{\'e}rsic index ($n\sim3.2$), with
structures similar to the passive population but considerably more compact (by $\sim40$ per cent).  In
contrast, at $z < 1$ the PSBs have significantly different structures.  At this epoch, PSBs are still
relatively compact [$r_{\rm e}\sim0.22$ arcsec ($\sim1.6\rm\,kpc$)], but exhibit much lower S{\'e}rsic
indices ($n\sim1.7$) than the PSBs at $z > 1$.

%% Best-fit overview %%------------------------------------------------------------------------------------%%

%% Fit results: High-z %%----------------------------------------------------------------------------------%%
With respect to these results, it is important to take into consideration the mass distributions of the
respective galaxy populations (see Fig.~\ref{mass-vs-z}), due to the well-established correlations between
mass and galaxy structure \cite[e.g.~the mass--size relation;][]{Shen_etal:2003}.  At high redshift
($z > 1$), our galaxies are all of high mass ($M_* > 10^{10}\rm\,M_{\odot}$) and have relatively similar mass
distributions.  Nonetheless, in our high-$z$ $\widetilde\mu(r)$ profiles it is possible that PSBs appear more
compact than the passive population due to slight differences in their respective mass distributions.  To
address this issue, we repeat our high-$z$ analysis using two narrower mass bins
($10^{10} < M_* < 10^{10.5}\rm\,M_{\odot}$ and $10^{10.5} < M_* < 10^{11}\rm\,M_{\odot}$).  In both cases,
PSBs remain significantly more compact than the passive population, and we observe the same trends in our
$\widetilde\mu(r)$ structural parameters (with respect to the errors).  Consequently, we conclude that any
slight differences in the mass distributions of our samples have no significant effect on our results for
$z > 1$.

%% Fit results: Low-z Low-M %%-----------------------------------------------------------------------------%%
In contrast, at lower redshift ($0.5 < z < 1$), there are more significant differences between the mass
distributions of our galaxy samples.  At this epoch, PSBs are generally of low stellar mass
($10^9 < M_* < 10^{10}\rm\,M_{\odot}$), while the passive and star-forming populations have a wide range of
masses ($10^9 < M_* < 10^{11.5}\rm\,M_{\odot}$).  Therefore, to perform a fair comparison at this epoch, we
need to match the passive and star-forming galaxies to a mass range comparable to the PSB population
(i.e.\ $10^9 < M_* < 10^{10}\rm\,M_{\odot}$).  Fig.~\ref{Sersic fits}\,(c) shows the structural parameters
for the resultant mass-matched $\widetilde\mu(r)$ profiles.  For the general star-forming population,
restricting the mass range has little effect on their typical structural properties.  However, there is a
significant change in the general structure of the passive population, which now resembles that of
PSBs [$r_{\rm e}\sim0.2$ arcsec ($\sim1.5\rm\,kpc$), $n\sim2$].  Therefore, we conclude that at this epoch,
PSBs have similar structures to those of the low-mass passive population (i.e.\ passive discs), the
population into which they will most likely evolve.  We explore this result in more detail in the following
sections.  We note that the low-mass SF3 population also resembles PSBs in structure, but since these
galaxies are unlikely to be PSB progenitors (due to their low sSFR; see Section~\ref{Sample selection}),
we do not consider this result any further.

%% Angular scale %%----------------------------------------------------------------------------------------%%
For each epoch studied ($0.5 < z < 1$ and $1 < z < 2$), it is also important to consider any differences in
the redshift distributions between the galaxy populations (see Fig.~\ref{mass-vs-z}), due to the potential
for structural evolution across the epoch.  A further consideration is the angular scale, which over
$1 < z < 2$ is relatively constant ($\sim8.5\rm\,kpc\,arcsec^{-1}$), but for $0.5 < z < 1$ varies more
significantly \mbox{($6$--$8\rm\,kpc\,arcsec^{-1}$)}.  To address these issues, we split both the
intermediate- and high-$z$ epochs into two narrower sub-epochs and assess the effect on our
$\widetilde\mu(r)$ profiles.  These sub-epochs are redshifts \mbox{$0.5$--$0.75$} and \mbox{$0.75$--$1.0$}
for the intermediate-$z$ epoch, and \mbox{$1.0$--$1.5$} and \mbox{$1.5$--$2.0$} for the high-$z$ epoch.  For
both the intermediate- and high-$z$ epochs, we find no significant differences in the structural parameters
between the respective sub-epochs for each galaxy population (with respect to the errors; see
Table~\ref{Sersic results}).  The only exceptions are: i)~high-$z$ passive and PSB galaxies, where there is a
slight indication of a higher S{\'e}rsic index $n$ at $1 < z < 1.5$ than at $1.5 < z  < 2$, but this has no
significant effect on the overall trends observed; and ii)~intermediate-$z$ galaxies, where we find some
minor differences in effective radius (in arcsec) between the two sub-epochs ($\delta{r_{\rm e}} < 0.05$
arcsec), as might be expected.  However, these differences equate to a change in the physical effective
radius of $<0.15\rm\,kpc$, and consequently have no significant effect on the overall trends for the
$0.5 < z <1$ epoch presented here.

%%%%%%%%%%%%%%%%%%%%%%%%%%%%%%%%%%%%%%%%%%%%%%%%%%%%%%%%%%%%%%%%%%%%%%%%%%%%%%%%%%%%%%%%%%%%%%%%%%%%%%%%%%%%%
\subsection[]{Further robustness tests}

\label{Further robustness tests}

%% PSF effects %%------------------------------------------------------------------------------------------%%
In this study, a careful treatment of the PSF is critical for determining structural properties ($r_{\rm e}$,
$n$), particularly for compact galaxy populations (e.g.\ PSBs).  In our profile fitting, we take account of
the PSF by using a library of PSF-convolved models (see Section~\ref{Profile fitting}).  Nonetheless, we
assess whether PSF effects could cause a bias in our fitted structural parameters by adopting a similar
method to that developed by \cite{Szomoru_etal:2010, Szomoru_etal:2012}, and deconvolve our
$\widetilde\mu(r)$ profiles for the PSF.  To achieve this, we first calculate a residual profile
$\widetilde\mu_{\rm res}(r)$ by subtracting our best fit profile (which is PSF-convolved) from the
median-stacked profile~$\widetilde\mu(r)$.  We then add the $\widetilde\mu_{\rm res}(r)$ profile to the
analytical form of the best fit (i.e.\ the PSF-deconvolved model) and obtain a corrected profile
$\widetilde\mu_{\rm corr}(r)$, that is effectively deconvolved for the PSF [at least to first order, since
the residual profile $\widetilde\mu_{\rm res}(r)$ remains PSF-convolved].  Finally, we perform an analytical
single S{\'e}rsic fit on these $\widetilde\mu_{\rm corr}(r)$ profiles to obtain structural parameters that
are corrected for PSF effects.  Using our median-stacked profiles $\widetilde\mu(r)$ (both $J_{\rm\,F125W}$
and $H_{\rm\,F160W}$; see~Fig.~\ref{Median profiles}), we find that for each galaxy population (both epochs),
the effect of the PSF correction on our structural parameters is minimal.  The effect on both $r_{\rm e}$ and
$n$ is typically $< 1$ per cent, and always $< 3$~per~cent, even for compact galaxy populations (e.g.\ PSBs).
Given that these differences are smaller than the stacking errors in our fitted structural parameters (see
Table~\ref{Sersic results}), we conclude that PSF effects have no significant impact on the results of this
study.  For a similar assessment of our optical profiles, see Section~\ref{Optical imaging}.

%% Sky subtraction error %%--------------------------------------------------------------------------------%%
Another important consideration is the robustness of our $\widetilde\mu(r)$ profiles, and their fitted
structural parameters ($r_{\rm e}$, $n$), to the error in the sky subtraction (see
Section~\ref{sky subtraction}).  To address this issue, we use the following Monte Carlo analysis for both
our $J_{\rm\,F125W}$ and $H_{\rm\,F160W}$ $\widetilde\mu(r)$ profiles.  For each galaxy population, we
generate $100$ median-stacked profiles $\widetilde\mu_{\rm sim}(r)$ using the same procedure as in
Section~\ref{Stacked light profiles}, but with random sky offsets applied to each of the individual profiles.
These offsets are generated by randomly sampling a Gaussian distribution with a standard deviation equal to
the typical $1\sigma$ error in the sky subtraction $\widetilde\sigma_{\rm sky}$ (see
Section~\ref{sky subtraction}).  The robustness of our results to the sky subtraction error is then
determined from the variance between profile fits performed on these $\widetilde\mu_{\rm sim}(r)$ profiles.
For each galaxy population, we find that the effect of the sky subtraction error is minimal, with
the effect on both $r_{\rm e}$ and $n$ typically $<5$ per cent (for both $J_{\rm\,F125W}$ and
$H_{\rm\,F160W}$).  We note that the differences observed in the structural parameters of our
$\widetilde\mu(r)$ profiles (see Fig.~\ref{Sersic fits}) are robust to uncertainties at this level.
Therefore, we conclude that errors in the sky subtraction have no significant effect on the results for our
near-infrared $\widetilde\mu(r)$ profiles.

%% GALFIT comparisons %%-----------------------------------------------------------------------------------%%
Finally, we note that in this study, the use of the stacked one-dimensional radial $\mu(r)$ profiles could
also introduce some uncertainty to the fitted structural parameters due to the loss of azimuthal information,
and an alternative would be to use two-dimensional analyses (i.e.\ stack galaxy images; see
Section~\ref{Radial light profiles}).  However, the difference in structural parameters obtained from these
two methods should be minimal, with a scatter at the $10$--$20$ per cent level
\citep[see e.g.][]{deJong:1996, MacArthur_etal:2003, McDonald_etal:2011}.  We confirm this expected
consistency by comparing the structural properties ($r_{\rm e}$, $n$) from our individual galaxy $\mu(r)$
profiles (determined from our profile fitting method; see Section~\ref{Profile fitting}), with those obtained
for the same galaxies in \cite{vanderWel_etal:2012}.  This work used two-dimensional S{\'e}rsic models (via
{\sc galfit}; \citealt{Peng_etal:2002}) to measure the near-infrared ($J_{\rm\,F125W}$ and $H_{\rm\,F160W}$)
structural properties of galaxies in CANDELS--UDS.  As expected, for both $J_{\rm\,F125W}$ and
$H_{\rm\,F160W}$, we find good agreement between the structural properties determined from the two fitting
methods, with a characteristic scatter ($\sigma_{\rm MAD}$) at the $10$--$20$~per~cent level (typically
$<10$~per~cent for $r_{\rm e}$ and $<20$~per~cent for $n$).  We note that the differences observed in the
structural properties of our $\widetilde\mu(r)$ profiles (see Fig.~\ref{Sersic fits}) are robust to
uncertainties at this level.

%%%%%%%%%%%%%%%%%%%%%%%%%%%%%%%%%%%%%%%%%%%%%%%%%%%%%%%%%%%%%%%%%%%%%%%%%%%%%%%%%%%%%%%%%%%%%%%%%%%%%%%%%%%%%
\subsection{Individual light $\mu(r)$ profiles}

\label{Individual profiles}

%% Structural properties %%--------------------------------------------------------------------------------%%
\begin{figure*}
\includegraphics[width=0.95\textwidth]{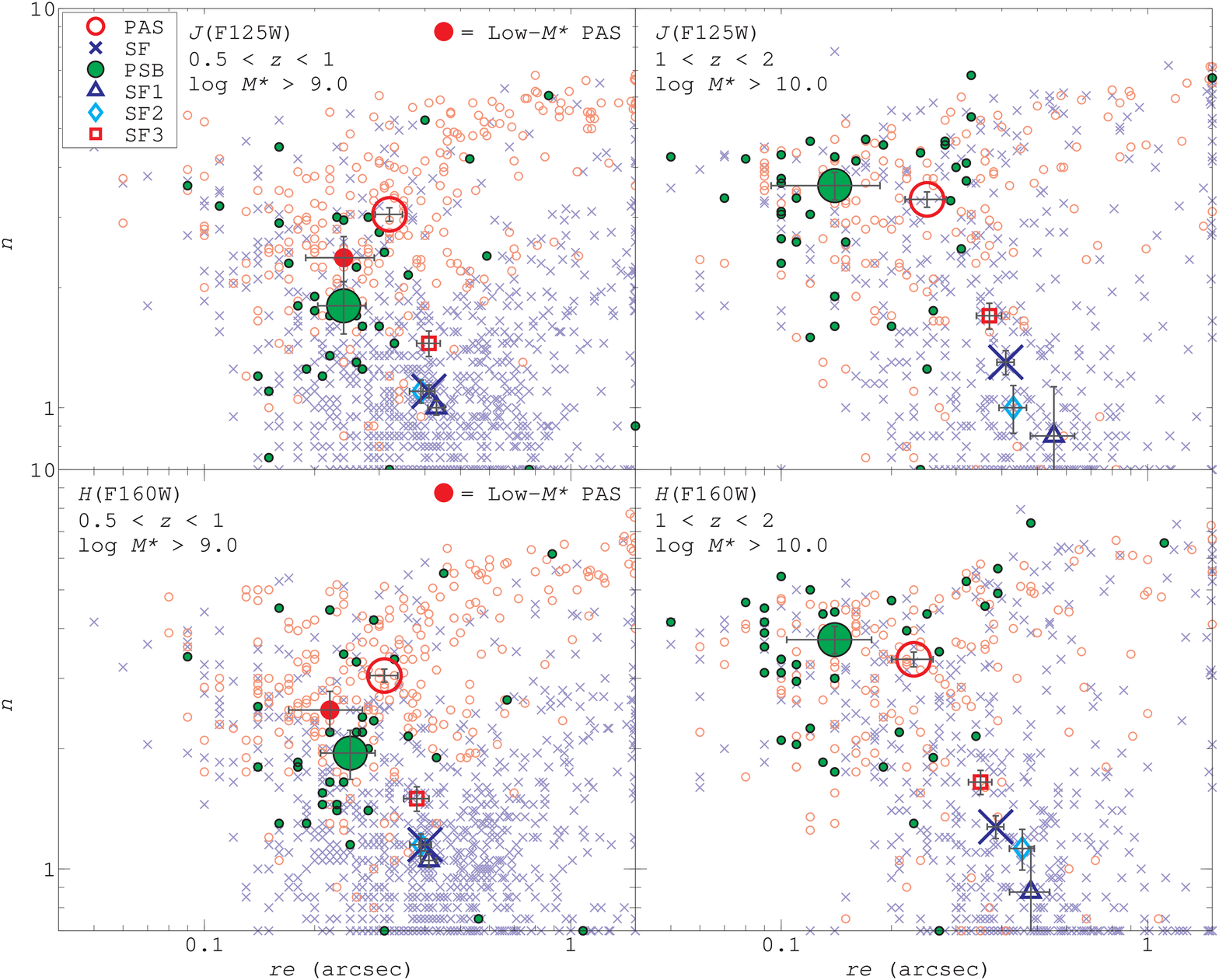}
\centering
\caption{\label{model 1 indv} Structural properties (effective radius $r_{\rm e}$, S{\'e}rsic index $n$) from
the single S{\'e}rsic fits to our individual $J_{\rm\,F125W}$ (top row) and $H_{\rm\,F160W}$ (bottom row) 
$\mu(r)$ profiles.  Results are shown for the passive (PAS), star-forming (SF) and PSB populations at two
epochs: $0.5 < z < 1$ (left-hand panels) and $1 < z < 2$ (right-hand panels).  Large symbols represent the
median structural properties for each population, with associated $1\sigma$ errors.  Median structural
properties for the star-forming sub-populations (SF1 $\rightarrow$ SF3; based on decreasing sSFR) and the
low-mass passive population ($10^{9} < M_* < 10^{10}\rm\,M_{\odot}$), are also shown.  In~all cases, the
median structural properties are very similar to those obtained from our $\widetilde\mu(r)$ profiles, with
the same trends observed between the different galaxy populations (compare to Fig.~\ref{Sersic fits}).  These
results support our findings that i) PSBs at $z > 1$ are extremely compact and spheroidal
[$r_{\rm e}\sim0.14$ arcsec ($\sim1.2\rm\,kpc$), $n\sim3.5$], with structures similar to massive passive
galaxies but more compact; and ii) PSBs at $z < 1$ are also compact but with low $n$
[$r_{\rm e}\sim0.24$~arcsec ($\sim1.5\rm\,kpc$), $n\sim1.8$], and have structures similar to the low-mass
passive population.}
\end{figure*}

%% Fit results %%------------------------------------------------------------------------------------------%%
In this study, the median-stacked profiles $\widetilde\mu(r)$ are the principal focus, providing
well-constrained typical structural properties for our galaxy populations (see
Section~\ref{Profile fitting: stacked light profiles}).  However, the S{\'e}rsic fits for our individual
galaxy $\mu(r)$ profiles, despite having greater uncertainty due to the lower signal-to-noise, can also
provide insight into the nature of our galaxy populations.   The resultant structural properties
($r_{\rm e}$,~$n$) for our $J_{\rm\,F125W}$ and $H_{\rm\,F160W}$ individual $\mu(r)$ profiles are presented
in Fig.~\ref{model 1 indv}.  In all cases, we find that the median structural properties are very similar to
those obtained from our $\widetilde\mu(r)$ profiles, with the same trends observed between the different
galaxy populations (compare to Fig.~\ref{Sersic fits}).  Similar trends in the structural properties are
also observed in both the $J_{\rm\,F125W}$ and $H_{\rm\,F160W}$ wavebands.  Furthermore, repeating these
analyses using the physical effective radius $r_{\rm e}({\rm kpc})$, as determined using the photometric
redshift for each galaxy, produces analogous distributions and trends in the structural parameters.  The
median $r_{\rm e}({\rm kpc})$ from these fits also confirm the physical $r_{\rm e}(\rm kpc)$ determined for
each galaxy population from our $\widetilde\mu(r)$ profiles (see
Section~\ref{Profile fitting: stacked light profiles}).  Furthermore, as with our $\widetilde\mu(r)$
profiles, we find PSF-effects and sky subtraction errors ($\pm\widetilde\sigma_{\rm sky}$) to have a minimal
influence on these individual fits, with a typical impact on the structural parameters (both $r_{\rm e}$ and
$n$) of $< 5$ per cent and $<10$ per cent, respectively.

With respect to these individual fits, we note that in some cases the profile fits have run into constraints
caused by limitations in the model grid (e.g. at $n = 0.7$).  These cases are rare in the passive and PSB
populations ($<5$ per cent), but more significant in the star-forming population ($\sim20$ per cent).  For
these galaxies, the {\sc galfit} structural parameters from \cite{vanderWel_etal:2012} suggest that the true
S{\'e}rsic index is actually $\sim0.7\pm0.1$ in most cases.  Consequently, we retain these fits in our median
analysis.  However, we note that removing these cases only affects the median S{\'e}rsic index~$n$ for the
star-forming galaxies at $1 < z < 2$ ($n\sim1.3\rightarrow1.8$), and the overall trends in the structural
parameters remain unaffected.

In addition to the median properties, these individual fits can also provide some further insight into the
nature of rare sub-populations.  For example, we find that the rare, high-mass PSBs at $z < 1$
($M_* > 10^{10}\rm\,M_{\odot}$; see Fig.~\ref{mass-vs-z}) have structures which are similar to PSBs at
$z > 1$, exhibiting analogous high~$n$ but also slightly larger $r_{\rm e}$ ($r_{\rm e} > 2\rm\,kpc$).  We
shall return to this result in Section~\ref{Discussion: intermediate redshift}.

In conclusion, these individual $\mu(r)$ profile fits support our findings that PSBs at $z > 1$ are extremely
compact and spheroidal [$r_{\rm e}\sim0.14$~arcsec ($\sim1.2\rm\,kpc$), $n\sim3.5$], while PSBs at $z < 1$
are generally compact but with more disc-like structures [$r_{\rm e}\sim0.24$ arcsec ($\sim1.5\rm\,kpc$),
$n\sim1.8$].  Furthermore, the consistency between these results and those from our median-stacked
$\widetilde\mu(r)$ profiles (see Section~\ref{Profile fitting: stacked light profiles}), confirms the
effectiveness of our stacking analysis and demonstrates that our $\widetilde\mu(r)$ profiles are truly
representative of their respective galaxy populations.

%%%%%%%%%%%%%%%%%%%%%%%%%%%%%%%%%%%%%%%%%%%%%%%%%%%%%%%%%%%%%%%%%%%%%%%%%%%%%%%%%%%%%%%%%%%%%%%%%%%%%%%%%%%%%
\section[]{Two-component fits}

\label{Two-component fits}

%% Overview %%---------------------------------------------------------------------------------------------%%
To explore the nature of PSBs in more detail, we extend our morphological analyses to allow for $\mu(r)$
profiles containing multiple components.  Such analyses complement our single S{\'e}rsic fits, and provide
further insight into the potential evolutionary histories of these galaxies.  For example, for high-$z$ PSBs
we can investigate i)~whether their stellar distributions are really compact, or if this is due to the point
source emission from either an AGN or unresolved decaying nuclear starburst; and ii) whether these galaxies
are genuinely spheroidally dominated, or if a bulge--disc system could equally account for their $\mu(r)$
profiles.  To address these issues, two models will be considered: i)~a S{\'e}rsic profile with a central
point source (see Section~\ref{Sersic profile + point source}); and ii)~a bulge--disc system comprising a
\cite{deVaucouleurs:1959} bulge plus an exponential disc (see~Section~\ref{BD decomposition}).  For these
fits, our stacking analysis is particularly important, enabling us to maximise signal-to-noise, particularly
for the outer galactic regions.  This is necessary for the identification of faint components (e.g.\ faint
outer discs) that may not be detected in our individual profiles.

%%%%%%%%%%%%%%%%%%%%%%%%%%%%%%%%%%%%%%%%%%%%%%%%%%%%%%%%%%%%%%%%%%%%%%%%%%%%%%%%%%%%%%%%%%%%%%%%%%%%%%%%%%%%%
\subsection[]{S{\'e}rsic profile + point source}

\label{Sersic profile + point source}

%% Two-component best-fit profiles %%----------------------------------------------------------------------%%
\begin{figure*}
\includegraphics[width=0.51\textwidth]{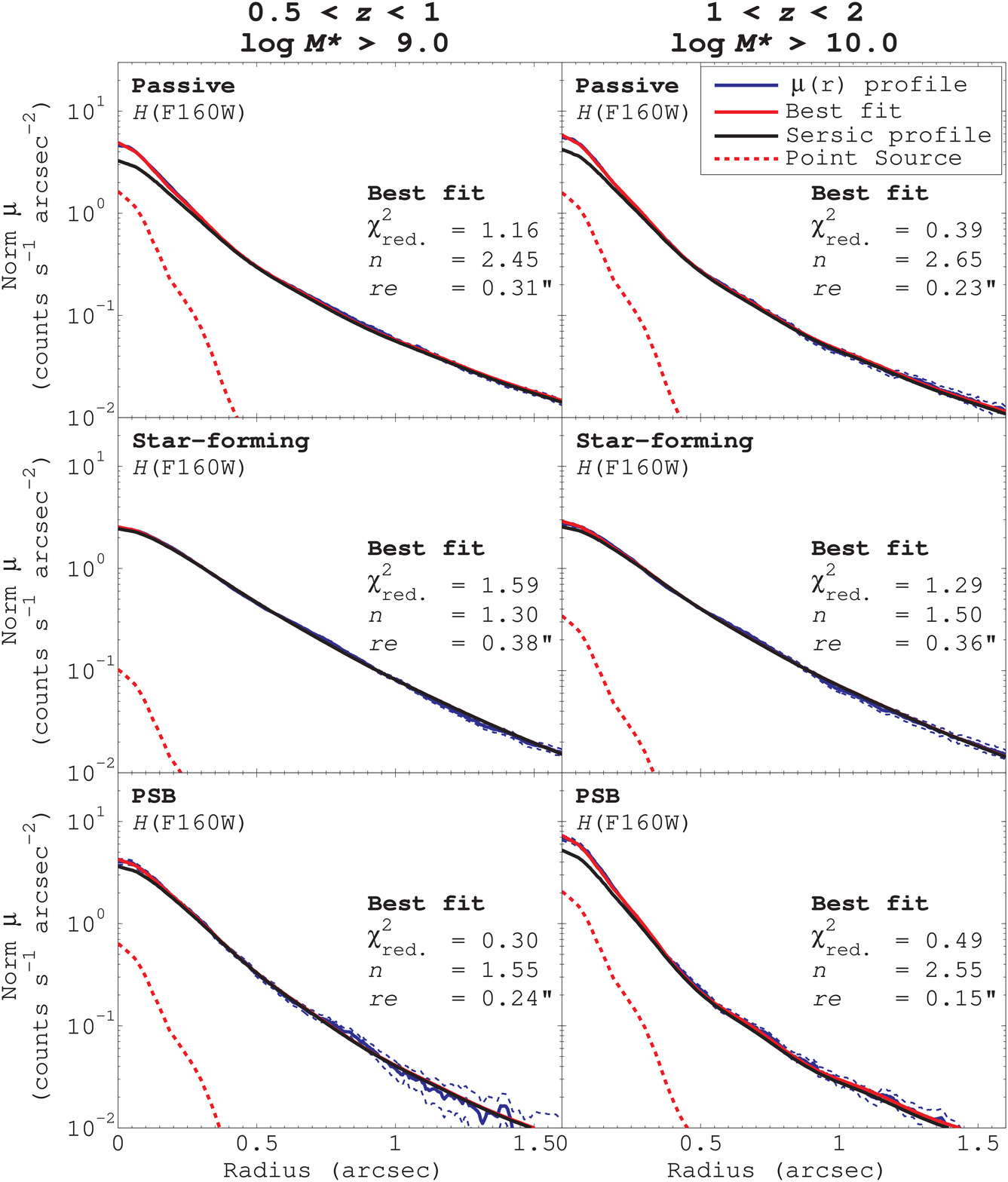}
\hspace{-0.5cm}
\includegraphics[width=0.51\textwidth]{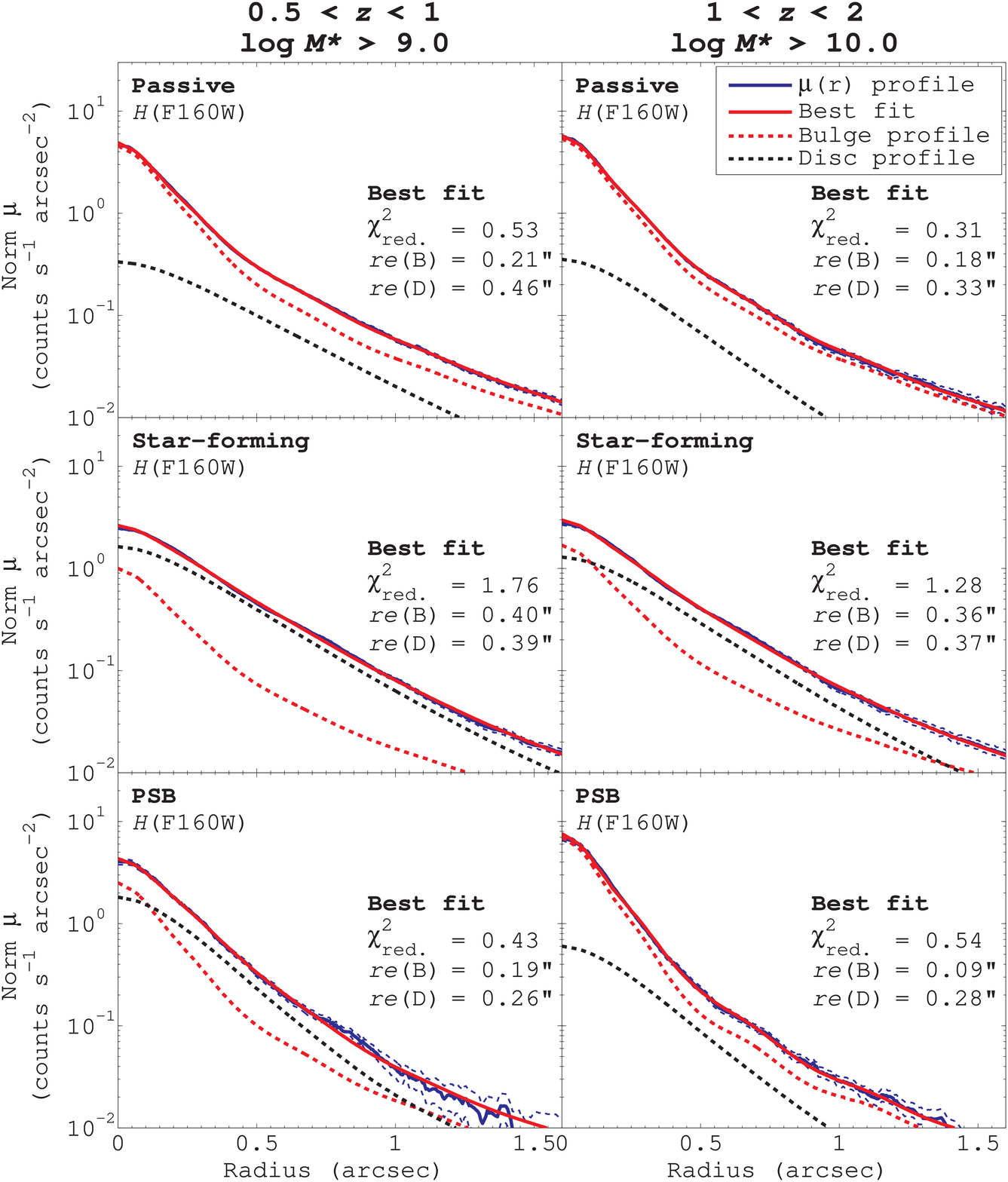}
\centering
\hspace{-0.5cm}
\caption{\label{two-component profiles} Two-component fits to our median-stacked $H_{\rm\,F160W}$ light
profiles $\widetilde\mu(r)$.  Left-hand panels: best-fits using a model comprising a S{\'e}rsic profile +
point source.  These fits show the {\em maximal} point source contribution, which is relatively minor for all
populations (see also Fig.~\ref{model 2 indv}).  Right-hand panels: best-fits using a model comprising a
de~Vaucouleurs ($r^{1/4}$) bulge + exponential disc.  These fits yield the {\em maximal} likely bulge
contribution, and show that at both epochs, passive galaxies are bulge-dominated, while star-forming galaxies
are disc-dominated.  In contrast, PSBs exhibit significantly different structures at different epochs:
bulge-dominated at $1 < z < 2$, but disc-dominated at $0.5 < z < 1$.
\vspace{-0.0cm}}
\end{figure*}

%% Best-fit structural parameters %%-----------------------------------------------------------------------%%
\begin{table*}
\centering
\begin{minipage}{175mm}
\centering
\caption{\label{model 2 fits} Near-infrared S{\'e}rsic + point source fits: the structural properties for the
S{\'e}rsic component of our median-stacked $J_{\rm\,F125W}$ and $H_{\rm\,F160W}$ light profiles
$\widetilde\mu(r)$.  Structural properties ($r_{\rm e}$,~$n$) are shown for different galaxy populations
(passive, star-forming, PSB) at two different epochs, $0.5 < z < 1$ and $1 < z < 2$.  Errors in the
structural parameters ($1\sigma$) are determined from the variance between fits performed on $100$ simulated
$\widetilde\mu(r)$ profile stacks generated via a bootstrap method.  For details of the corresponding point
source component, see Table~\ref{two-component results}.}
\begin{tabular}{lccccccccccc}
\hline
Galaxy		&\multicolumn{5}{c}{$0.5 < z < 1$}							&{}	&\multicolumn{5}{c}{$1 < z < 2$}							\\
Population	&\multicolumn{2}{c}{$J_{\rm\,F125W}$}	&{}	&\multicolumn{2}{c}{$H_{\rm\,F160W}$}	&{}	&\multicolumn{2}{c}{$J_{\rm\,F125W}$}	&{}	&\multicolumn{2}{c}{$H_{\rm\,F160W}$}	\\
{}		&{$n$}		&{$r_{\rm e}$}		&{}	&{$n$}		&{$r_{\rm e}$}		&{}	&{$n$}		&{$r_{\rm e}$}		&{}	&{$n$}		&{$r_{\rm e}$}		\\
{}		&{}		&{(arcsec)}		&{}	&{}		&{(arcsec)}		&{}	&{}		&{(arcsec)}		&{}	&{}		&{(arcsec)}		\\[0.5ex]
\hline																									\\[-1.5ex]
Passive		&{$2.50\pm0.19$}&{$0.32\pm0.01$}	&{}	&{$2.45\pm0.19$}&{$0.31\pm0.01$}	&{}	&{$2.70\pm0.29$}&{$0.23\pm0.02$}	&{}	&{$2.65\pm0.26$}&{$0.23\pm0.02$}	\\[1ex]	
Star-forming	&{$1.25\pm0.06$}&{$0.39\pm0.01$}	&{}	&{$1.30\pm0.05$}&{$0.38\pm0.01$}	&{}	&{$1.60\pm0.10$}&{$0.38\pm0.01$}	&{}	&{$1.50\pm0.08$}&{$0.36\pm0.01$}	\\[1ex]
\ \ SF1		&{$1.10\pm0.05$}&{$0.40\pm0.01$}	&{}	&{$1.10\pm0.05$}&{$0.39\pm0.01$}	&{}	&{$1.20\pm0.16$}&{$0.52\pm0.06$}	&{}	&{$1.20\pm0.15$}&{$0.47\pm0.06$}	\\
\ \ SF2		&{$1.35\pm0.08$}&{$0.39\pm0.02$}	&{}	&{$1.30\pm0.08$}&{$0.38\pm0.02$}	&{}	&{$1.45\pm0.13$}&{$0.43\pm0.02$}	&{}	&{$1.30\pm0.11$}&{$0.41\pm0.02$}	\\
\ \ SF3		&{$1.50\pm0.15$}&{$0.39\pm0.02$}	&{}	&{$1.70\pm0.14$}&{$0.36\pm0.02$}	&{}	&{$1.70\pm0.12$}&{$0.34\pm0.01$}	&{}	&{$1.60\pm0.12$}&{$0.32\pm0.01$}	\\[1ex]
Post-starburst	&{$1.55\pm0.29$}&{$0.23\pm0.02$}	&{}	&{$1.55\pm0.27$}&{$0.24\pm0.01$}	&{}	&{$3.50\pm0.45$}&{$0.14\pm0.02$}	&{}	&{$2.55\pm0.46$}&{$0.15\pm0.03$}	\\
\hline
\end{tabular}
\end{minipage}
\end{table*}

%% Motivation %%-------------------------------------------------------------------------------------------%%
In this work, we find PSBs to be extremely compact, particularly at $z > 1$ (see Figs.~\ref{Sersic fits} and
\ref{model 1 indv}).  One potential explanation is that these PSBs contain significant point source emission,
from either an AGN or unresolved decaying nuclear starburst.  This scenario would result in a $\mu(r)$
profile that would be inadequately modelled by a single S{\'e}rsic profile, and structural parameters biased
towards low effective radii $r_{\rm e}$ and high S{\'e}rsic index $n$.  Considering PSBs are recently
quenched, the presence of an AGN might actually be expected \citep[see e.g.][]{Hopkins:2012}, and may cause
their host galaxies to appear compact.  Alternatively, a decaying nuclear starburst may also be expected in
PSBs, since many quenching processes are expected to result in gas being funnelled into the central regions
of the galaxy.  This could potentially trigger a nuclear starburst, and lead to a central concentration in
the stellar distribution of the quenched system.

%% Profile fitting %%--------------------------------------------------------------------------------------%%
To address this issue, we include point source emission in our profile-fitting model, and assess the effect
on the structure ($r_{\rm e}$,~$n$) of our galaxy populations.  For profile fitting, the point source is
modelled using the $\mu(r)$ profile of the relevant PSF (see Section~\ref{PSF determination}), and added to
the S{\'e}rsic profiles of the model library $\{\mu_{\rm mock}(r)\}$ (see Section~\ref{Profile fitting}).  To
account for varying strengths of point source emission, we use more than $100$ variations per model profile,
with point source emission accounting for between $0$--$100$ per cent of the peak/central flux.  Each model
$\mu(r)$ profile ($>2000\,000$ in total) is then compared with the measured $\mu(r)$ profile and a $\chi^2$
minimisation used to obtain the best fit.  Note that since our $\widetilde\mu(r)$ profiles are already
well-defined by a single S{\'e}rsic profile (see Fig.~\ref{Sersic profiles}), the addition of a point source
will not significantly improve the quality of the fit.  Therefore, these fits are not intended to yield the
actual point source contribution, but the {\em maximal} likely contribution to the $\mu(r)$ profile.  The
resultant best-fits for the $H_{\rm\,F160W}$ $\widetilde\mu(r)$ profiles are presented in
Fig.~\ref{two-component profiles}.  These profiles show that for all galaxy populations (both epochs), the
{\em maximal} point source contribution is relatively minor and has little effect on their structural
properties.

\begin{figure*}
\includegraphics[width=0.8\textwidth]{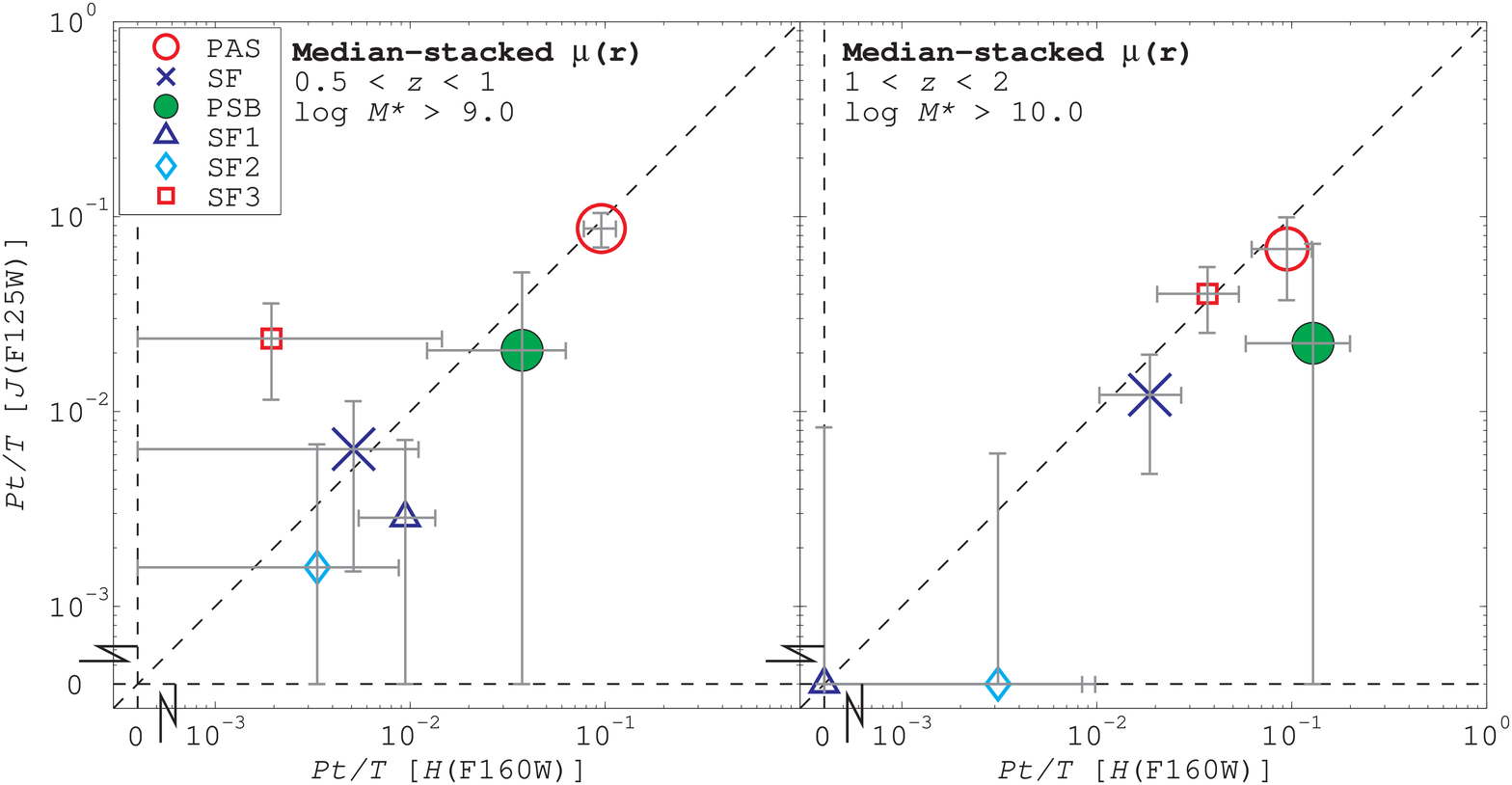}
\includegraphics[width=0.8\textwidth]{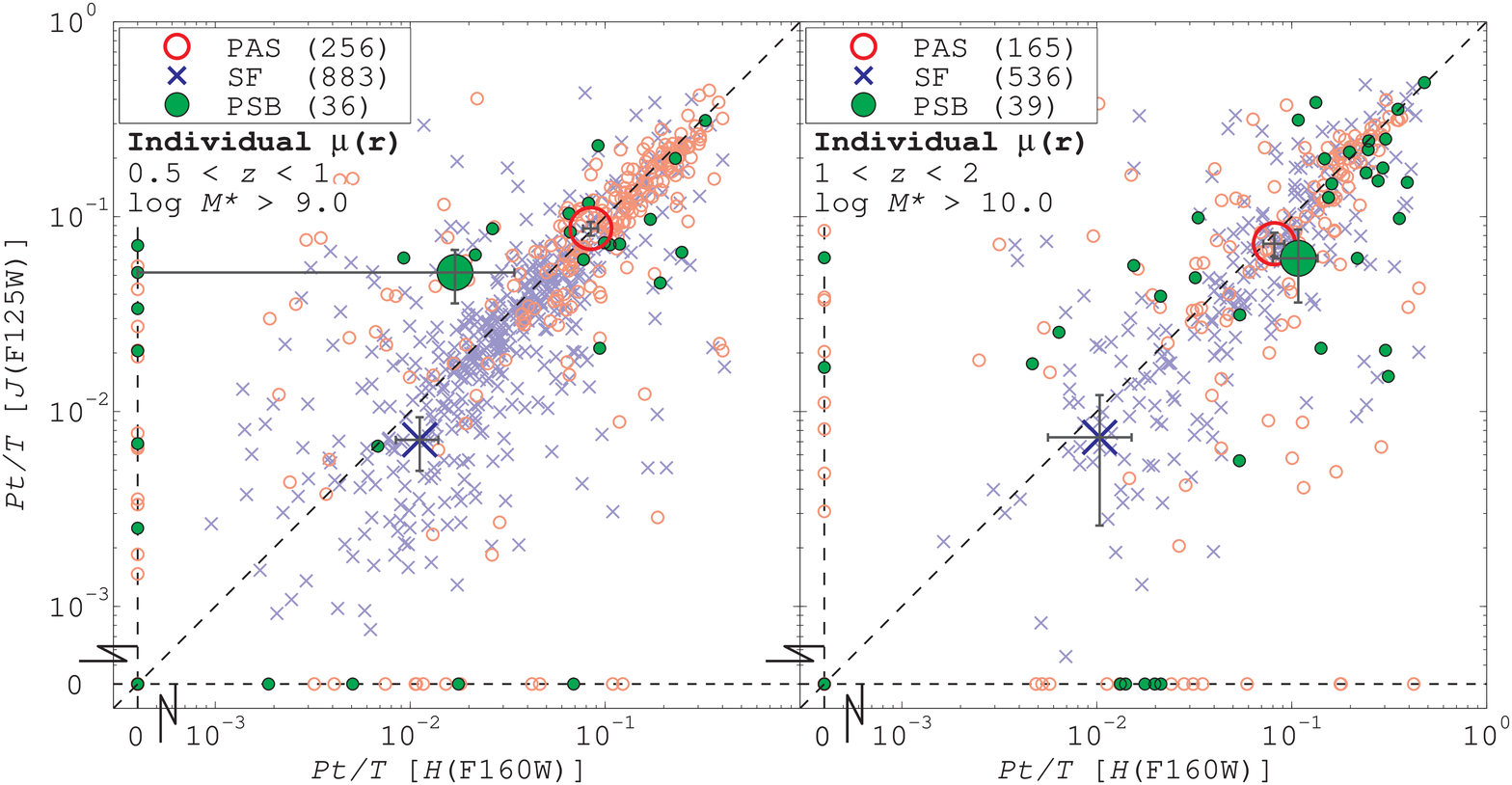}
\centering
\caption{\label{model 2 indv} The {\em maximal} point source emission for our galaxy populations at two
epochs: $0.5 < z < 1$ (left-hand panels) and $1 < z < 2$ (right-hand panels).  A comparison of the maximal
point source contribution ($\mathit{Pt}/T$) in $J_{\rm\,F125W}$ and $H_{\rm\,F160W}$ for the median-stacked
light profiles $\widetilde\mu(r)$ (top panels), and for our individual $\mu(r)$ profiles (bottom panels).
In the bottom panels, large symbols represent the median $\mathit{Pt}/T$ in each population, with associated
$1\sigma$ errors in the median position.  For each population, typically $<15$ per cent of the total light
emitted can be attributed to a potential point source.}
\end{figure*}

\begin{table*}
\centering
\begin{minipage}{175mm}
\centering
\caption{\label{two-component results} Near-infrared multiple component fits: the {\em maximal} contribution
of point-source/bulge light to the total light from the galaxy ($\mathit{Pt}/T$ and $B/T$, respectively) for
our median-stacked $J_{\rm\,F125W}$ and $H_{\rm\,F160W}$ light profiles $\widetilde\mu(r)$.  Results are
shown for different galaxy populations (passive, star-forming, PSB) at two different epochs, $0.5 < z < 1$
and $1 < z < 2$.  Errors in these measurements ($1\sigma$) are determined from the variance between fits
performed on $100$ simulated $\widetilde\mu(r)$ profile stacks generated via a bootstrap method.}
\begin{tabular}{lccccccccccc}
\hline
Galaxy		&\multicolumn{5}{c}{$0.5 < z < 1$}							&{}	&\multicolumn{5}{c}{$1 < z < 2$}							\\
Population	&\multicolumn{2}{c}{$J_{\rm\,F125W}$}	&{}	&\multicolumn{2}{c}{$H_{\rm\,F160W}$}	&{}	&\multicolumn{2}{c}{$J_{\rm\,F125W}$}	&{}	&\multicolumn{2}{c}{$H_{\rm\,F160W}$}	\\
{}		&{$\mathit{Pt}/T$}	&{$B/T$}		&{}	&{$\mathit{Pt}/T$}	&{$B/T$}		&{}	&{$\mathit{Pt}/T$}	&{$B/T$}		&{}	&{$\mathit{Pt}/T$}	&{$B/T$}		\\
{}		&{($\times10^{-2}$)}	&{}		&{}	&{($\times10^{-2}$)}	&{}		&{}	&{($\times10^{-2}$)}	&{}		&{}	&{($\times10^{-2}$)}	&{}		\\[0.5ex]
\hline																									\\[-1.5ex]
Passive		&{$8.70\pm1.75$}&{$0.78\pm0.04$}	&{}	&{$9.56\pm1.79$}&{$0.76\pm0.04$}	&{}	&{$6.83\pm3.10$}&{$0.83\pm0.04$}	&{}	&{$9.46\pm3.23$}&{$0.84\pm0.04$}	\\[1ex]	
Star-forming	&{$0.64\pm0.49$}&{$0.19\pm0.03$}	&{}	&{$0.51\pm0.59$}&{$0.21\pm0.02$}	&{}	&{$1.22\pm0.74$}&{$0.41\pm0.04$}	&{}	&{$1.87\pm0.84$}&{$0.36\pm0.03$}	\\[1ex]
\ \ SF1		&{$0.29\pm0.43$}&{$0.07\pm0.03$}	&{}	&{$0.94\pm0.40$}&{$0.10\pm0.03$}	&{}	&{$0.00\pm0.79$}&{$0.15\pm0.09$}	&{}	&{$0.00\pm0.80$}&{$0.15\pm0.09$}	\\
\ \ SF2		&{$0.16\pm0.52$}&{$0.24\pm0.04$}	&{}	&{$0.33\pm0.54$}&{$0.22\pm0.04$}	&{}	&{$0.00\pm0.57$}&{$0.31\pm0.07$}	&{}	&{$0.31\pm0.67$}&{$0.28\pm0.06$}	\\
\ \ SF3		&{$2.37\pm1.22$}&{$0.37\pm0.07$}	&{}	&{$0.19\pm1.26$}&{$0.42\pm0.06$}	&{}	&{$4.02\pm1.49$}&{$0.48\pm0.05$}	&{}	&{$3.70\pm1.66$}&{$0.45\pm0.05$}	\\[1ex]
Post-starburst	&{$2.06\pm3.12$}&{$0.40\pm0.12$}	&{}	&{$3.75\pm2.53$}&{$0.40\pm0.12$}	&{}	&{$2.24\pm5.02$}&{$0.90\pm0.07$}	&{}	&{$12.86\pm7.05$}&{$0.77\pm0.08$}	\\
\hline
\end{tabular}
\end{minipage}
\vspace{0.2cm}
\end{table*}

%% Pt/T comparisons: average profiles %%-------------------------------------------------------------------%%
For each galaxy population, we determine the {\em maximal} point source contribution to the total light
emitted by the galaxy ($Pt/T$).  We compare these $\mathit{Pt}/T$ measurements for the $J_{\rm\,F125W}$ and
$H_{\rm\,F160W}$ $\widetilde\mu(r)$ profiles in Fig.~\ref{model 2 indv}.  These results show that for all
galaxy populations, typically $<15$ per cent of the total light emitted can be attributed to a potential
point source.  Consequently, point source emission is not a major component in our $\widetilde\mu(r)$
profiles.  The $\mathit{Pt}/T$ for both the $J_{\rm\,F125W}$ and $H_{\rm\,F160W}$ $\widetilde\mu(r)$
profiles are presented in Table~\ref{two-component results}, with $1\sigma$ errors determined from a similar
bootstrap analysis to that used for our single S{\'e}rsic fits (see
Section~\ref{Profile fitting: stacked light profiles}).  Furthermore, using a similar analysis to that
described in Section~\ref{Further robustness tests}, we find these measurements to be robust to sky
subtraction errors ($\pm\widetilde\sigma_{\rm sky}$), with typical effects on $\mathit{Pt}/T$ of $<10$ per
cent.

With respect to the S{\'e}rsic component, the resultant structural parameters ($r_{\rm e}$, $n$) are
presented in Table~\ref{model 2 fits}.  For each galaxy population, these are very similar to those produced
by our single S{\'e}rsic fits (see Table~\ref{Sersic results}), with the effective radius $r_{\rm e}$ being
relatively unchanged and the S{\'e}rsic index~$n$ only decreasing slightly by the inclusion of a point
source.  The decrease in $n$ is notably the strongest where the point source contribution is the most
significant, i.e.\ passive galaxies and \mbox{high-$z$} PSBs (see Figs.~\ref{two-component profiles} and
\ref{model 2 indv}).  However, the differences between the structural properties of each galaxy population
which are observed in our single S{\'e}rsic fits, all remain present (see Fig.~\ref{Sersic fits}).  High-$z$
PSBs ($z > 1$) remain compact [$r_{\rm e}\sim0.15$ arcsec ($\sim1.2\rm\,kpc$)] and of relatively high
S{\'e}rsic index ($n > 2.5$), even when the maximal contribution from a point source is taken into account.
These PSBs also remain considerably more compact than the passive population.  Consequently, point source
emission, from either an AGN or unresolved decaying nuclear starburst, is not sufficient to explain the
compact nature of massive PSBs at this epoch.

%% Pt/T comparisons: individual profiles %%----------------------------------------------------------------%%
For our individual $\mu(r)$ profiles, we also perform analogous two-component fits.  We note that these fits
have greater uncertainty than those for our $\widetilde\mu(r)$ profiles, due to the lower signal-to-noise.
Nonetheless, they may offer further insight into the nature of our galaxy populations.  The resultant
distributions of $\mathit{Pt}/T$ in both $J_{\rm\,F125W}$ and $H_{\rm\,F160W}$, are presented in
Fig.~\ref{model 2 indv}.  In general, we find the same trends in the $\mathit{Pt}/T$ of the galaxy
populations as observed for our $\widetilde\mu(r)$ profiles.  For both epochs, $\mathit{Pt}/T$ is generally
more significant in the passive population ($\mathit{Pt}/T_{\rm median}\sim0.1$) than the star-forming
population ($\mathit{Pt}/T_{\rm median}<0.05$).  This may indicate either an AGN or decaying nuclear
starburst in a significant fraction of passive galaxies at these epochs (see also
\citealt{Whitaker_etal:2013}, who report similar findings for passive galaxies at $z > 1.4$).  For PSBs at
both epochs, the $\mathit{Pt}/T$ is generally low ($\mathit{Pt}/T_{\rm median} < 0.15$).  However, we also
find that at $1 < z < 2$ there is a population of PSBs that show evidence for a moderate {\em maximal} point
source contribution ($\mathit{Pt}/T > 0.1$; $\sim40$~per cent), and that these cases are much rarer at
$0.5 < z < 1$ ($\mathit{Pt}/T > 0.1$; $\sim15$ per cent).  Taken together, these results indicate that while
in general, point source emission is not driving the compact nature of PSBs, at high redshift ($z > 1$) we
cannot rule out that a fraction of these galaxies may host either an AGN or unresolved decaying nuclear
starburst.  These results also suggest that PSBs at $z >1$ may have experienced a different evolutionary
history to those at lower redshifts.

%%%%%%%%%%%%%%%%%%%%%%%%%%%%%%%%%%%%%%%%%%%%%%%%%%%%%%%%%%%%%%%%%%%%%%%%%%%%%%%%%%%%%%%%%%%%%%%%%%%%%%%%%%%%%
\subsection[]{Bulge--disc decomposition}

\label{BD decomposition}

%% Motivation %%-------------------------------------------------------------------------------------------%%
In this work, we find a significant difference in the structure of PSBs at different epochs (see
Figs.~\ref{Sersic fits} and \ref{model 1 indv}).  PSBs at $z > 1$ are typically massive
($M_* > 10^{10}\rm\,M_{\odot}$), very compact and of high~$n$; while at $z < 1$, they are generally of lower
mass ($M_* < 10^{10}\rm\,M_{\odot}$) and exhibit compact but less concentrated profiles (i.e.\ lower $n$).
This suggests that PSBs at $z > 1$ are typically spheroidal systems, while at $z < 1$ they contain a
significant stellar disc.  However, although S{\'e}rsic index~$n$ is generally considered a good proxy for
bulge--disc structure, recent works have shown that this is not necessarily the case
\citep[e.g.][]{Bruce_etal:2014b}.  Furthermore, since single S{\'e}rsic fits are largely driven by the
central regions of the profile, the presence of a faint outer disc can easily be missed.  Therefore, to
investigate the structure of these galaxies in more detail, we perform bulge--disc (B--D) decomposition and
assess the contribution of these two main structural components to the overall structure.

%% Profile fitting %%--------------------------------------------------------------------------------------%%
To perform B--D decomposition, we use a two-component model comprising a de Vaucouleurs ($r^{1/4}$) bulge and
single exponential disc.  The motivation behind adopting this bulge profile, instead of the more realistic
free S{\'e}rsic profile, is to i)~avoid the degeneracy and instability issues inherent to adding more degrees
of freedom to the models; and ii)~restrict the range of parameter space that needs to be explored in the
fitting.  We note that in adopting this bulge profile, we do not necessarily obtain its actual contribution,
and that many of our galaxies will have less concentrated bulges (i.e.\ pseudo-bulges), particularly at
low-mass \mbox{($M_* < 10^{10}\rm\,M_{\odot}$; see e.g.\ \citealt{Fisher&Drory:2011})}.  Therefore, by design
these fits will not yield the actual bulge components, but the {\em maximal} likely bulge contribution.  For
profile fitting, the sum of a wide range of bulge and disc profiles are compared to the measured $\mu(r)$
profile and a $\chi^2$ minimisation used to find the best fit.  In this process, the S{\'e}rsic index of the
bulge and disc are fixed at $n = 4$ and $n = 1$, respectively, but the effective radius $r_{\rm e}$ and
normalisation of each component are free to vary.  For both components, the full $r_{\rm e}$ parameter space
probed by the model library is analysed, ensuring a global minimum solution is obtained.

\begin{figure*}
\includegraphics[width=0.80\textwidth]{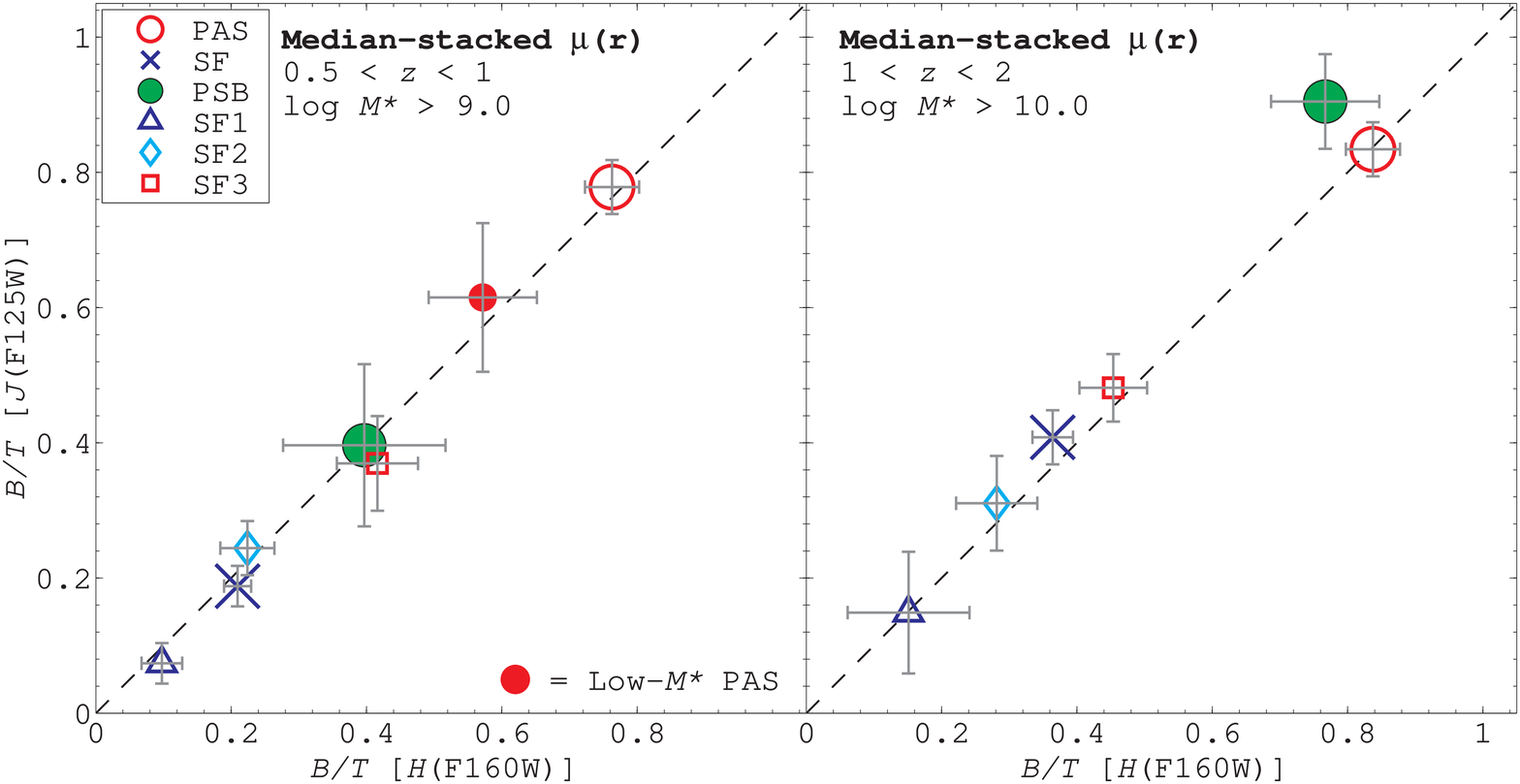}
\includegraphics[width=0.80\textwidth]{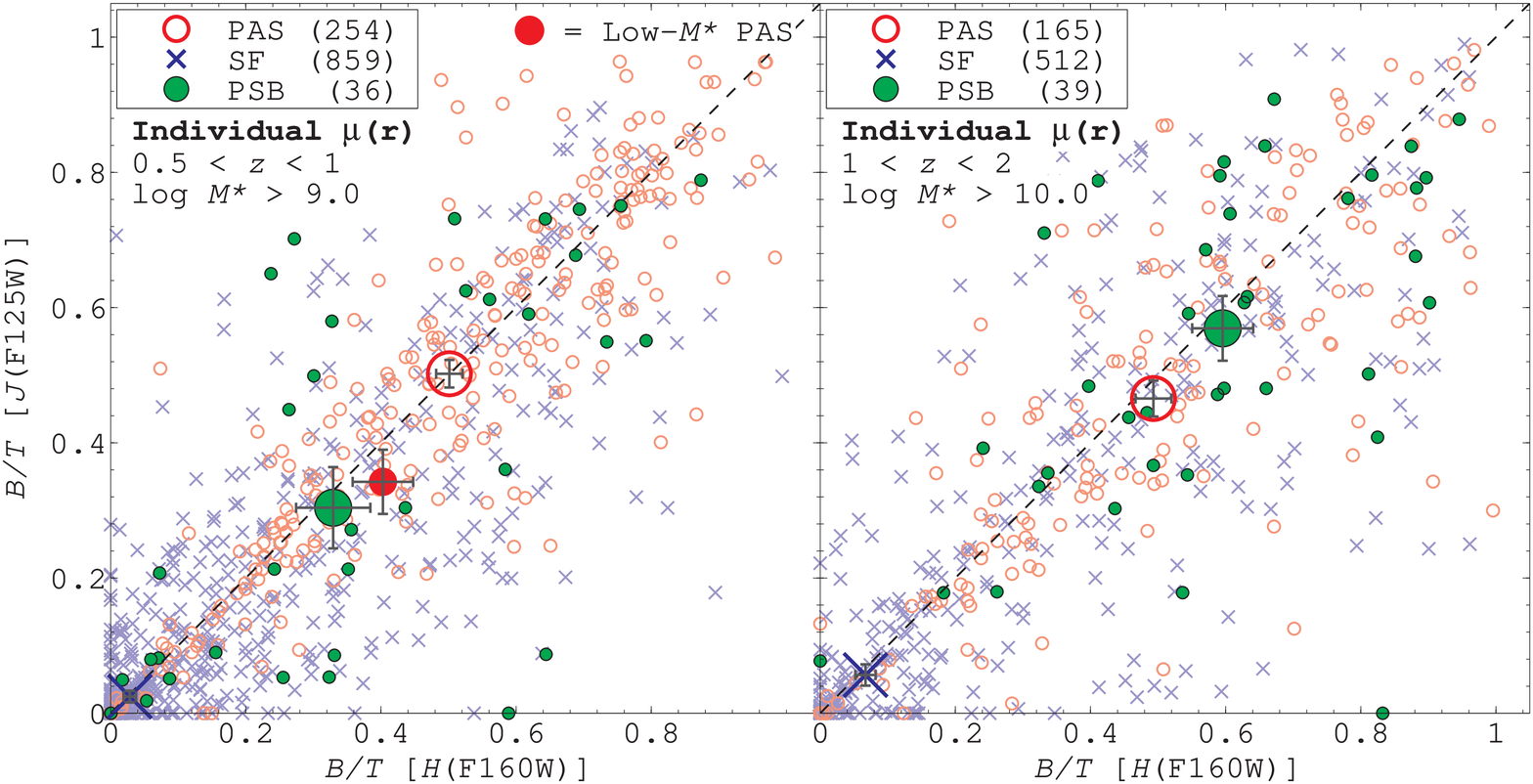}
\centering
\vspace{0.0cm}
\caption{\label{BD results} Bulge--disc (B--D) decompositions for our galaxy populations at two epochs:
$0.5 < z < 1$ (left-hand panels) and $1 < z < 2$ (right-hand panels).  A~comparison of the {\em maximal}
bulge-to-total ratio ($B/T$) in $J_{\rm\,F125W}$ and $H_{\rm\,F160W}$ for the median-stacked light profiles
$\widetilde\mu(r)$ (top panels), and for our individual $\mu(r)$ profiles (bottom panels).  In the bottom
panels, large symbols represent the median $B/T$ in each population, with associated $1\sigma$ errors.
The typical $B/T$ for the low-mass passive population ($10^{9} < M_* < 10^{10}\rm\,M_{\odot}$) is also
indicated.  We find PSBs to exhibit significantly different structures at different epochs: bulge-dominated
at $1 < z < 2$, but disc-dominated at $0.5 < z < 1$.}
\vspace{-0.1cm}
\end{figure*}

%% B/T comparisons: average profiles %%--------------------------------------------------------------------%%
For each population, the resultant best-fits for the $H_{\rm\,F160W}$ $\widetilde\mu(r)$ profiles are
presented in Fig.~\ref{two-component profiles}.  In all cases, we find the $\widetilde\mu(r)$ profile to be
well-described by a B--D system \mbox{(i.e.~$\chi_{\rm red}^2\sim1$)}.  However, since these profiles are
already well-described by a single S{\'e}rsic profile (see Fig.~\ref{Sersic profiles}), the adoption of a
B--D model does not necessarily improve the quality of the fit.  Therefore, it is important to note that
these B--D decompositions only yield the most likely structure, assuming a two-component B--D system.  A
comparison of the $J_{\rm\,F125W}$ and $H_{\rm\,F160W}$ bulge-to-total light ratios ($B/T$) is presented in
Fig.~\ref{BD results} and Table~\ref{two-component results}.  The $1\sigma$ errors in $B/T$ are determined
using a similar bootstrap analysis to that used for our single S{\'e}rsic fits (see
Section~\ref{Profile fitting: stacked light profiles}).  Furthermore, using a similar analysis to that
described in Section~\ref{Further robustness tests}, we find these measurements to be robust to sky
subtraction errors ($\pm\widetilde\sigma_{\rm sky}$), with typical effects of $<10$ per cent.  For passive
and star-forming galaxies, we find similar results at both epochs ($0.5 < z < 1$ and $1 < z < 2$), with the
passive population being bulge-dominated ($B/T\sim0.8$) and the star-forming populations being generally
disc-dominated ($B/T < 0.5$).  For the star-forming sub-populations (SF1--3), we find that at both epochs,
all are relatively disc-dominated but there is an increase in $B/T$ from SF1 $\rightarrow$ SF3 (i.e.\ with
increasing mean stellar age).  Interestingly, for PSBs we observe a significant difference in $B/T$ at
different epochs.  At $z > 1$, the PSBs exhibit bulge-dominated profiles ($B/T\sim0.8$), and with B--D
structures similar to the massive passive population.  It is also clear from
Fig.~\ref{two-component profiles}, that these galaxies contain no significant faint outer disc.  In contrast
at $z < 1$, PSBs have completely different structures with much more significant disc components
\mbox{($B/T\sim0.4$)}, and with B--D structures not dissimilar to those of the low-mass passive population
($M_* < 10^{10}\rm\,M_{\odot}$).

%% B/T comparisons: individual profiles %%-----------------------------------------------------------------%%
For our individual $\mu(r)$ profiles, we also perform analogous B--D decompositions.  We note that these fits
have much greater uncertainty than those for the $\widetilde\mu(r)$ profiles, but nonetheless can still
provide insight into the nature of our galaxy populations.  The resultant $B/T$ distributions in both
$J_{\rm\,F125W}$ and $H_{\rm\,F160W}$, are presented in Fig.~\ref{BD results}.  In general, although there is
significant scatter, we find similar results as before for our $\widetilde\mu(r)$ profiles, with PSBs having
bulge-dominated profiles at $z > 1$, and more significant disc components at $z < 1$.

With respect to these results, recall that our B--D decompositions yield the {\em maximal} bulge contribution
and consequently will likely overestimate this component, especially at low masses
\mbox{($M_* < 10^{10}\rm\,M_{\odot}$}; see e.g.~\citealt{Fisher&Drory:2011}).  This is particularly relevant
for PSBs at $z < 1$, which will be even more disc-dominated than our results suggest, but less of an issue at
$z > 1$.  We therefore conclude that there is a significant difference in the \mbox{B--D} structure of PSBs
at different epochs, which suggests that PSBs at $z > 1$ have undergone a completely different evolutionary
history compared to their counterparts at lower redshifts.

%%%%%%%%%%%%%%%%%%%%%%%%%%%%%%%%%%%%%%%%%%%%%%%%%%%%%%%%%%%%%%%%%%%%%%%%%%%%%%%%%%%%%%%%%%%%%%%%%%%%%%%%%%%%%
\section[]{Optical imaging}

\label{Optical imaging}

\begin{figure*}
\includegraphics[width=0.995\textwidth]{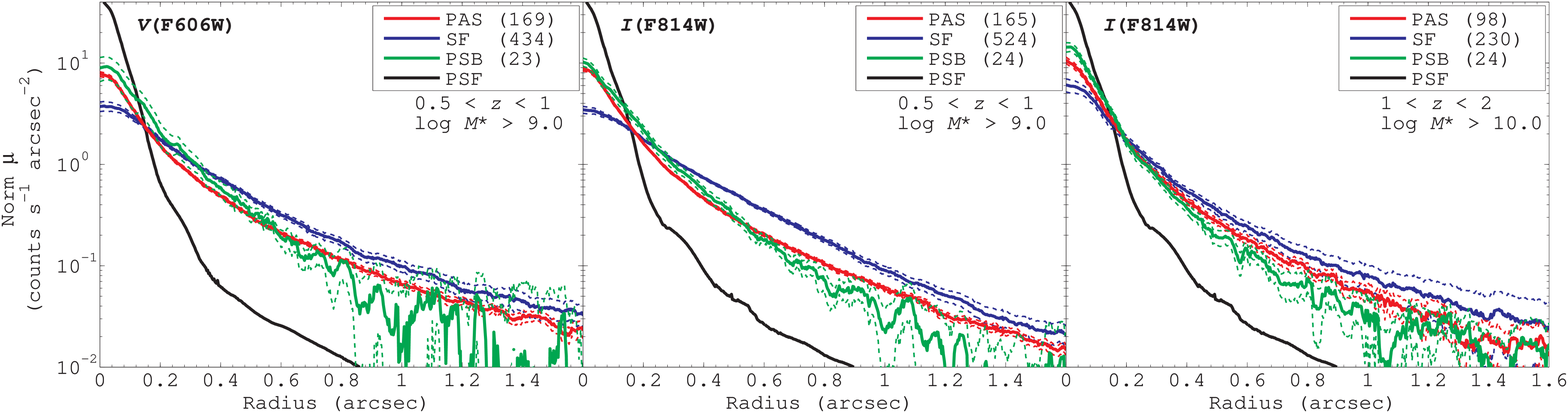}\\
\hspace{0.00cm}
\includegraphics[width=0.960\textwidth]{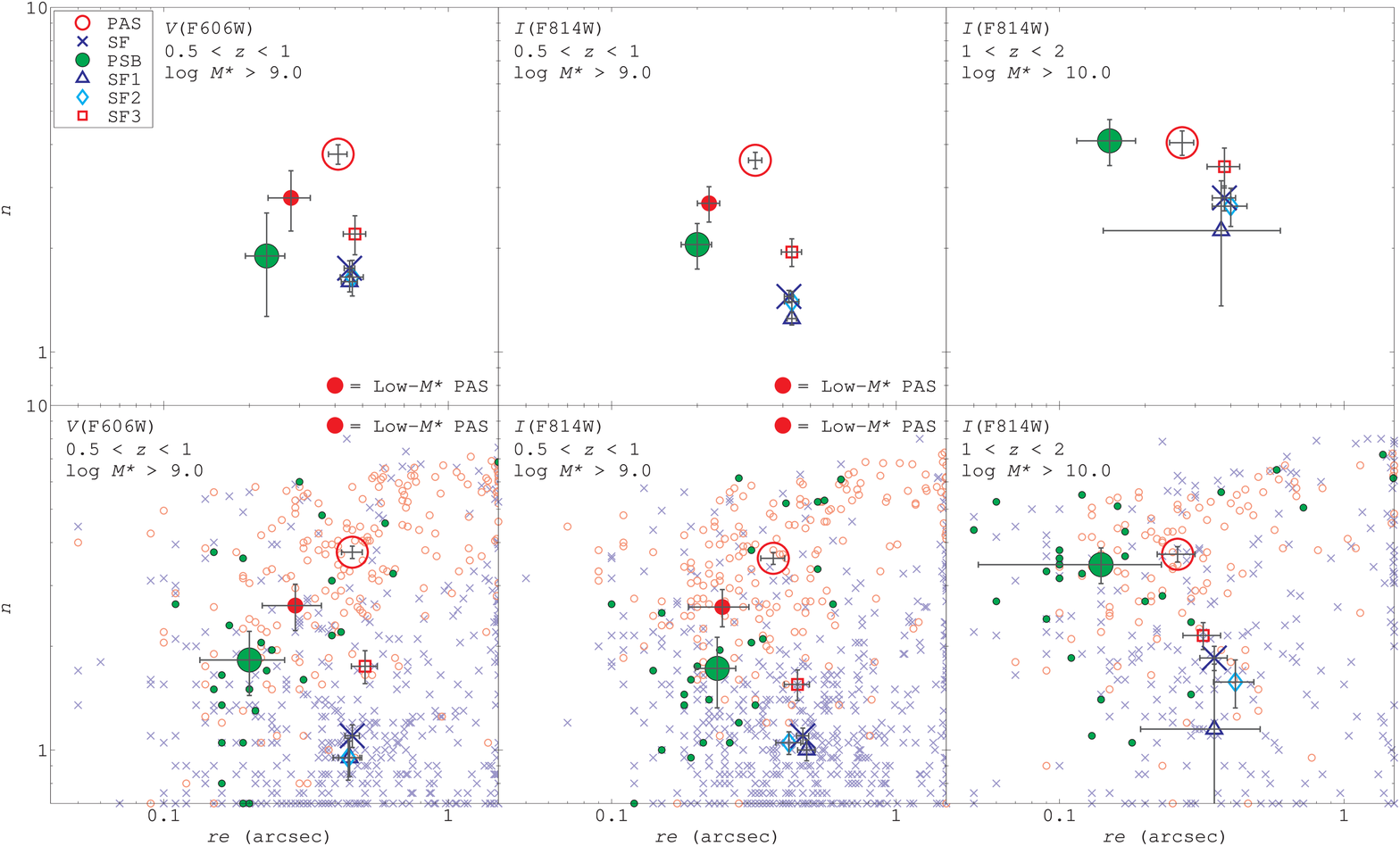}
\centering
\vspace{-0.1cm}
\caption{\label{ACS results} Structural analyses for our optical $\mu(r)$ profiles (ACS -- $I_{\rm\,F814W}$,
$V_{\rm\,F606W}$).  Top row: median-stacked optical light profiles $\widetilde\mu(r)$ for different galaxy
populations and at different epochs: $V_{\rm\,F606W}$ at $0.5 < z < 1$ (left-hand panel), $I_{\rm\,F814W}$
at $0.5 < z < 1$ (centre panel), and $I_{\rm\,F814W}$ at $1 < z < 2$ (right-hand panel).  Middle row:
the corresponding structural parameters ($r_{\rm e}$, $n$) from the single S{\'e}rsic fits to these optical
$\widetilde\mu(r)$ profiles.  Bottom row: structural parameters from corresponding analogous fits to the
individual optical $\mu(r)$ profiles, with the large symbols showing the median structural properties of each
population (including associated $1\sigma$ errors).  The typical structural properties for the low-mass
passive population ($10^{9} < M_* < 10^{10}\rm\,M_{\odot}$) are also indicated.  The results of these
structural analyses are further summarised in Fig.~\ref{multiband summary}, and compared to the corresponding
results from our near-infrared $\mu(r)$ profiles.
\vspace{-0.15cm}}
\end{figure*}

%% Best-fit structural parameters %%-----------------------------------------------------------------------%%
\begin{table*}
\centering
\begin{minipage}{175mm}
\centering
\caption{\label{ACS fits} Optical single S{\'e}rsic fits: the structural properties for our median-stacked
$V_{\rm\,F606W}$ and $I_{\rm\,F814W}$ light profiles $\widetilde\mu(r)$.  Structural properties ($r_{\rm e}$,
$n$) are shown for different galaxy populations (passive, star-forming, PSB) at two different epochs,
$0.5 < z < 1$ and $1 < z < 2$.  Errors in the structural parameters ($1\sigma$) are determined from the
variance between fits performed on $100$ simulated $\widetilde\mu(r)$ profile stacks generated via a
bootstrap method.}
\begin{tabular}{lccccccccccc}
\hline
Galaxy		&\multicolumn{5}{c}{$0.5 < z < 1$}							&{}	&\multicolumn{5}{c}{$1 < z < 2$}												\\
Population	&\multicolumn{2}{c}{$V_{\rm\,F606W}$}	&{}	&\multicolumn{2}{c}{$I_{\rm\,F814W}$}	&{}	&\multicolumn{2}{c}{$V_{\rm\,F606W}$}						&{}	&\multicolumn{2}{c}{$I_{\rm\,F814W}$}	\\
{}		&{$n$}		&{$r_{\rm e}$}		&{}	&{$n$}		&{$r_{\rm e}$}		&{}	&{$n$}					&{$r_{\rm e}$}				&{}	&{$n$}		&{$r_{\rm e}$}		\\
{}		&{}		&{(arcsec)}		&{}	&{}		&{(arcsec)}		&{}	&{}					&{(arcsec)}				&{}	&{}		&{(arcsec)}		\\[0.5ex]
\hline																														\\[-1.5ex]
Passive		&{$3.75\pm0.24$}&{$0.41\pm0.03$}	&{}	&{$3.60\pm0.20$}&{$0.32\pm0.02$}	&{}	&{\hspace{0.6cm}-\hspace{0.6cm}}	&{\hspace{0.6cm}-\hspace{0.6cm}}	&{}	&{$4.05\pm0.33$}&{$0.27\pm0.03$}	\\[1ex]	
Star-forming	&{$1.75\pm0.09$}&{$0.45\pm0.02$}	&{}	&{$1.45\pm0.06$}&{$0.42\pm0.02$}	&{}	&{-}					&{-}					&{}	&{$2.80\pm0.23$}&{$0.38\pm0.04$}	\\[1ex]
\ \ SF1		&{$1.60\pm0.11$}&{$0.45\pm0.03$}	&{}	&{$1.25\pm0.05$}&{$0.43\pm0.02$}	&{}	&{-}					&{-}					&{}	&{$2.25\pm0.89$}&{$0.37\pm0.23$}	\\
\ \ SF2		&{$1.65\pm0.20$}&{$0.46\pm0.04$}	&{}	&{$1.40\pm0.10$}&{$0.43\pm0.03$}	&{}	&{-}					&{-}					&{}	&{$2.65\pm0.34$}&{$0.40\pm0.06$}	\\
\ \ SF3		&{$2.20\pm0.28$}&{$0.47\pm0.04$}	&{}	&{$1.95\pm0.18$}&{$0.43\pm0.04$}	&{}	&{-}					&{-}					&{}	&{$3.45\pm0.46$}&{$0.38\pm0.05$}	\\[1ex]
Post-starburst	&{$1.90\pm0.63$}&{$0.23\pm0.04$}	&{}	&{$2.05\pm0.31$}&{$0.20\pm0.02$}	&{}	&{-}					&{-}					&{}	&{$4.10\pm0.62$}&{$0.15\pm0.04$}	\\
\hline
\end{tabular}
\end{minipage}
\vspace{0.0cm}
\end{table*}

%% Overview %%---------------------------------------------------------------------------------------------%%
In this paper, we mainly focus on the structural analyses for our near-infrared $\mu(r)$ profiles.  At the
redshifts studied here \mbox{($0.5 < z < 2$)}, these profiles generally trace the distribution of the old
stellar component (i.e.~$\lambda_{\rm rest} > 4000$\rm\,\AA), which comprises the bulk of the stellar mass.
However, an important addition to this study is the structural analyses for our optical $\mu(r)$ profiles
($V_{\rm\,F606W}$, $I_{\rm\,F814W}$), which can be used to probe the distribution of younger, more recently
formed stellar populations (i.e.\ O\,B\,A\,F stars; see later discussion for further details).  Such
analyses can be used to determine whether these younger stars trace the stellar mass, or whether they are
more centrally located, which for PSBs can place useful constraints on their evolutionary history (e.g.\
whether the preceding starburst was strongly centralized).  These structural analyses are analogous to those
presented for our near-infrared profiles in Section~\ref{Profile fitting}.  However, since our optical
profiles are generally of poorer quality than those in the near-infrared (i.e.\ fainter and with
more significant noise), these structural analyses are limited to single S{\'e}rsic fits only.

\begin{figure*}
\includegraphics[width=1\textwidth]{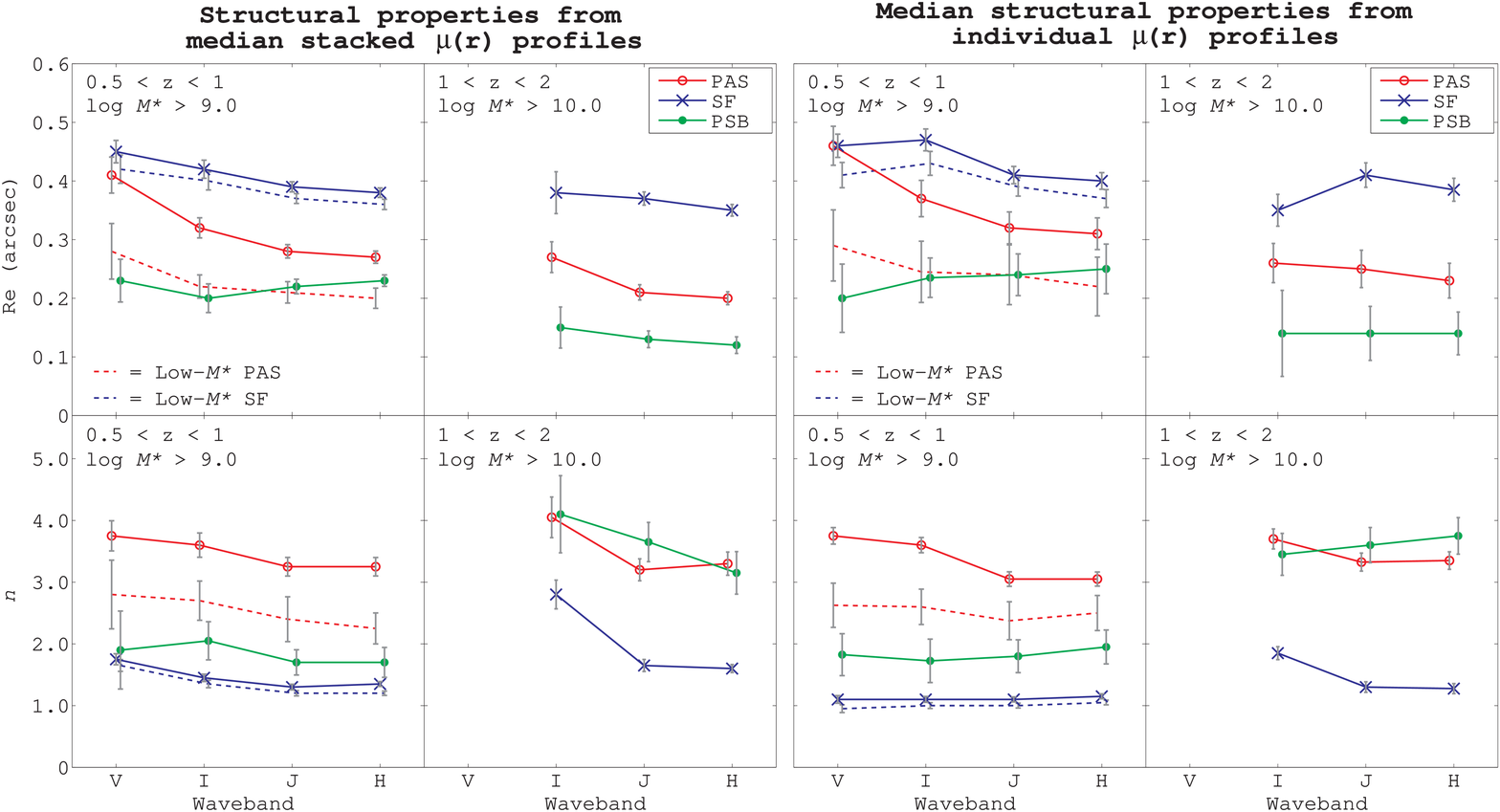}\\
\vspace{-0.15cm}
\caption{\label{multiband summary} The wavelength dependence of structural parameters ($r_{\rm e}$, $n$) for
our galaxy populations at two epochs ($0.5 < z < 1$ and $1 < z < 2$).  Left-hand panels: a summary of results
from the single S{\'e}rsic fits to our median-stacked profiles $\widetilde\mu(r)$ across all four CANDELS
wavebands ($V_{\rm\,F606W}$, $I_{\rm\,F814W}$, $J_{\rm\,F125W}$, $H_{\rm\,F160W}$).  Right-hand panels: an
analogous summary for the median structural properties ($r_{\rm e}$, $n$) of our individual $\mu(r)$
profiles.  Both the median-stacked $\widetilde\mu(r)$ and individual $\mu(r)$ profile analyses yield
consistent results.  For both epochs, PSBs show consistent structural parameters across all wavebands,
regardless of whether the light being traced is dominated by the old stellar component
($\lambda_{\rm rest} > 4000$\,\AA) or younger stellar populations ($\lambda_{\rm rest} < 4000$\,\AA).}
\vspace{-0.05cm}
\end{figure*}

%% Optical profiles %%-------------------------------------------------------------------------------------%%
For each galaxy population, the median stacked profiles $\widetilde\mu(r)$ from both the CANDELS
$V_{\rm\,F606W}$ and $I_{\rm\,F814W}$ imaging are presented in Fig.~\ref{ACS results}.  The $V_{\rm\,F606W}$
$\widetilde\mu(r)$ profiles are limited to the intermediate-$z$ epoch ($0.5 < z < 1$), due to the inadequate
signal-to-noise in this waveband at $z > 1$.  As with our near-infrared profiles, we perform single
S{\'e}rsic fits on these $\widetilde\mu(r)$ profiles, and obtain the typical structural properties
($r_{\rm e}$,~$n$) of each galaxy population (see Fig.~\ref{ACS results} and Table~\ref{ACS fits}).  The
$1\sigma$ errors in these structural parameters are determined from the variance between fits performed on
$100$ simulated $\widetilde\mu(r)$ stacks generated via a bootstrap analysis.  We also find that in all
cases, the $\widetilde\mu(r)$ profiles are well described by a single S{\'e}rsic profile, with the best fit
having a reduced chi-squared $\chi_{\rm red}^2\sim1$.  Analogous fits were also performed on the individual
$\mu(r)$ profiles (see Fig.~\ref{ACS results}).  Finally, as with our near-infrared profiles (see
Section~\ref{Further robustness tests}), we find PSF-effects and sky subtraction errors to have a minimal
influence on both $r_{\rm e}$ and $n$ (typically $<10$ per cent).  In the following, we compare the results
of these fits to those obtained from the near-infrared wavebands in Section~\ref{Profile fitting}.  A summary
of these comparisons is presented in Fig.~\ref{multiband summary}.

%% High redshift %%----------------------------------------------------------------------------------------%%
At high redshift ($1 < z < 2$), both passive and PSB galaxies show no significant variation in their
structural properties between the optical ($I_{\rm\,F814W}$) and near-infrared wavebands (see
Fig.~\ref{multiband summary}).  In both regimes, these populations exhibit compact and spheroidal structures;
passive [$n\sim3.5$, $r_{\rm e}\sim0.25\rm\,arcsec$ ($\sim2.1\rm\,kpc$)]; PSB [\mbox{$n\sim3.5$},
$r_{\rm e}\sim0.15\rm\,arcsec$ ($\sim1.3\rm\,kpc$)].  In contrast, star-forming galaxies have significantly
larger $n$ in $I_{\rm\,F814W}$ compared to the near-infrared ($n_{\rm\,F814W}\sim3$ vs.\
$n_{\rm\,near-IR}\sim1.5$ from the $\widetilde\mu(r)$ profiles).  Note that at these redshifts,
$I_{\rm\,F814W}$ probes a different stellar population to the near-infrared
(i.e.~$\lambda_{\rm rest}<4000$\,\AA), and generally traces younger stellar populations (O\,B\,A\,F stars).
Therefore, these comparisons indicate that at $z > 1$: i)~younger stars in passive/PSB galaxies trace the
structure of the old stellar population (i.e.\ stellar mass), which for PSBs suggests that the preceding
starburst and/or quenching was not strongly centralized and occurred throughout the stellar distribution; and
ii)~in star-forming galaxies, younger stars are more centralized than the old stellar population (i.e.~more
prominent in the central bulge, than the outer disc).

%% Intermediate redshift %%--------------------------------------------------------------------------------%%
At intermediate redshift ($0.5 < z < 1$), for both star-forming and PSB galaxies, we also find no significant
variation in structure between the optical ($I_{\rm\,F814W}$, $V_{\rm\,F606W}$) and near-infrared wavebands
(see Fig.~\ref{multiband summary}).  Star-forming galaxies are extended and disc-like
[\mbox{$r_{\rm e}\sim0.45$}~arcsec ($\sim3.3\rm\,kpc$), $1 < n < 2$], and PSBs are compact and of low $n$
[$r_{\rm e}\sim0.2$ arcsec ($\sim1.5\rm\,kpc$), $n\sim2$].  Furthermore, across all wavebands, PSBs have
structures that are similar to the low-mass passive population, the population into which they will most
likely evolve.  Note that at this epoch, $I_{\rm\,F814W}$ will trace a similar stellar population as the
near-infrared, but $V_{\rm\,F606W}$ will generally trace younger populations
(i.e.~$\lambda_{\rm rest}<4000$\rm\,\AA; O\,B\,A\,F stars).  Therefore, these results indicate that at
$0.5 < z < 1$, younger stars in star-forming/PSB galaxies trace the structure of the old stellar population
(i.e.\ stellar mass).  For PSBs, this again suggests that the preceding starburst and/or quenching was not
strongly centralized (i.e.\ it was global in nature).  In contrast, the general passive population exhibits
significantly larger $r_{\rm e}$ in the optical wavebands compared to the near-infrared, especially in
$V_{\rm\,F606W}$ ($\Delta r_{\rm e}\sim40$~per cent).  Interestingly, this trend could indicate younger stars
in the outskirts of passive galaxies at this epoch ($z < 1$).  This might be expected if they were quenched
from the `inside--out' \citep[see e.g.][]{Tacchella_etal:2016}, or if a minor merger resulted in the
accretion of younger stars to an outer envelope \citep[e.g.][]{Naab_etal:2009}.

%% Population models %%------------------------------------------------------------------------------------%%
Finally, we note that our conclusions above assume that the majority of light emitted at
$\lambda_{\rm rest} < 4000$\,\AA\ originates from young stellar populations (O\,B\,A\,F stars).  In order to
quantify this for our PSBs we create simple mock spectra using \cite{Bruzual&Charlot:2003} models, assuming
solar metallicity, Chabrier IMF, and a moderate amount of dust attenuation (effective attenuation
$\tau_{\,V} = 1.0$ and fraction of dust in the interstellar medium $\mu = 0.3$, following the
\citealt{Charlot&Fall:2000} dust model as adapted by \citealt{Wild_etal:2011}).  We assume two underlying
star-formation histories, both $6\rm\,Gyr$ old and exponentially declining with a timescale of $0.1$ or
$3\rm\,Gyr$, to represent an underlying quiescent or star-forming population.  Superimposed on this is a
$500\rm\,Myr$ old burst population, with an exponentially declining star-formation history of timescale
$0.3\rm\,Gyr$, and varying burst mass fraction.  For a burst mass fraction of $10$ per cent, the minimum
expected for our photometrically-selected PSBs, we calculate the fraction of light from the burst population
in both $V_{\rm\,F606W}$ and $H_{\rm\,F160W}$ at $z = 0.75$, and $I_{\rm\,F814W}$ and $H_{\rm\,F160W}$ at
$z = 1.5$ (the central redshifts of the epochs studied).  We find $\sim70$--$80$ per cent of the light in the
optical wavebands is from the burst population, compared to $\sim40$ per cent in $H_{\rm\,F160W}$ for both
epochs.  Consequently, this shows that the optical/near-infrared wavebands used in this study are able to
broadly differentiate between the young and old stellar populations in our PSB galaxies.

%%%%%%%%%%%%%%%%%%%%%%%%%%%%%%%%%%%%%%%%%%%%%%%%%%%%%%%%%%%%%%%%%%%%%%%%%%%%%%%%%%%%%%%%%%%%%%%%%%%%%%%%%%%%%
\section[]{Summary and Discussion}

\label{Discussion}

%% Discussion of results %%--------------------------------------------------------------------------------%%
In this study, we have explored in detail the structure of PSBs at $0.5 < z < 2$.  For this we have used a
combination of near-infrared and optical $\mu(r)$ profiles, probing both the old stellar component as well as
younger, and more recently formed stellar populations.  Various structural analyses have also been performed,
including single S{\'e}rsic and multiple component fits, which have revealed significant differences in the
structure of PSBs at different epochs.  At $z > 1$, PSBs are typically massive ($M_*>10^{10}\rm\,M_{\odot}$),
very compact and exhibit high S{\'e}rsic indices~$n$, with structures that differ significantly from their
star-forming progenitors, but are similar to massive passive galaxies.  In contrast at lower redshift
\mbox{($0.5 < z < 1$)}, PSBs are generally of low mass ($M_* < 10^{10}\rm\,M_{\odot}$) and exhibit compact
but less concentrated profiles (i.e.\ lower $n$), with structures similar to low-mass passive galaxies
(i.e.\ passive discs).

Taken together, these results suggest that PSBs at $z > 1$ are an intrinsically different population to those
at $z < 1$, indicating different quenching routes are active at different epochs, with the PSB phase being
triggered by different processes.  Furthermore, for both epochs, we find a remarkable consistency in PSB
structure across the optical/near-infrared wavebands, which suggests that the old/intermediate--young aged
stellar populations probed follow the same distribution.  This implies that any preceding starburst and/or
quenching in these galaxies was not strongly centralized, and therefore occurred globally.  In this section,
we present a more in depth discussion of these results and their implications for the potential quenching
mechanisms experienced by PSBs at different epochs.  To complement this discussion, we refer the reader to
Fig.~\ref{multiband summary}, which provides a summary of our structural analyses across each of the four
CANDELS wavebands ($V_{\rm\,F606W}$, $I_{\rm\,F814W}$, $J_{\rm\,F125W}$, $H_{\rm\,F160W}$).

\vspace{0.3cm}

%%%%%%%%%%%%%%%%%%%%%%%%%%%%%%%%%%%%%%%%%%%%%%%%%%%%%%%%%%%%%%%%%%%%%%%%%%%%%%%%%%%%%%%%%%%%%%%%%%%%%%%%%%%%%
\subsection{Post-starburst galaxies at $1 < z < 2$}

\label{Discussion: high redshift}

%% Summary %%----------------------------------------------------------------------------------------------%%
For the high-redshift epoch ($1 < z < 2$), the main results from our structural analyses can be summarised as
follows:

\begin{enumerate}

\item PSBs at $z > 1$ are of high mass ($M_* > 10^{10}\rm\,M_{\odot}$), and exhibit structures that are
extremely compact [$r_{\rm e}\sim0.13\rm\,arcsec$, ($\sim1.1\rm\,kpc$)] and of high S{\'e}rsic index
($n\sim3.5$).  In general, their structures differ from those of their star-forming progenitors, and are
more similar to those of the old massive passive population, although considerably more compact (by $\sim40$
per cent).  These results confirm the recent findings of \cite{Almaini_etal:2017}, who find that massive PSBs
at $z > 1$ are compact proto-spheroids.  This implies that morphological/structural transformation must have
occurred prior to the post-starburst phase, and therefore before (or during) the event that quenched the
galaxy's star formation (see Sections~\ref{Profile fitting: stacked light profiles} and
\ref{Individual profiles}).

\vspace{0.2cm}

\item Point source emission from either an AGN or unresolved decaying nuclear starburst is not sufficient to
explain the compact nature of massive PSBs at this epoch.  Even when the {\em maximal} emission from a
potential point source is taken into account, these PSBs remain compact [$r_{\rm e}\sim0.15\rm\,arcsec$,
($\sim1.2\rm\,kpc$)] and of relatively high S{\'e}rsic index ($n > 2.5$).  They also remain significantly
more compact than the massive passive population.  However, we note that while point source emission cannot
explain their compact nature, we cannot rule out the presence of an AGN, or unresolved decaying nuclear
starburst in a fraction ($<40$~per~cent) of PSBs at this epoch (see
Section~\ref{Sersic profile + point source}).

\vspace{0.2cm}

\item  Bulge--disc decomposition indicates that massive PSBs at $z > 1$ are generally bulge-dominated systems
($B/T\sim0.8$), with little or no residual disc component.  Their $B/T$ is similar to those of the old
massive passive population (see Section~\ref{BD decomposition}).

\vspace{0.2cm}

\item Massive PSBs at $z > 1$ exhibit consistent structural parameters ($r_{\rm e}$,~$n$) between all three
wavebands studied at this epoch (see Fig.~\ref{multiband summary}).  This consistency between wavebands
probing both the old stellar component ($\lambda_{\rm rest} > 4000$\,\AA; $J_{\rm\,F125W}$ and
$H_{\rm\,F160W}$) and younger populations (O\,B\,A\,F stars; $\lambda_{\rm rest} < 4000$\,\AA;
$I_{\rm\,F814W}$) indicates that younger stars are tracing the old stellar population (i.e.~stellar mass)
in these galaxies.  This suggests that any preceding starburst and/or quenching was not strongly centralized
within the existing stellar distribution (i.e.\ it was global in nature).  In contrast, massive star-forming
galaxies show a significant increase in S{\'e}rsic index moving from the near-infrared to the optical
wavebands (i.e.~old $\rightarrow$ younger stellar populations), potentially indicating centralized
star-formation in these galaxies and the build-up of galactic bulges at this epoch (see
Fig.~\ref{multiband summary} and Section~\ref{Optical imaging}).

\end{enumerate}

%% Discussion %%-------------------------------------------------------------------------------------------%%
The results presented here suggest that high-$z$ PSBs ($z > 1$) have experienced a major disruptive event
that quenched their star formation and led to a `compaction' of the stellar distribution.  Such an event
could be a gas-rich major merger \citep[e.g.][]{Hopkins_etal:2009, Wellons_etal:2015} or a dissipative
`protogalactic collapse': gas inflow to a massive disc, which then destabilises and collapses
\citep[e.g.][]{Dekel_etal:2009, Zolotov_etal:2015}.  In both cases, gas would be driven into the central
galactic regions, triggering a starburst, and lead to the formation of a compact remnant.  We note that in
our observations the lack of excess young (O\,B\,A\,F) stars in the central regions of PSBs does not
necessarily rule out these scenarios (see later discussion, for more details).  Following this `compaction'
event, any subsequent star formation would be rapidly quenched via feedback from either an AGN or the
starburst itself, both of which would result in the characteristic post-starburst spectral features
(i.e.~strong Balmer absorption).  These scenarios would also naturally lead to the destruction of the
stellar disc and the formation of a compact spheroidally-dominated stellar distribution (i.e.~high~$n$ and
high $B/T$), both of which match our observations.  Furthermore, since these scenarios lead to significant
structural transformations during the quenching event, they are consistent with our findings that PSBs at
this epoch already exhibit structures similar to the massive passive galaxies into which they will most
likely evolve.

At low redshift ($z < 0.1$), gas-rich major mergers have also been linked to PSBs in low-density environments
\cite[e.g.][]{Zabludoff_etal:1996, Blake_etal:2004, Pawlik_etal:2016, Pawlik_etal:2018}.  However, in
contrast to our results at $z > 1$, several studies have reported centrally-concentrated young stellar
populations in these galaxies \citep[e.g.][]{Norton_etal:2001, Yamauchi&Goto:2005, Pracy_etal:2013},
indicative of a merger-induced centralized starburst.  Despite these low redshift results, we note that the
lack of a stellar-age gradient in our high-$z$ PSBs does not necessarily rule out a gas-rich merger scenario
for their origin, since the remnant structure will be strongly dependent on the nature of these mergers at
high/low redshift.  At low redshift, gas-rich mergers will funnel gas into the central regions of the galaxy
and trigger a nucleated starburst prior to the PSB phase.  In contrast, at $z > 1$ these events will be
significantly more gas rich than their local counterparts, leading to a more substantial starburst and the
formation of a compact remnant (see discussion above).  This would potentially lead to either: i) the bulk of
the stellar mass being formed during a centralized starburst (e.g.~monolithic collapse); or ii) a compaction
of the original structure to subsequently match that of the starburst itself.  Both of these scenarios would
result in little or no radial age gradient in the PSB phase, matching our observations.

In comparison to previous works, we find that our results confirm those of the recent study by
\cite{Almaini_etal:2017}, who also performed a detailed structural analysis of high-redshift PSBs ($z > 1$)
in the UDS field.  Using both ground-/space-based near-infrared imaging (UDS-$K$ and
CANDELS-$H_{\rm\,F160W}$), they use two-dimensional S{\'e}rsic models to examine the stellar structure of
massive ($M_* > 10^{10}\rm\,M_{\odot}$) PSBs at $z > 1$.  They also conclude that PSBs at this epoch are
exceptionally compact and with structures similar to the old massive passive population (i.e.~high S{\'e}rsic
indices; spheroidally-dominated).  Furthermore, they find evidence for massive PSBs being smaller on average
than comparable passive galaxies at the same epoch, which is also consistent with our structural analyses
(see Fig.~\ref{multiband summary}).  Similar results have also been reported at $z > 1$ by
\cite{Whitaker_etal:2012a} and \cite{Yano_etal:2016}, where younger passive galaxies are found to be more
compact than their older counterparts.  However, we note that at more modest redshifts ($z\sim1$) previous
studies have found conflicting results on the relationship between stellar age and the compactness of
passive/recently-quenched galaxies \citep[e.g.][]{Keating_etal:2015, Williams_etal:2017}, and further study
is still required.

In conclusion, these results indicate that PSBs at high-redshift ($z > 1$) are quenched in a relatively
violent event (e.g.\ a gas-rich major merger or protogalactic collapse), that led to a `compaction' of the
stellar distribution, and that this may be followed by a gradual growth in size as the galaxy evolves into a
more established passive system (e.g.\ via minor `dry' mergers; \citealt{Naab_etal:2009}).

%%%%%%%%%%%%%%%%%%%%%%%%%%%%%%%%%%%%%%%%%%%%%%%%%%%%%%%%%%%%%%%%%%%%%%%%%%%%%%%%%%%%%%%%%%%%%%%%%%%%%%%%%%%%%
\subsection{Post-starburst galaxies at $0.5 < z < 1$}

\label{Discussion: intermediate redshift}

%% Summary %%----------------------------------------------------------------------------------------------%%
At intermediate redshifts ($0.5 < z < 1$), our structural analyses reveal that PSBs have significantly
different structures to their conterparts at $z > 1$.  The main results from our structural analyses can be
summarised as follows:

\begin{enumerate}

\item PSBs at intermediate redshift ($0.5 < z < 1$) are generally of low mass
($M_* < 10^{10}\rm\,M_{\odot}$), and exhibit structures that are still relatively compact
[$r_{\rm e}\sim0.2\rm\,arcsec$, ($1.4\rm\,kpc$)] but of much lower S{\'e}rsic index ($n\sim1.7$) than the
massive PSBs at $z > 1$.  These PSBs are more compact than the general low-mass star-forming population, but
have structures similar to those of low-mass passive galaxies (i.e.~passive discs), the population into which
they will most likely evolve.  We note that more massive PSBs ($M_* > 10^{10}\rm\,M_{\odot}$) do exist at
this epoch, but these galaxies are rare and interestingly exhibit high $n$ values similar to the massive PSBs
at $z > 1$.  This suggests that the quenching process producing massive PSBs at $z > 1$
still occurs at lower redshifts but at a much lower frequency (see Sections~\ref{Profile fitting: stacked
light profiles} and \ref{Individual profiles}).  Finally, we note that the presence of a known supercluster
in the CANDELS--UDS field at this epoch ($z\sim0.65$; \citealt{vanBreukelen_etal:2006, Galametz_etal:2018})
appears to cause no significant bias in these results.  Using the $K$-band structural parameters of
\cite{Almaini_etal:2017}, which were determined for all galaxies in our parent sample (i.e.\ the full UDS
field; see Section~\ref{Sample selection}), we find entirely consistent results for each galaxy population.

\vspace{0.2cm}

\item  PSBs at this epoch do not show any evidence for significant point source emission.  This suggests that
neither an AGN, nor an unresolved decaying nuclear starburst are significant during the post-starburst phase.
However, we cannot rule out that these events were related to the quenching of these galaxies
(see Section~\ref{Sersic profile + point source}).

\vspace{0.2cm}

\item Bulge--disc decomposition indicates that PSBs at this epoch contain a significant disc component
($B/T < 0.4$), which has survived the event that quenched the star-formation.  Their $B/T$ is similar to
those of the low-mass passive population (see Section~\ref{BD decomposition}).  This result is consistent
with previous works at this epoch, which find that although PSBs are a morphologically heterogeneous
population, they generally exhibit disc-like morphologies \citep[e.g.][]{Dressler_etal:1999,
Caldwell_etal:1999,Tran_etal:2003,Poggianti_etal:2009,Vergani_etal:2010, Pawlik_etal:2016}.

\vspace{0.2cm}

\item  PSBs at $0.5 < z < 1$ exhibit consistent structural parameters ($r_{\rm e}$, $n$) between all four
wavebands studied at this epoch.  This simlarity between wavebands probing both the old stellar component
($\lambda_{\rm rest} > 4000$\,\AA; $I_{\rm\,F814W}$, $J_{\rm\,F125W}$ and $H_{\rm\,F160W}$) and younger
stellar populations (O\,B\,A\,F stars; $\lambda_{\rm rest} < 4000$\,\AA; $V_{\rm\,F606W}$) indicates that
younger stars are tracing the old stellar population (i.e.\ stellar mass) in these galaxies.  As with PSBs at
$z > 1$, this suggests that any preceding starburst, extended star-formation episode and/or quenching was not
strongly centralized, and occurred throughout the stellar distribution (i.e.\ globally; see
Section~\ref{Optical imaging}).

\end{enumerate}

%% Discussion %%-------------------------------------------------------------------------------------------%%
Taken together, these results suggest that intermediate-$z$ PSBs ($0.5 < z < 1$) have not experienced a major
disruption to their stellar distribution (e.g.\ major merger or disc collapse), and that consequently the
quenching mechanism responsible must be a relatively gentle process.  We note that although PSBs at this
epoch are generally more compact than analogous star-forming galaxies (i.e.~those of similar mass), this does
not necessarily imply that these galaxies have experienced a violent `compaction' event.  In fact the low
S{\'e}rsic indices of this population would suggest that this is not the case.  With respect to major
mergers, we also note that while a new disc may eventually reform, the timescale involved is expected to be
longer than that of the PSB phase \citep[$>1\rm\,Gyr$, see e.g.][]{Athanassoula_etal:2016}.  Consequently,
these events are unlikely to be the origin of the disc-dominated PSBs at this epoch.  Furthermore, given that
at this epoch not all star-forming galaxies are expected to experience a PSB phase \citep[e.g.][]
{Wild_etal:2016, Socolovsky_etal:2018}, the general star-forming population may not be representative of the
true progenitors of these PSBs.  We explore this issue in more detail in a forthcoming publication
(Socolovsky et al., in preparation).  Finally, we note that since these intermediate-$z$ PSBs have structures
very similar to low-mass passive galaxies (i.e.~passive discs), it is likely that any significant structural
changes related to the quenching process have already taken place, and that these galaxies are simply quietly
transitioning into established passive discs (i.e.\ S0s).  The resultant fading of the stellar disc leading
to the slight increase in $n$ and $B/T$ observed (see Fig.~\ref{multiband summary}).

In comparison to previous works, we note that gas-rich major mergers have been linked to PSBs at
$0.5 < z < 1$ \cite[e.g.][]{Wild_etal:2009, Wu_etal:2014}, which is in apparent contrast to our findings.
However, these previous studies focus on massive PSBs ($M* > 10^{10}\rm\,M_{\odot}$), which are rare in the
CANDELS--UDS field at this epoch (see Fig.~\ref{mass-vs-z}).  Consequently, our findings are not in
contradiction to these previous works, but suggest an alternative, less disruptive process is primarily
responsible for PSBs at lower masses ($M_* < 10^{10}\rm\,M_{\odot}$).  Furthermore, we note that at this
epoch, the rare, massive PSBs in the CANDELS--UDS field do exhibit the high S{\'e}rsic indices expected for
the remnant of a gas-rich major merger, which is consistent with these previous studies.

With respect to the dominant quenching mechanism, our results suggest two scenarios for PSBs at this epoch:
i)~these galaxies experience a weaker disruptive event to the PSBs at $z > 1$ which allowed their
disc-dominated structures to survive, e.g.~minor mergers; or ii)~they are a sub-population of disc galaxies
that have experienced gas stripping/removal (e.g.\ via AGN/stellar feedback or environmental processes)
and a subsequent disc fading.  Since the PSBs at this epoch are typically of low mass
\mbox{($M_* < 10^{10}\rm\,M_{\odot}$)}, such processes would have a strong potential to cause the rapid
quenching of star formation, necessary to produce the characteristic PSB spectral features (i.e.~strong
Balmer absorption), without significant structural influence.  Disentangling these quenching scenarios
is beyond the scope of this work, but the role of environment in quenching PSBs at $0.5 < z < 1$ is explored
in detail by the recent study of \cite{Socolovsky_etal:2018}, which also uses the PSBs identified from the
full UDS field.  Furthermore, we note that the lack of excess intermediate--young aged stars (O\,B\,A\,F) in
the central regions of these PSBs might place useful constraints on the quenching process, as it suggests the
resultant star-burst was either very weak, or global in nature.  We shall explore this issue in future work.
Finally, with respect to the potential quenching processes, we note that recent gas measurements for
both local PSBs \citep[$z\lesssim0.1$;][]{French_etal:2015, Rowlands_etal:2015} and two PSBs at higher
redshift \citep[$z\sim0.7$;][]{Suess_etal:2017} suggest that the complete removal or depletion of the
molecular gas reservoir is not necessarily required to terminate star-formation.  We also explore this issue,
and the cold interstellar medium (ISM) content of PSBs in the full UDS field across a wide redshift range
($0.5 < z < 2$), in a forthcoming publication (Rowlands et al., in preparation).

%%%%%%%%%%%%%%%%%%%%%%%%%%%%%%%%%%%%%%%%%%%%%%%%%%%%%%%%%%%%%%%%%%%%%%%%%%%%%%%%%%%%%%%%%%%%%%%%%%%%%%%%%%%%%
\section[]{Conclusions}

\label{Conclusions}

%% Summary of results %%-----------------------------------------------------------------------------------%%
In this study, we present a detailed analysis of the structure of PSBs at $0.5 < z < 2$ using data from the
UDS and CANDELS.  Using a large sample of photometrically-selected PSBs recently identified in the UDS field
\citep{Wild_etal:2016},  we examine the structure of $\sim80$ of these recently-quenched systems, and compare
to a large sample of $\sim2000$ passive and star-forming galaxies.  For our analysis we use a combination of
near-infrared and optical $\mu(r)$ profiles, obtained from CANDELS {\em HST} imaging, which probe both the
old stellar component as well as younger, and more recently formed stellar populations (i.e.\ O\,B\,A\,F
stars).  Using both stacked and individual $\mu(r)$ profiles, various structural analyses have been
performed, including single S{\'e}rsic and multiple component fits, which have revealed significant
differences in the structure of PSBs at different epochs.

%% High-z %%-----------------------------------------------------------------------------------------------%%
At high redshift ($1 < z < 2$), PSBs are typically massive ($M_* > 10^{10}\rm\,M_{\odot}$), ultra compact,
bulge-dominated and have high S{\'e}rsic indices.  In general, the structure of these PSBs differs
significantly from their star-forming progenitors and is very similar to those of the old massive passive
population, but considerably more compact.  These results indicate that these galaxies were quenched in a
relatively violent event (e.g.\ gas-rich major merger or dissipative `protogalactic' collapse) that produced
a very compact, centrally-condensed remnant.  Furthermore, we also find consistent structures for these PSBs
across all the wavebands studied ($I_{\rm\,F814W}$, $J_{\rm\,F125W}$ and $H_{\rm\,F160W}$), regardless of
whether the old stellar component or younger (O\,B\,A\,F) stellar populations are being principally traced.
Our results suggest that for most PSBs at this epoch, any preceding starburst and/or quenching was not
strongly centralised and occurred throughout the stellar distribution (i.e.\ it was global in nature).

%% Low-z %%------------------------------------------------------------------------------------------------%%
In contrast, at lower redshifts ($0.5 < z < 1$), the structure of PSBs is significantly different.  At this
epoch, PSBs are generally of low mass ($M_* < 10^{10}\rm\,M_{\odot}$), and exhibit structures that are still
relatively compact, but disc-dominated and of much lower S{\'e}rsic index than PSBs at $z > 1$.  Their
structures are similar to the low-mass passive population (i.e.\ passive discs), the population into which
they will most likely evolve.  These results suggest that these galaxies have been quenched by a more gentle
process that did not significantly disrupt the stellar distribution, and allowed their disc structures to
survive (e.g.\ environmental processes such as gas stripping and/or minor mergers).  Furthermore, we also
find consistent structures for these PSBs in all the wavebands studied ($V_{\rm\,F606W}$, $I_{\rm\,F814W}$,
$J_{\rm\,F125W}$ and $H_{\rm\,F160W}$), regardless of whether the old stellar component or younger
(O\,B\,A\,F) stellar populations are being principally traced.  Consequently, as with PSBs at $z > 1$, our
results suggest that any preceding starburst and/or quenching was not strongly centralized and occurred
throughout the stellar distribution (i.e.\ globally).

In conclusion, we find that PSBs (i.e.\ recently-quenched galaxies) at $z > 1$ are an intrinsically different
population to those at lower redshifts.  Our results indicate that different quenching routes are active at
different epochs, with the PSB phase being triggered by different evolutionary processes.

%%%%%%%%%%%%%%%%%%%%%%%%%%%%%%%%%%%%%%%%%%%%%%%%%%%%%%%%%%%%%%%%%%%%%%%%%%%%%%%%%%%%%%%%%%%%%%%%%%%%%%%%%%%%%
\section[]{Acknowledgements}

This work is based on observations taken by the CANDELS Multi-Cycle Treasury Program with the NASA/ESA HST,
which is operated by the Association of Universities for Research in Astronomy, Inc., under NASA contract
NAS5-26555.  DTM acknowledges support from STFC.  NAH acknowledges support from STFC through an Ernest
Rutherford Fellowship.

%%%%%%%%%%%%%%%%%%%%%%%%%%%%%%%%%%%%%%%%%%%%%%%%%%%%%%%%%%%%%%%%%%%%%%%%%%%%%%%%%%%%%%%%%%%%%%%%%%%%%%%%%%%%%

\bibliographystyle{mnras} \bibliography{DTM_bibtex}

\begin{thebibliography}{}
\makeatletter
\relax
\def\mn@urlcharsother{\let\do\@makeother \do\$\do\&\do\#\do\^\do\_\do\%\do\~}
\def\mn@doi{\begingroup\mn@urlcharsother \@ifnextchar [ {\mn@doi@}
  {\mn@doi@[]}}
\def\mn@doi@[#1]#2{\def\@tempa{#1}\ifx\@tempa\@empty \href
  {http://dx.doi.org/#2} {doi:#2}\else \href {http://dx.doi.org/#2} {#1}\fi
  \endgroup}
\def\mn@eprint#1#2{\mn@eprint@#1:#2::\@nil}
\def\mn@eprint@arXiv#1{\href {http://arxiv.org/abs/#1} {{\tt arXiv:#1}}}
\def\mn@eprint@dblp#1{\href {http://dblp.uni-trier.de/rec/bibtex/#1.xml}
  {dblp:#1}}
\def\mn@eprint@#1:#2:#3:#4\@nil{\def\@tempa {#1}\def\@tempb {#2}\def\@tempc
  {#3}\ifx \@tempc \@empty \let \@tempc \@tempb \let \@tempb \@tempa \fi \ifx
  \@tempb \@empty \def\@tempb {arXiv}\fi \@ifundefined
  {mn@eprint@\@tempb}{\@tempb:\@tempc}{\expandafter \expandafter \csname
  mn@eprint@\@tempb\endcsname \expandafter{\@tempc}}}

\bibitem[\protect\citeauthoryear{{Almaini} et~al.}{{Almaini}
  et~al.}{2017}]{Almaini_etal:2017}
{Almaini} O.,  et~al., 2017, \mn@doi [\mnras] {10.1093/mnras/stx1957}, \href
  {http://adsabs.harvard.edu/abs/2017MNRAS.472.1401A} {472, 1401}

\bibitem[\protect\citeauthoryear{{Athanassoula}, {Rodionov}, {Peschken}  \&
  {Lambert}}{{Athanassoula} et~al.}{2016}]{Athanassoula_etal:2016}
{Athanassoula} E.,  {Rodionov} S.~A.,  {Peschken} N.,   {Lambert} J.~C.,  2016,
  \mn@doi [\apj] {10.3847/0004-637X/821/2/90}, \href
  {http://adsabs.harvard.edu/abs/2016ApJ...821...90A} {821, 90}

\bibitem[\protect\citeauthoryear{{Bell} et~al.}{{Bell}
  et~al.}{2004}]{Bell_etal:2004}
{Bell} E.~F.,  et~al., 2004, \mn@doi [\apj] {10.1086/420778}, \href
  {http://ukads.nottingham.ac.uk/abs/2004ApJ...608..752B} {608, 752}

\bibitem[\protect\citeauthoryear{{Belli}, {Newman}  \& {Ellis}}{{Belli}
  et~al.}{2015}]{Belli_etal:2015}
{Belli} S.,  {Newman} A.~B.,   {Ellis} R.~S.,  2015, \mn@doi [\apj]
  {10.1088/0004-637X/799/2/206}, \href
  {http://adsabs.harvard.edu/abs/2015ApJ...799..206B} {799, 206}

\bibitem[\protect\citeauthoryear{{Bertin} \& {Arnouts}}{{Bertin} \&
  {Arnouts}}{1996}]{Bertin&Arnouts:1996}
{Bertin} E.,  {Arnouts} S.,  1996, \mn@doi [\aaps] {10.1051/aas:1996164}, \href
  {http://adsabs.harvard.edu/abs/1996A%26AS..117..393B} {117, 393}

\bibitem[\protect\citeauthoryear{{Best}, {Kauffmann}, {Heckman}, {Brinchmann},
  {Charlot}, {Ivezi{\'c}}  \& {White}}{{Best} et~al.}{2005}]{Best_etal:2005}
{Best} P.~N.,  {Kauffmann} G.,  {Heckman} T.~M.,  {Brinchmann} J.,  {Charlot}
  S.,  {Ivezi{\'c}} {\v Z}.,   {White} S.~D.~M.,  2005, \mn@doi [\mnras]
  {10.1111/j.1365-2966.2005.09192.x}, \href
  {http://adsabs.harvard.edu/abs/2005MNRAS.362...25B} {362, 25}

\bibitem[\protect\citeauthoryear{{Best}, {Kaiser}, {Heckman}  \&
  {Kauffmann}}{{Best} et~al.}{2006}]{Best_etal:2006}
{Best} P.~N.,  {Kaiser} C.~R.,  {Heckman} T.~M.,   {Kauffmann} G.,  2006,
  \mn@doi [\mnras] {10.1111/j.1745-3933.2006.00159.x}, \href
  {http://adsabs.harvard.edu/abs/2006MNRAS.368L..67B} {368, L67}

\bibitem[\protect\citeauthoryear{{Bezanson}, {van Dokkum}, {van de Sande},
  {Franx}  \& {Kriek}}{{Bezanson} et~al.}{2013}]{Bezanson_etal:2013}
{Bezanson} R.,  {van Dokkum} P.,  {van de Sande} J.,  {Franx} M.,   {Kriek} M.,
   2013, \mn@doi [\apjl] {10.1088/2041-8205/764/1/L8}, \href
  {http://adsabs.harvard.edu/abs/2013ApJ...764L...8B} {764, L8}

\bibitem[\protect\citeauthoryear{{Blake} et~al.}{{Blake}
  et~al.}{2004}]{Blake_etal:2004}
{Blake} C.,  et~al., 2004, \mn@doi [\mnras] {10.1111/j.1365-2966.2004.08351.x},
  \href {http://adsabs.harvard.edu/abs/2004MNRAS.355..713B} {355, 713}

\bibitem[\protect\citeauthoryear{{Brammer} et~al.}{{Brammer}
  et~al.}{2011}]{Brammer_etal:2011}
{Brammer} G.~B.,  et~al., 2011, \mn@doi [\apj] {10.1088/0004-637X/739/1/24},
  \href {http://adsabs.harvard.edu/abs/2011ApJ...739...24B} {739, 24}

\bibitem[\protect\citeauthoryear{{Bruce} et~al.}{{Bruce}
  et~al.}{2014}]{Bruce_etal:2014b}
{Bruce} V.~A.,  et~al., 2014, \mn@doi [\mnras] {10.1093/mnras/stu1537}, \href
  {http://adsabs.harvard.edu/abs/2014MNRAS.444.1660B} {444, 1660}

\bibitem[\protect\citeauthoryear{{Bruzual} \& {Charlot}}{{Bruzual} \&
  {Charlot}}{2003}]{Bruzual&Charlot:2003}
{Bruzual} G.,  {Charlot} S.,  2003, \mn@doi [\mnras]
  {10.1046/j.1365-8711.2003.06897.x}, \href
  {http://adsabs.harvard.edu/abs/2003MNRAS.344.1000B} {344, 1000}

\bibitem[\protect\citeauthoryear{{Buitrago}, {Trujillo}, {Conselice}  \&
  {H{\"a}u{\ss}ler}}{{Buitrago} et~al.}{2013}]{Buitrago_etal:2013}
{Buitrago} F.,  {Trujillo} I.,  {Conselice} C.~J.,   {H{\"a}u{\ss}ler} B.,
  2013, \mn@doi [\mnras] {10.1093/mnras/sts124}, \href
  {http://adsabs.harvard.edu/abs/2013MNRAS.428.1460B} {428, 1460}

\bibitem[\protect\citeauthoryear{{Caldwell}, {Rose}  \& {Dendy}}{{Caldwell}
  et~al.}{1999}]{Caldwell_etal:1999}
{Caldwell} N.,  {Rose} J.~A.,   {Dendy} K.,  1999, \mn@doi [\aj]
  {10.1086/300679}, \href {http://adsabs.harvard.edu/abs/1999AJ....117..140C}
  {117, 140}

\bibitem[\protect\citeauthoryear{{Carollo} et~al.,}{{Carollo}
  et~al.}{2013}]{Carollo_etal:2013}
{Carollo} C.~M.,  et~al., 2013, \mn@doi [\apj] {10.1088/0004-637X/773/2/112},
  \href {http://adsabs.harvard.edu/abs/2013ApJ...773..112C} {773, 112}

\bibitem[\protect\citeauthoryear{{Chabrier}}{{Chabrier}}{2003}]{Chabrier:2003}
{Chabrier} G.,  2003, \mn@doi [\pasp] {10.1086/376392}, \href
  {http://adsabs.harvard.edu/abs/2003PASP..115..763C} {115, 763}

\bibitem[\protect\citeauthoryear{{Charlot} \& {Fall}}{{Charlot} \&
  {Fall}}{2000}]{Charlot&Fall:2000}
{Charlot} S.,  {Fall} S.~M.,  2000, \mn@doi [\apj] {10.1086/309250}, \href
  {http://adsabs.harvard.edu/abs/2000ApJ...539..718C} {539, 718}

\bibitem[\protect\citeauthoryear{{Cirasuolo} et~al.}{{Cirasuolo}
  et~al.}{2007}]{Cirasuolo_etal:2007}
{Cirasuolo} M.,  et~al., 2007, \mn@doi [\mnras]
  {10.1111/j.1365-2966.2007.12038.x}, \href
  {http://adsabs.harvard.edu/abs/2007MNRAS.380..585C} {380, 585}

\bibitem[\protect\citeauthoryear{{Dekel} \& {Birnboim}}{{Dekel} \&
  {Birnboim}}{2006}]{Dekel&Birnboim:2006}
{Dekel} A.,  {Birnboim} Y.,  2006, \mn@doi [\mnras]
  {10.1111/j.1365-2966.2006.10145.x}, \href
  {http://adsabs.harvard.edu/abs/2006MNRAS.368....2D} {368, 2}

\bibitem[\protect\citeauthoryear{{Dekel} et~al.}{{Dekel}
  et~al.}{2009}]{Dekel_etal:2009}
{Dekel} A.,  et~al., 2009, \mn@doi [\nat] {10.1038/nature07648}, \href
  {http://adsabs.harvard.edu/abs/2009Natur.457..451D} {457, 451}

\bibitem[\protect\citeauthoryear{{Diamond-Stanic}, {Moustakas}, {Tremonti},
  {Coil}, {Hickox}, {Robaina}, {Rudnick}  \& {Sell}}{{Diamond-Stanic}
  et~al.}{2012}]{Diamond-Stanic_etal:2012}
{Diamond-Stanic} A.~M.,  {Moustakas} J.,  {Tremonti} C.~A.,  {Coil} A.~L.,
  {Hickox} R.~C.,  {Robaina} A.~R.,  {Rudnick} G.~H.,   {Sell} P.~H.,  2012,
  \mn@doi [\apjl] {10.1088/2041-8205/755/2/L26}, \href
  {http://adsabs.harvard.edu/abs/2012ApJ...755L..26D} {755, L26}

\bibitem[\protect\citeauthoryear{{Dressler} \& {Gunn}}{{Dressler} \&
  {Gunn}}{1983}]{Dressler&Gunn:1983}
{Dressler} A.,  {Gunn} J.~E.,  1983, \mn@doi [\apj] {10.1086/161093}, \href
  {http://adsabs.harvard.edu/abs/1983ApJ...270....7D} {270, 7}

\bibitem[\protect\citeauthoryear{{Dressler}, {Smail}, {Poggianti}, {Butcher},
  {Couch}, {Ellis}  \& {Oemler}}{{Dressler} et~al.}{1999}]{Dressler_etal:1999}
{Dressler} A.,  {Smail} I.,  {Poggianti} B.~M.,  {Butcher} H.,  {Couch} W.~J.,
  {Ellis} R.~S.,   {Oemler} Jr. A.,  1999, \mn@doi [\apjs] {10.1086/313213},
  \href {http://adsabs.harvard.edu/abs/1999ApJS..122...51D} {122, 51}

\bibitem[\protect\citeauthoryear{{Faber} et~al.}{{Faber}
  et~al.}{2007}]{Faber_etal:2007}
{Faber} S.~M.,  et~al., 2007, \mn@doi [\apj] {10.1086/519294}, \href
  {http://adsabs.harvard.edu/abs/2007ApJ...665..265F} {665, 265}

\bibitem[\protect\citeauthoryear{{Fisher} \& {Drory}}{{Fisher} \&
  {Drory}}{2011}]{Fisher&Drory:2011}
{Fisher} D.~B.,  {Drory} N.,  2011, \mn@doi [\apjl]
  {10.1088/2041-8205/733/2/L47}, \href
  {http://adsabs.harvard.edu/abs/2011ApJ...733L..47F} {733, L47}

\bibitem[\protect\citeauthoryear{{French}, {Yang}, {Zabludoff}, {Narayanan},
  {Shirley}, {Walter}, {Smith}  \& {Tremonti}}{{French}
  et~al.}{2015}]{French_etal:2015}
{French} K.~D.,  {Yang} Y.,  {Zabludoff} A.,  {Narayanan} D.,  {Shirley} Y.,
  {Walter} F.,  {Smith} J.-D.,   {Tremonti} C.~A.,  2015, \mn@doi [\apj]
  {10.1088/0004-637X/801/1/1}, \href
  {http://adsabs.harvard.edu/abs/2015ApJ...801....1F} {801, 1}

\bibitem[\protect\citeauthoryear{{Furusawa} et~al.}{{Furusawa}
  et~al.}{2008}]{Furusawa_etal:2008}
{Furusawa} H.,  et~al., 2008, \mn@doi [\apjs] {10.1086/527321}, \href
  {http://adsabs.harvard.edu/abs/2008ApJS..176....1F} {176, 1}

\bibitem[\protect\citeauthoryear{{Galametz} et~al.,}{{Galametz}
  et~al.}{2018}]{Galametz_etal:2018}
{Galametz} A.,  et~al., 2018, \mn@doi [\mnras] {10.1093/mnras/sty095}, \href
  {http://adsabs.harvard.edu/abs/2018MNRAS.475.4148G} {475, 4148}

\bibitem[\protect\citeauthoryear{{Goto}}{{Goto}}{2007}]{Goto:2007}
{Goto} T.,  2007, \mn@doi [\mnras] {10.1111/j.1365-2966.2007.12227.x}, \href
  {http://adsabs.harvard.edu/abs/2007MNRAS.381..187G} {381, 187}

\bibitem[\protect\citeauthoryear{{Grogin} et~al.}{{Grogin}
  et~al.}{2011}]{Grogin_etal:2011}
{Grogin} N.~A.,  et~al., 2011, \mn@doi [\apjs] {10.1088/0067-0049/197/2/35},
  \href {http://adsabs.harvard.edu/abs/2011ApJS..197...35G} {197, 35}

\bibitem[\protect\citeauthoryear{{Gunn} \& {Gott}}{{Gunn} \&
  {Gott}}{1972}]{Gunn&Gott:1972}
{Gunn} J.~E.,  {Gott} III J.~R.,  1972, \mn@doi [\apj] {10.1086/151605}, \href
  {http://adsabs.harvard.edu/abs/1972ApJ...176....1G} {176, 1}

\bibitem[\protect\citeauthoryear{{Hartley} et~al.}{{Hartley}
  et~al.}{2013}]{Hartley_etal:2013}
{Hartley} W.~G.,  et~al., 2013, \mn@doi [\mnras] {10.1093/mnras/stt383}, \href
  {http://adsabs.harvard.edu/abs/2013MNRAS.431.3045H} {431, 3045}

\bibitem[\protect\citeauthoryear{{Hopkins}}{{Hopkins}}{2012}]{Hopkins:2012}
{Hopkins} P.~F.,  2012, \mn@doi [\mnras] {10.1111/j.1745-3933.2011.01179.x},
  \href {http://adsabs.harvard.edu/abs/2012MNRAS.420L...8H} {420, L8}

\bibitem[\protect\citeauthoryear{{Hopkins}, {Hernquist}, {Cox}, {Di Matteo},
  {Martini}, {Robertson}  \& {Springel}}{{Hopkins}
  et~al.}{2005}]{Hopkins_etal:2005}
{Hopkins} P.~F.,  {Hernquist} L.,  {Cox} T.~J.,  {Di Matteo} T.,  {Martini} P.,
   {Robertson} B.,   {Springel} V.,  2005, \mn@doi [\apj] {10.1086/432438},
  \href {http://adsabs.harvard.edu/abs/2005ApJ...630..705H} {630, 705}

\bibitem[\protect\citeauthoryear{{Hopkins}, {Cox}, {Younger}  \&
  {Hernquist}}{{Hopkins} et~al.}{2009}]{Hopkins_etal:2009}
{Hopkins} P.~F.,  {Cox} T.~J.,  {Younger} J.~D.,   {Hernquist} L.,  2009,
  \mn@doi [\apj] {10.1088/0004-637X/691/2/1168}, \href
  {http://adsabs.harvard.edu/abs/2009ApJ...691.1168H} {691, 1168}

\bibitem[\protect\citeauthoryear{{Jedrzejewski}}{{Jedrzejewski}}{1987}]{Jedrzejewski:1987}
{Jedrzejewski} R.~I.,  1987, \mnras, \href
  {http://adsabs.harvard.edu/abs/1987MNRAS.226..747J} {226, 747}

\bibitem[\protect\citeauthoryear{{Keating}, {Abraham}, {Schiavon}, {Graves},
  {Damjanov}, {Yan}, {Newman}  \& {Simard}}{{Keating}
  et~al.}{2015}]{Keating_etal:2015}
{Keating} S.~K.,  {Abraham} R.~G.,  {Schiavon} R.,  {Graves} G.,  {Damjanov}
  I.,  {Yan} R.,  {Newman} J.,   {Simard} L.,  2015, \mn@doi [\apj]
  {10.1088/0004-637X/798/1/26}, \href
  {http://adsabs.harvard.edu/abs/2015ApJ...798...26K} {798, 26}

\bibitem[\protect\citeauthoryear{{Koekemoer} et~al.}{{Koekemoer}
  et~al.}{2011}]{Koekemoer_etal:2011}
{Koekemoer} A.~M.,  et~al., 2011, \mn@doi [\apjs] {10.1088/0067-0049/197/2/36},
  \href {http://adsabs.harvard.edu/abs/2011ApJS..197...36K} {197, 36}

\bibitem[\protect\citeauthoryear{{Kron}}{{Kron}}{1980}]{Kron:1980}
{Kron} R.~G.,  1980, \mn@doi [\apjs] {10.1086/190669}, \href
  {http://ukads.nottingham.ac.uk/abs/1980ApJS...43..305K} {43, 305}

\bibitem[\protect\citeauthoryear{{Lani} et~al.}{{Lani}
  et~al.}{2013}]{Lani_etal:2013}
{Lani} C.,  et~al., 2013, \mn@doi [\mnras] {10.1093/mnras/stt1275}, \href
  {http://adsabs.harvard.edu/abs/2013MNRAS.435..207L} {435, 207}

\bibitem[\protect\citeauthoryear{{Larson}, {Tinsley}  \& {Caldwell}}{{Larson}
  et~al.}{1980}]{Larson_etal:1980}
{Larson} R.~B.,  {Tinsley} B.~M.,   {Caldwell} C.~N.,  1980, \mn@doi [\apj]
  {10.1086/157917}, \href {http://adsabs.harvard.edu/abs/1980ApJ...237..692L}
  {237, 692}

\bibitem[\protect\citeauthoryear{{Lawrence} et~al.}{{Lawrence}
  et~al.}{2007}]{Lawrence_etal:2007}
{Lawrence} A.,  et~al., 2007, \mn@doi [\mnras]
  {10.1111/j.1365-2966.2007.12040.x}, \href
  {http://adsabs.harvard.edu/abs/2007MNRAS.379.1599L} {379, 1599}

\bibitem[\protect\citeauthoryear{{MacArthur}, {Courteau}  \&
  {Holtzman}}{{MacArthur} et~al.}{2003}]{MacArthur_etal:2003}
{MacArthur} L.~A.,  {Courteau} S.,   {Holtzman} J.~A.,  2003, \mn@doi [\apj]
  {10.1086/344506}, \href
  {http://ukads.nottingham.ac.uk/abs/2003ApJ...582..689M} {582, 689}

\bibitem[\protect\citeauthoryear{{Maltby} et~al.}{{Maltby}
  et~al.}{2012a}]{Maltby_etal:2012a}
{Maltby} D.~T.,  et~al., 2012a, \mn@doi [\mnras]
  {10.1111/j.1365-2966.2011.19727.x}, \href
  {http://adsabs.harvard.edu/abs/2012MNRAS.419..669M} {419, 669}

\bibitem[\protect\citeauthoryear{{Maltby}, {Hoyos}, {Gray},
  {Arag{\'o}n-Salamanca}  \& {Wolf}}{{Maltby}
  et~al.}{2012b}]{Maltby_etal:2012b}
{Maltby} D.~T.,  {Hoyos} C.,  {Gray} M.~E.,  {Arag{\'o}n-Salamanca} A.,
  {Wolf} C.,  2012b, \mn@doi [\mnras] {10.1111/j.1365-2966.2011.20211.x}, \href
  {http://adsabs.harvard.edu/abs/2012MNRAS.420.2475M} {420, 2475}

\bibitem[\protect\citeauthoryear{{Maltby}, {Arag{\'o}n-Salamanca}, {Gray},
  {Hoyos}, {Wolf}, {Jogee}  \& {B{\"o}hm}}{{Maltby}
  et~al.}{2015}]{Maltby_etal:2015}
{Maltby} D.~T.,  {Arag{\'o}n-Salamanca} A.,  {Gray} M.~E.,  {Hoyos} C.,  {Wolf}
  C.,  {Jogee} S.,   {B{\"o}hm} A.,  2015, \mn@doi [\mnras]
  {10.1093/mnras/stu2536}, \href
  {http://adsabs.harvard.edu/abs/2015MNRAS.447.1506M} {447, 1506}

\bibitem[\protect\citeauthoryear{{Maltby} et~al.}{{Maltby}
  et~al.}{2016}]{Maltby_etal:2016}
{Maltby} D.~T.,  et~al., 2016, \mn@doi [\mnras] {10.1093/mnrasl/slw057}, \href
  {http://adsabs.harvard.edu/abs/2016MNRAS.459L.114M} {459, L114}

\bibitem[\protect\citeauthoryear{{Martig}, {Bournaud}, {Teyssier}  \&
  {Dekel}}{{Martig} et~al.}{2009}]{Martig_etal:2009}
{Martig} M.,  {Bournaud} F.,  {Teyssier} R.,   {Dekel} A.,  2009, \mn@doi
  [\apj] {10.1088/0004-637X/707/1/250}, \href
  {http://adsabs.harvard.edu/abs/2009ApJ...707..250M} {707, 250}

\bibitem[\protect\citeauthoryear{{McDonald}, {Courteau}, {Tully}  \&
  {Roediger}}{{McDonald} et~al.}{2011}]{McDonald_etal:2011}
{McDonald} M.,  {Courteau} S.,  {Tully} R.~B.,   {Roediger} J.,  2011, \mn@doi
  [\mnras] {10.1111/j.1365-2966.2011.18519.x}, \href
  {http://ukads.nottingham.ac.uk/abs/2011MNRAS.414.2055M} {414, 2055}

\bibitem[\protect\citeauthoryear{{Mortlock} et~al.}{{Mortlock}
  et~al.}{2013}]{Mortlock_etal:2013}
{Mortlock} A.,  et~al., 2013, \mn@doi [\mnras] {10.1093/mnras/stt793}, \href
  {http://adsabs.harvard.edu/abs/2013MNRAS.433.1185M} {433, 1185}

\bibitem[\protect\citeauthoryear{{Muzzin} et~al.}{{Muzzin}
  et~al.}{2013}]{Muzzin_etal:2013}
{Muzzin} A.,  et~al., 2013, \mn@doi [\apj] {10.1088/0004-637X/777/1/18}, \href
  {http://adsabs.harvard.edu/abs/2013ApJ...777...18M} {777, 18}

\bibitem[\protect\citeauthoryear{{Naab}, {Johansson}  \& {Ostriker}}{{Naab}
  et~al.}{2009}]{Naab_etal:2009}
{Naab} T.,  {Johansson} P.~H.,   {Ostriker} J.~P.,  2009, \mn@doi [\apjl]
  {10.1088/0004-637X/699/2/L178}, \href
  {http://adsabs.harvard.edu/abs/2009ApJ...699L.178N} {699, L178}

\bibitem[\protect\citeauthoryear{{Newman}, {Belli}  \& {Ellis}}{{Newman}
  et~al.}{2015}]{Newman_etal:2015}
{Newman} A.~B.,  {Belli} S.,   {Ellis} R.~S.,  2015, \mn@doi [\apjl]
  {10.1088/2041-8205/813/1/L7}, \href
  {http://adsabs.harvard.edu/abs/2015ApJ...813L...7N} {813, L7}

\bibitem[\protect\citeauthoryear{{Norton}, {Gebhardt}, {Zabludoff}  \&
  {Zaritsky}}{{Norton} et~al.}{2001}]{Norton_etal:2001}
{Norton} S.~A.,  {Gebhardt} K.,  {Zabludoff} A.~I.,   {Zaritsky} D.,  2001,
  \mn@doi [\apj] {10.1086/321668}, \href
  {http://adsabs.harvard.edu/abs/2001ApJ...557..150N} {557, 150}

\bibitem[\protect\citeauthoryear{{Pawlik}, {Wild}, {Walcher}, {Johansson},
  {Villforth}, {Rowlands}, {Mendez-Abreu}  \& {Hewlett}}{{Pawlik}
  et~al.}{2016}]{Pawlik_etal:2016}
{Pawlik} M.~M.,  {Wild} V.,  {Walcher} C.~J.,  {Johansson} P.~H.,  {Villforth}
  C.,  {Rowlands} K.,  {Mendez-Abreu} J.,   {Hewlett} T.,  2016, \mn@doi
  [\mnras] {10.1093/mnras/stv2878}, \href
  {http://adsabs.harvard.edu/abs/2016MNRAS.456.3032P} {456, 3032}

\bibitem[\protect\citeauthoryear{{Pawlik} et~al.}{{Pawlik}
  et~al.}{2018}]{Pawlik_etal:2018}
{Pawlik} M.~M.,  et~al., 2018, \mn@doi [\mnras] {10.1093/mnras/sty589}, \href
  {http://adsabs.harvard.edu/abs/2018MNRAS.477.1708P} {477, 1708}

\bibitem[\protect\citeauthoryear{{Peng}, {Ho}, {Impey}  \& {Rix}}{{Peng}
  et~al.}{2002}]{Peng_etal:2002}
{Peng} C.~Y.,  {Ho} L.~C.,  {Impey} C.~D.,   {Rix} H.-W.,  2002, \mn@doi [\aj]
  {10.1086/340952}, \href
  {http://ukads.nottingham.ac.uk/abs/2002AJ....124..266P} {124, 266}

\bibitem[\protect\citeauthoryear{{Poggianti} et~al.}{{Poggianti}
  et~al.}{2009}]{Poggianti_etal:2009}
{Poggianti} B.~M.,  et~al., 2009, \mn@doi [\apj] {10.1088/0004-637X/693/1/112},
  \href {http://adsabs.harvard.edu/abs/2009ApJ...693..112P} {693, 112}

\bibitem[\protect\citeauthoryear{{Pozzetti} et~al.}{{Pozzetti}
  et~al.}{2010}]{Pozzetti_etal:2010}
{Pozzetti} L.,  et~al., 2010, \mn@doi [\aap] {10.1051/0004-6361/200913020},
  \href {http://adsabs.harvard.edu/abs/2010A%26A...523A..13P} {523, A13}

\bibitem[\protect\citeauthoryear{{Pracy} et~al.}{{Pracy}
  et~al.}{2013}]{Pracy_etal:2013}
{Pracy} M.~B.,  et~al., 2013, \mn@doi [\mnras] {10.1093/mnras/stt666}, \href
  {http://adsabs.harvard.edu/abs/2013MNRAS.432.3131P} {432, 3131}

\bibitem[\protect\citeauthoryear{{Rowlands}, {Wild}, {Nesvadba}, {Sibthorpe},
  {Mortier}, {Lehnert}  \& {da Cunha}}{{Rowlands}
  et~al.}{2015}]{Rowlands_etal:2015}
{Rowlands} K.,  {Wild} V.,  {Nesvadba} N.,  {Sibthorpe} B.,  {Mortier} A.,
  {Lehnert} M.,   {da Cunha} E.,  2015, \mn@doi [\mnras]
  {10.1093/mnras/stu2714}, \href
  {http://adsabs.harvard.edu/abs/2015MNRAS.448..258R} {448, 258}

\bibitem[\protect\citeauthoryear{{Shen}, {Mo}, {White}, {Blanton}, {Kauffmann},
  {Voges}, {Brinkmann}  \& {Csabai}}{{Shen} et~al.}{2003}]{Shen_etal:2003}
{Shen} S.,  {Mo} H.~J.,  {White} S.~D.~M.,  {Blanton} M.~R.,  {Kauffmann} G.,
  {Voges} W.,  {Brinkmann} J.,   {Csabai} I.,  2003, \mn@doi [\mnras]
  {10.1046/j.1365-8711.2003.06740.x}, \href
  {http://ukads.nottingham.ac.uk/abs/2003MNRAS.343..978S} {343, 978}

\bibitem[\protect\citeauthoryear{{Silk} \& {Rees}}{{Silk} \&
  {Rees}}{1998}]{Silk&Rees:1998}
{Silk} J.,  {Rees} M.~J.,  1998, \aap, \href
  {http://adsabs.harvard.edu/abs/1998A%26A...331L...1S} {331, L1}

\bibitem[\protect\citeauthoryear{{Simpson} et~al.}{{Simpson}
  et~al.}{2012}]{Simpson_etal:2012}
{Simpson} C.,  et~al., 2012, \mn@doi [\mnras]
  {10.1111/j.1365-2966.2012.20529.x}, \href
  {http://adsabs.harvard.edu/abs/2012MNRAS.421.3060S} {421, 3060}

\bibitem[\protect\citeauthoryear{{Simpson}, {Westoby}, {Arumugam}, {Ivison},
  {Hartley}  \& {Almaini}}{{Simpson} et~al.}{2013}]{Simpson_etal:2013}
{Simpson} C.,  {Westoby} P.,  {Arumugam} V.,  {Ivison} R.,  {Hartley} W.,
  {Almaini} O.,  2013, \mn@doi [\mnras] {10.1093/mnras/stt940}, \href
  {http://adsabs.harvard.edu/abs/2013MNRAS.433.2647S} {433, 2647}

\bibitem[\protect\citeauthoryear{{Socolovsky}, {Almaini}, {Hatch}, {Wild},
  {Maltby}, {Hartley}  \& {Simpson}}{{Socolovsky}
  et~al.}{2018}]{Socolovsky_etal:2018}
{Socolovsky} M.,  {Almaini} O.,  {Hatch} N.~A.,  {Wild} V.,  {Maltby} D.~T.,
  {Hartley} W.~G.,   {Simpson} C.,  2018, \mn@doi [\mnras]
  {10.1093/mnras/sty312}, \href
  {http://adsabs.harvard.edu/abs/2018MNRAS.476.1242S} {476, 1242}

\bibitem[\protect\citeauthoryear{{Strateva} et~al.}{{Strateva}
  et~al.}{2001}]{Strateva_etal:2001}
{Strateva} I.,  et~al., 2001, \mn@doi [\aj] {10.1086/323301}, \href
  {http://ukads.nottingham.ac.uk/abs/2001AJ....122.1861S} {122, 1861}

\bibitem[\protect\citeauthoryear{{Suess}, {Bezanson}, {Spilker}, {Kriek},
  {Greene}, {Feldmann}, {Hunt}  \& {Narayanan}}{{Suess}
  et~al.}{2017}]{Suess_etal:2017}
{Suess} K.~A.,  {Bezanson} R.,  {Spilker} J.~S.,  {Kriek} M.,  {Greene} J.~E.,
  {Feldmann} R.,  {Hunt} Q.,   {Narayanan} D.,  2017, \mn@doi [\apjl]
  {10.3847/2041-8213/aa85dc}, \href
  {http://adsabs.harvard.edu/abs/2017ApJ...846L..14S} {846, L14}

\bibitem[\protect\citeauthoryear{{Szomoru} et~al.,}{{Szomoru}
  et~al.}{2010}]{Szomoru_etal:2010}
{Szomoru} D.,  et~al., 2010, \mn@doi [\apjl] {10.1088/2041-8205/714/2/L244},
  \href {http://adsabs.harvard.edu/abs/2010ApJ...714L.244S} {714, L244}

\bibitem[\protect\citeauthoryear{{Szomoru}, {Franx}  \& {van Dokkum}}{{Szomoru}
  et~al.}{2012}]{Szomoru_etal:2012}
{Szomoru} D.,  {Franx} M.,   {van Dokkum} P.~G.,  2012, \mn@doi [\apj]
  {10.1088/0004-637X/749/2/121}, \href
  {http://adsabs.harvard.edu/abs/2012ApJ...749..121S} {749, 121}

\bibitem[\protect\citeauthoryear{{Tacchella}, {Dekel}, {Carollo}, {Ceverino},
  {DeGraf}, {Lapiner}, {Mandelker}  \& {Primack}}{{Tacchella}
  et~al.}{2016}]{Tacchella_etal:2016}
{Tacchella} S.,  {Dekel} A.,  {Carollo} C.~M.,  {Ceverino} D.,  {DeGraf} C.,
  {Lapiner} S.,  {Mandelker} N.,   {Primack} J.~R.,  2016, \mn@doi [\mnras]
  {10.1093/mnras/stw303}, \href
  {http://adsabs.harvard.edu/abs/2016MNRAS.458..242T} {458, 242}

\bibitem[\protect\citeauthoryear{{Tran}, {Franx}, {Illingworth}, {Kelson}  \&
  {van Dokkum}}{{Tran} et~al.}{2003}]{Tran_etal:2003}
{Tran} K.-V.~H.,  {Franx} M.,  {Illingworth} G.,  {Kelson} D.~D.,   {van
  Dokkum} P.,  2003, \mn@doi [\apj] {10.1086/379804}, \href
  {http://adsabs.harvard.edu/abs/2003ApJ...599..865T} {599, 865}

\bibitem[\protect\citeauthoryear{{Trujillo} et~al.}{{Trujillo}
  et~al.}{2006}]{Trujillo_etal:2006}
{Trujillo} I.,  et~al., 2006, \mn@doi [\mnras]
  {10.1111/j.1745-3933.2006.00238.x}, \href
  {http://adsabs.harvard.edu/abs/2006MNRAS.373L..36T} {373, L36}

\bibitem[\protect\citeauthoryear{{Vergani} et~al.}{{Vergani}
  et~al.}{2010}]{Vergani_etal:2010}
{Vergani} D.,  et~al., 2010, \mn@doi [\aap] {10.1051/0004-6361/200912802},
  \href {http://adsabs.harvard.edu/abs/2010A%26A...509A..42V} {509, A42}

\bibitem[\protect\citeauthoryear{{Wellons} et~al.}{{Wellons}
  et~al.}{2015}]{Wellons_etal:2015}
{Wellons} S.,  et~al., 2015, \mn@doi [\mnras] {10.1093/mnras/stv303}, \href
  {http://adsabs.harvard.edu/abs/2015MNRAS.449..361W} {449, 361}

\bibitem[\protect\citeauthoryear{{Whitaker}, {Kriek}, {van Dokkum}, {Bezanson},
  {Brammer}, {Franx}  \& {Labb{\'e}}}{{Whitaker}
  et~al.}{2012}]{Whitaker_etal:2012a}
{Whitaker} K.~E.,  {Kriek} M.,  {van Dokkum} P.~G.,  {Bezanson} R.,  {Brammer}
  G.,  {Franx} M.,   {Labb{\'e}} I.,  2012, \mn@doi [\apj]
  {10.1088/0004-637X/745/2/179}, \href
  {http://adsabs.harvard.edu/abs/2012ApJ...745..179W} {745, 179}

\bibitem[\protect\citeauthoryear{{Whitaker} et~al.}{{Whitaker}
  et~al.}{2013}]{Whitaker_etal:2013}
{Whitaker} K.~E.,  et~al., 2013, \mn@doi [\apjl] {10.1088/2041-8205/770/2/L39},
  \href {http://adsabs.harvard.edu/abs/2013ApJ...770L..39W} {770, L39}

\bibitem[\protect\citeauthoryear{{Wild}, {Walcher}, {Johansson}, {Tresse},
  {Charlot}, {Pollo}, {Le F{\`e}vre}  \& {de Ravel}}{{Wild}
  et~al.}{2009}]{Wild_etal:2009}
{Wild} V.,  {Walcher} C.~J.,  {Johansson} P.~H.,  {Tresse} L.,  {Charlot} S.,
  {Pollo} A.,  {Le F{\`e}vre} O.,   {de Ravel} L.,  2009, \mn@doi [\mnras]
  {10.1111/j.1365-2966.2009.14537.x}, \href
  {http://adsabs.harvard.edu/abs/2009MNRAS.395..144W} {395, 144}

\bibitem[\protect\citeauthoryear{{Wild} et~al.}{{Wild}
  et~al.}{2011}]{Wild_etal:2011}
{Wild} V.,  et~al., 2011, \mn@doi [\mnras] {10.1111/j.1365-2966.2010.17536.x},
  \href {http://adsabs.harvard.edu/abs/2011MNRAS.410.1593W} {410, 1593}

\bibitem[\protect\citeauthoryear{{Wild} et~al.}{{Wild}
  et~al.}{2014}]{Wild_etal:2014}
{Wild} V.,  et~al., 2014, \mn@doi [\mnras] {10.1093/mnras/stu212}, \href
  {http://adsabs.harvard.edu/abs/2014MNRAS.440.1880W} {440, 1880}

\bibitem[\protect\citeauthoryear{{Wild}, {Almaini}, {Dunlop}, {Simpson},
  {Rowlands}, {Bowler}, {Maltby}  \& {McLure}}{{Wild}
  et~al.}{2016}]{Wild_etal:2016}
{Wild} V.,  {Almaini} O.,  {Dunlop} J.,  {Simpson} C.,  {Rowlands} K.,
  {Bowler} R.,  {Maltby} D.,   {McLure} R.,  2016, \mn@doi [\mnras]
  {10.1093/mnras/stw1996}, \href
  {http://adsabs.harvard.edu/abs/2016MNRAS.463..832W} {463, 832}

\bibitem[\protect\citeauthoryear{{Williams}, {Quadri}, {Franx}, {van Dokkum},
  {Toft}, {Kriek}  \& {Labb{\'e}}}{{Williams}
  et~al.}{2010}]{Williams_etal:2010}
{Williams} R.~J.,  {Quadri} R.~F.,  {Franx} M.,  {van Dokkum} P.,  {Toft} S.,
  {Kriek} M.,   {Labb{\'e}} I.,  2010, \mn@doi [\apj]
  {10.1088/0004-637X/713/2/738}, \href
  {http://adsabs.harvard.edu/abs/2010ApJ...713..738W} {713, 738}

\bibitem[\protect\citeauthoryear{{Williams} et~al.}{{Williams}
  et~al.}{2017}]{Williams_etal:2017}
{Williams} C.~C.,  et~al., 2017, \mn@doi [\apj] {10.3847/1538-4357/aa662f},
  \href {http://adsabs.harvard.edu/abs/2017ApJ...838...94W} {838, 94}

\bibitem[\protect\citeauthoryear{{Wu}, {Gal}, {Lemaux}, {Kocevski}, {Lubin},
  {Rumbaugh}  \& {Squires}}{{Wu} et~al.}{2014}]{Wu_etal:2014}
{Wu} P.-F.,  {Gal} R.~R.,  {Lemaux} B.~C.,  {Kocevski} D.~D.,  {Lubin} L.~M.,
  {Rumbaugh} N.,   {Squires} G.~K.,  2014, \mn@doi [\apj]
  {10.1088/0004-637X/792/1/16}, \href
  {http://adsabs.harvard.edu/abs/2014ApJ...792...16W} {792, 16}

\bibitem[\protect\citeauthoryear{{Yamauchi} \& {Goto}}{{Yamauchi} \&
  {Goto}}{2005}]{Yamauchi&Goto:2005}
{Yamauchi} C.,  {Goto} T.,  2005, \mn@doi [\mnras]
  {10.1111/j.1365-2966.2005.08996.x}, \href
  {http://adsabs.harvard.edu/abs/2005MNRAS.359.1557Y} {359, 1557}

\bibitem[\protect\citeauthoryear{{Yano}, {Kriek}, {van der Wel}  \&
  {Whitaker}}{{Yano} et~al.}{2016}]{Yano_etal:2016}
{Yano} M.,  {Kriek} M.,  {van der Wel} A.,   {Whitaker} K.~E.,  2016, \mn@doi
  [\apjl] {10.3847/2041-8205/817/2/L21}, \href
  {http://adsabs.harvard.edu/abs/2016ApJ...817L..21Y} {817, L21}

\bibitem[\protect\citeauthoryear{{Zabludoff}, {Zaritsky}, {Lin}, {Tucker},
  {Hashimoto}, {Shectman}, {Oemler}  \& {Kirshner}}{{Zabludoff}
  et~al.}{1996}]{Zabludoff_etal:1996}
{Zabludoff} A.~I.,  {Zaritsky} D.,  {Lin} H.,  {Tucker} D.,  {Hashimoto} Y.,
  {Shectman} S.~A.,  {Oemler} A.,   {Kirshner} R.~P.,  1996, \mn@doi [\apj]
  {10.1086/177495}, \href {http://adsabs.harvard.edu/abs/1996ApJ...466..104Z}
  {466, 104}

\bibitem[\protect\citeauthoryear{{Zolotov} et~al.}{{Zolotov}
  et~al.}{2015}]{Zolotov_etal:2015}
{Zolotov} A.,  et~al., 2015, \mn@doi [\mnras] {10.1093/mnras/stv740}, \href
  {http://adsabs.harvard.edu/abs/2015MNRAS.450.2327Z} {450, 2327}

\bibitem[\protect\citeauthoryear{{de Jong}}{{de Jong}}{1996}]{deJong:1996}
{de Jong} R.~S.,  1996, \aaps, \href
  {http://ukads.nottingham.ac.uk/abs/1996A%26AS..118..557D} {118, 557}

\bibitem[\protect\citeauthoryear{{de Vaucouleurs}}{{de
  Vaucouleurs}}{1959}]{deVaucouleurs:1959}
{de Vaucouleurs} G.,  1959, Handbuch der Physik, \href
  {http://adsabs.harvard.edu/abs/1959HDP....53..311D} {53, 311}

\bibitem[\protect\citeauthoryear{{van Breukelen} et~al.}{{van Breukelen}
  et~al.}{2006}]{vanBreukelen_etal:2006}
{van Breukelen} C.,  et~al., 2006, \mn@doi [\mnras]
  {10.1111/j.1745-3933.2006.00236.x}, \href
  {http://adsabs.harvard.edu/abs/2006MNRAS.373L..26V} {373, L26}

\bibitem[\protect\citeauthoryear{{van de Sande} et~al.,}{{van de Sande}
  et~al.}{2013}]{vandeSande_etal:2013}
{van de Sande} J.,  et~al., 2013, \mn@doi [\apj] {10.1088/0004-637X/771/2/85},
  \href {http://adsabs.harvard.edu/abs/2013ApJ...771...85V} {771, 85}

\bibitem[\protect\citeauthoryear{{van der Wel} et~al.}{{van der Wel}
  et~al.}{2011}]{vanderWel_etal:2011}
{van der Wel} A.,  et~al., 2011, \mn@doi [\apj] {10.1088/0004-637X/730/1/38},
  \href {http://adsabs.harvard.edu/abs/2011ApJ...730...38V} {730, 38}

\bibitem[\protect\citeauthoryear{{van der Wel} et~al.}{{van der Wel}
  et~al.}{2012}]{vanderWel_etal:2012}
{van der Wel} A.,  et~al., 2012, \mn@doi [\apjs] {10.1088/0067-0049/203/2/24},
  \href {http://adsabs.harvard.edu/abs/2012ApJS..203...24V} {203, 24}

\bibitem[\protect\citeauthoryear{{van der Wel} et~al.}{{van der Wel}
  et~al.}{2014}]{vanderWel_etal:2014}
{van der Wel} A.,  et~al., 2014, \mn@doi [\apj] {10.1088/0004-637X/788/1/28},
  \href {http://adsabs.harvard.edu/abs/2014ApJ...788...28V} {788, 28}

\makeatother
\end{thebibliography}
\bsp

\label{lastpage}

\end{document}